\documentclass[aps,a4paper, amsfonts, amssymb, amsmath, reprint, showkeys, nofootinbib, twoside]{revtex4-1}

\usepackage[english]{babel}
\usepackage[utf8]{inputenc}
\usepackage[colorinlistoftodos, color=green!40, prependcaption]{todonotes}
\usepackage{graphicx}
\usepackage{bm}

\usepackage{xcolor}

\begin{document}

\title{Longitudinal Modes in Diffusion and Localization of Light}


\author{B.A. van Tiggelen}
\email[]{Bart.Van-Tiggelen@lpmmc.cnrs.fr}
\affiliation{Univ. Grenoble Alpes, CNRS, LPMMC, 38000 Grenoble, France}
\author{S.E. Skipetrov}
\email[]{Sergey.Skipetrov@lpmmc.cnrs.fr}
\affiliation{Univ. Grenoble Alpes, CNRS, LPMMC, 38000 Grenoble, France}

\date{\today} 

\begin{abstract}
In this work we include the elastic scattering of longitudinal electromagnetic waves in transport theory using a medium filled with point-like, electric dipoles.
The interference between longitudinal and transverse waves creates two new channels among which
one allows energy transport. This picture is worked out
by extending the independent scattering framework of radiative transfer to include binary dipole-dipole interactions.
We calculate the diffusion constant of light in the new transport channel
and investigate  the role of longitudinal waves in other aspects of
light diffusion by considering the density of states, equipartition, and Lorentz local field.
In the strongly scattering regime, the different transport mechanisms couple and impose a minimum conductivity of electromagnetic waves,
thereby preventing Anderson localization of light in the medium. We extend the self-consistent theory of localization and compare the predictions to extensive numerical simulations.
\end{abstract}


\maketitle

\section{Introduction}

The traditional and widely used  picture of elastic multiple scattering of  light is one of a  plane wave that exponentially decays on the length scale of a mean free path  while
propagating from one particle  to another, with an electric field orthogonal to the direction of propagation, and with subsequent scattering into a  different direction in space,
with exactly the same frequency.
It is well known that transversality in real space ($\mathbf{r} \cdot \mathbf{E}(\mathbf{r})=0$) is only valid in the far field of the scatterers,
at distances much larger than the wavelength.
In the near field of a dielectric object, the electric field achieves a ``dipolar'' structure, with a component directed along the propagation direction,
while still being divergence-free, i.e. $\bm{\nabla} \cdot\mathbf{ E }(\mathbf{r})= 0$. In many approaches of multiple light scattering,
 these longitudinal modes are widely appreciated, yet considered ``virtual", in the sense that they do not carry a Poynting vector so that they
 cannot transport electromagnetic energy themselves, though they can mediate the propagation of other waves, such as mechanical \cite{levitov} or matter \cite{matterdipole} waves.

However, in inhomogeneous media, the
dielectric constant
$\varepsilon(\mathbf{r})$ of the matter varies in space, and
Gauss' equation imposes $\bm{\nabla} \cdot [\varepsilon(\mathbf{r}) \mathbf{ E }]= 0$. As a result, true longitudinal electric fields exist,
with $\bm{\nabla} \cdot\mathbf{ E }\neq  0$ and a finite density of states (DOS) in phase space, to which elastic scattering could take place.
Induced polarization charges possess
 Coulomb energy, and also stock dipole-dipole energy
among different scatterers but have no Poynting vector, so how can they transport energy?  In atomic physics, the well-known process of F\"{o}rster coupling  \cite{forster}  facilitates a non-radiative
transport mechanism to exchange quantum states and to move  Coulomb energy from one atom to another. Like spontaneous emission, this process  is  inherently inelastic and incoherent,
and  is \emph{de facto} excluded in a picture where only elastic multiple scattering, including interferences, is allowed. Much in the spirit of F\"{o}rster coupling,
 Ref.~\cite{theoL} added explicitly the quasi-static  dipole-dipole coupling as a new channel in transport theory of electromagnetic waves.
In this work we will show that this transport channel
naturally emerges from a rigorous electromagnetic transport theory. The finite Poynting vector of this channel is shown to originate from the  interference between longitudinal and transverse modes.

The transverse picture of electromagnetic waves emerges naturally in the so-called ``independent scattering approximation'' (ISA) of diffuse transport. In this approximation, the
 longitudinal waves are usually ignored, and only transverse, propagating states are counted, associated with damped plane waves with
 wave numbers close to the frequency
 shell $p \approx k = \omega/ c_0 $ in phase space. A fundamental question is whether this picture is significantly altered, \emph{within and  beyond} the ISA,
 or if  just quantitative modifications occur.
Longitudinal  states have a finite density of states (DOLS), proportional to the imaginary part of the (longitudinal) dielectric
constant of the effective medium. Being mainly confined to scatterers,
they exist far from the frequency shell in phase space, typically at very large wave vectors $p \gg k$.
We will show that, due to the dipole size that is much smaller than the optical wavelength, excitations with large wave numbers can scatter and mode-convert to both
transverse and longitudinal states. As such they take fully part in the diffuse transport.

\section{road map of this work} \label{roadmap}

The purpose of this work is to elucidate the role of longitudinal waves in the
transport of electromagnetic waves propagating in disordered media at scales well beyond the mean free path. Our basic starting point is that light propagation over a distance
$\mathbf{r}$ is described by the full vector Green's function $\mathbf{G}(k,\mathbf{ r})$. Its longitudinal part dominates, at any frequency $\omega = kc_0$, for small distances, or in Fourier space
at large wave numbers $p$, typically
$p \gg k$. This affects many aspects of multiple scattering. The consideration of the full vector Green's function in both the scattering and the propagation
guarantees energy conservation in all orders of scattering, and the neglect of longitudinal fields would violate this principle fundamental for long-range diffusion.
The following  4  main sections revisit each  a well-known result of standard multiple scattering theory to include the
longitudinal waves.

In section~\ref{sectionEM}  we revisit the effective medium theory of electromagnetic waves, associated with the average electromagnetic field. It is fully characterized  by a complex self-energy tensor $\mathbf{\Sigma}(k, \mathbf{p})$
that depends on both circular frequency $\omega= k c_0$ and wave vector $\mathbf{p}$, including its direction. We discuss the role of the longitudinal fields in several issues that are directly related to
the effective medium:
independent and recurrent scattering, density of states and
equipartition between longitudinal and transverse electromagnetic waves.  The longitudinal component is related to the  subtle Lorentz cavity described in many text books
\cite{bw,jackson}.
An electric dipole scatterer that is impenetrable to the light (arguably a simple ``atom") is the simplest model that highlights longitudinal excitations in resonant scattering.
In this model, the energy stored by the
radiating dipole is large and \emph{entirely }of longitudinal nature.
We study the effective medium of
a 3D space filled with a volume density $n$ of such dipoles. Longitudinal fields  give rise to
 singularities at large wave numbers, and show up later in diffusion and localization of light. They are treated
while respecting the conservation of energy. The longitudinal self-energy has a non-trivial behavior for $p\rightarrow \infty$ that will play a role in
long-range diffusion discussed later. It gives rise to a new complex wave number
$K_L$ associated with the density of states of longitudinal excitations (DOLS) with large wave numbers. The contribution of  binary
dipole-dipole coupling to DOLS will be calculated analytically, and is obtained numerically in all orders
of the dipole density.  Finally we show that for multiple scattering  in this model, longitudinal excitations rapidly dominate the total density of states.

Section~\ref{kubosection} deals with  the transport theory of the average light intensity and in particular with the role of longitudinal fields in the rigorous
Kubo formalism for the diffusion constant. At length scales well beyond the mean free path, the exact ``Bethe-Salpeter" transport equation simplifies to the diffusion current tensor
 $\mathbf{J}(k, \mathbf{p},\mathbf{q})$ that determines
the flow of electromagnetic energy in phase space for excitations near the resonant frequency of the dipoles  and for arbitrary  wave number and polarization, induced by an energy gradient (described by $\mathbf{q}$ in
Fourier space).
We demonstrate the presence of \emph{four} different channels that
affect this flow differently, among which\emph{ two} directly affect the diffusion constant. The first so-called $J_0$ channel is the widely studied diffusion of transverse electromagnetic waves, with wave numbers close
to the frequency shell
$p \approx k $ of the effective medium. The second $J_2$ channel  has - to our knowledge - never
been discussed before, and originates from the interference between
longitudinal and transverse excitations.
Because longitudinal waves alone do not possess a Poynting vector they need this interference to produce long-range diffusion.  Like weak localization in the $J_0$ channel ,
the $J_2$ channel is a high-order
effect in perturbation theory and is absent in the Boltzmann approximation valid for weak disorder. The leading contributions to $J_2$ come from the
Drude diffusion (defined as the contribution of the effective medium to diffusion \cite{mahan})  and the weak localization (interference of counter-propagating waves) induced by two close dipoles. In the channel $J_0$ the latter is
known to be of order $-1/k\ell$
(with $\ell$ the mean free path) \cite{bartpre, cherro,kwong}, the weak localization in the $J_2$ channel turns out to be \emph{positive}  and of order
$+1/(k\ell)^2$. We  demonstrate that by summing all diagrams involving two dipoles, large contributions to $J_2$ diffusion stemming from large wave vectors  cancel rigourously. This is a very convenient albeit non-trivial conclusion.

In Section~\ref{sectionrad} we recall the  radiative force exerted by a diffuse electromagnetic flow,  well-know and widely used in astrophysics. We demonstrate that the stored
longitudinal excitations, although not contributing to the Poynting vector, resonantly  enhance the radiative force density in the disordered medium.

Finally in section~\ref{sectionAL} we make a first attempt to incorporate the longitudinal transport channels into the self-consistent theory of localization. For scalar waves (acoustic waves or spinless electrons)
this theory predicts without much effort a mobility edge in disordered  three-dimensional media when $k\ell \approx 1$ \cite{vw,zhang}. Static electric dipole coupling was already identified as a possible source of
delocalization of mechanical  waves \cite{levitov}. Recent numerical simulations with electromagnetic wave scattering from point-like electric dipoles
revealed the absence of a mobility edge \cite{sergey0} and are difficult to explain within the
traditional picture that only acknowledges the transverse field as a mechanism for diffuse transport.
 {Also experiments \cite{naraghi} have revealed that diffusion of electromagnetic waves in dense media cannot be explained by a
the familiar scenario of a  transition from diffusion to localization, most likely due to near-field couplings.    }
 We demonstrate that the standard self-consistent theory, when applied with the usual approximations
to electromagnetic waves,
couples the newly identified diffusion channel $J_2$ to the channel $J_3$. This channel does not affect the Poynting vector, proportional to  $\mathrm{Re}\, ( {\mathbf{E} }\times \bar{\mathbf{B}}) $
but rather induces a non-zero value for $\mathrm{Im}\, ({\mathbf{E} }\times \bar{\mathbf{B}})$, a vector discussed e.g. by Jackson \cite{jackson}.
Its coupling to $J_2$ in the self-consistent equations leads to a minimum value
for the diffusion constant. We have worked out this theory assuming the effective medium to be the same  for transverse and longitudinal waves, and
characterized by a single complex wave number $k_e + i/2 \ell$ that was calculated numerically for frequencies near the resonance.
 An unfortunate technical complication is that the self-consistent theory - even in its scalar version applied to point-like particles - suffers from a genuine
singularity at large wave numbers that was not
discussed before in literature. This divergency  is actually identical  to the one encountered in Sec.~\ref{kubosection} where it was seen to cancel in low orders of the density when
 summing all diagrams.
We postulate that this singularity is artificial and for scalar waves recover the usual results in literature. The resulting self-consistent theory for electromagnetic waves
is seen to be in good agreement with the
exact numerical results.

\section{Effective medium theory of electromagnetic waves}\label{sectionEM}

In standard transport theory \cite{PR}, the dispersion and extinction of waves are described by a complex self-energy $\mathbf{\Sigma} (k, \mathbf{p})$, associated with the effective medium. For
electromagnetic waves this is a second-rank tensor,
depending on frequency and wave
vector $\mathbf{p}$.  Scattering between two states in phase space is described  by the four-rank scattering vertex $\mathbf{U}_{\mathbf{p}\mathbf{p}'}(k)$.
 {In this work we disregard optical absorption and assume throughout the conservation
of electromagnetic energy in multiple scattering, as expressed }
by the Ward identity \cite{PR,sheng},
\begin{equation}\label{ward}
    {-\mathrm{Im}\,  \mathbf{\Sigma}(k+i\epsilon, \mathbf{p})}= \sum_{\mathbf{p}'} \mathbf{U}_{\mathbf{p}\mathbf{p}'}(k) \cdot -\mathrm{Im} \, \mathbf{G}(k+i\epsilon,
    \mathbf{p}')
\end{equation}
with the notation $\mathrm{Im}\,\mathbf{ A}  \equiv (\mathbf{A} - \mathbf{A}^*)/2i$
{where $\mathbf{A}^*$ denotes the Hermitian conjugate of a $3 \times 3$ matrix $\mathbf{A}$ ($A^*_{ij} =  \bar{A}_{ji}$)}.
The left hand side stands for the extinction of an electromagnetic excitation at wave vector $\mathbf{p}$, the right hand side puts this equal to the elastic  scattering of the
same excitation from $\mathbf{p}$ towards all other accessible states $\mathbf{p}'$ in the phase space.
The ``spectral tensor''  $-\mathrm{Im} \, \mathbf{G}(k+i\epsilon, \mathbf{p}')$ is positive (as $\epsilon \downarrow 0$, for positive frequencies)
and determines the availability  of  microstates
 at the wave vector $\mathbf{p}'$, given the frequency $\omega = k c_0$ that is conserved in elastic scattering. For convenience we will drop explicit reference to $\epsilon$ and assume its presence in $k +i\epsilon$
 implicitly.  Both $\mathbf{\Sigma}(k, \mathbf{p})$ and
 $ \mathbf{U}_{\mathbf{p}\mathbf{p}'}(k)$
 will be discussed in more detail below.

 \subsection{Dyson Green's function}\label{greensection}

 In Fourier space the  Dyson Green's tensor of an electromagnetic ``quasiexcitation''
 with frequency $\omega = kc_0$ and wave vector $\mathbf{p}$ of the
 effective medium is given by \cite{PR}
 \begin{eqnarray}\label{dyson}
    \mathbf{G}(k, \mathbf{p}) &=& \left[k^2 - p^2\mathbf{ \Delta}_p -\mathbf{\Sigma}(k, \mathbf{p})\right]^{-1} \nonumber \\
    &=& \frac{\mathbf{\hat{p}}\mathbf{\hat{p}} }{k^2 -{\Sigma_L}(k, {p})} + \frac{\mathbf{ \Delta}_p}{k^2 -p^2 -{\Sigma_T}(k,p)}
 \end{eqnarray}
split up into a longitudinal and a
 transverse part, with $\mathbf{\Sigma}(k, \mathbf{p}) = {\Sigma_L}(k, {p}) \mathbf{\hat{p}}\mathbf{\hat{p}}
 +  {\Sigma_T}(k, {p})\mathbf{ \Delta}_p$,  with $\mathbf{ \Delta}_p = \mathbf{1} - \mathbf{\hat{p}}\mathbf{\hat{p}} $ the  projection tensor to transverse states.
 In transport theory, the tensor $\mathbf{G }(k,\mathbf{p})\otimes \mathbf{G}^*(k,\mathbf{p}') $ is the building block of multiple scattering, and it is important to understand
$ \mathbf{G }(k,\mathbf{p})$ in great detail on all scales.

 The  longitudinal part of $\mathbf{G}(k, \mathbf{p})$ is associated with local Coulomb interactions between
 induced charges inside scatterers, often referred to as ``non-radiative, static'', dipole-dipole coupling at a distance. The transverse part describes propagating waves. In the following we investigate both components in real space for small and large distances. We demonstrate that at small distances, the longitudinal part
 of the Dyson Green's function dominates very generally
 and takes the form of
 dipole-dipole coupling with the usual Lorentz contact term \cite{jackson}, and surprisingly, is seen not to be static.  At large distances,
 only transverse excitations contribute and  $\mathbf{G}(k,\mathbf{r}) $ is under very general conditions equal to an exponentially small,
 propagating  excitation with a polarization transverse to the direction of propagation $\mathbf{r}$.
 This implies that  $\mathbf{G}(k,\mathbf{r})$ contains the familiar near and far fields of electromagnetism, without the need to add the first
 by hand \cite{theoL}. At large distances the traditional picture, described earlier, emerges.

In real space, the Green's tensor $\mathbf{G}(k,\mathbf{r})$ is the Fourier transform of Eq.~(\ref{dyson}) and describes the propagation of electromagnetic
 waves over a distance $\mathbf{r}$  in the effective medium.
The near-field component is ``non-radiative'' in the sense that a longitudinal field $\mathbf{E} \parallel \mathbf{k}$   induces no magnetic field as
$k \mathbf{B } \sim \mathbf{k }\times \mathbf{E}=0$. Alone, it carries therefore no Poynting vector. However, we will show later in this work that
the interference of longitudinal and transverse components in the tensor product $\mathbf{G } \otimes \mathbf{G}^* $ does carry a Poynting vector
and facilitates a new channel to transport energy.

 With  $K_L^2(p) \equiv k^2 - {\Sigma_L}(k,p)$ the square of a complex longitudinal wave vector, one obtains in real space,
 \begin{eqnarray}\label{GL}
    \mathbf{G}_L(k,\mathbf{r}) &=& \sum_\mathbf{p}  \frac{\mathbf{\hat{p}}\mathbf{\hat{p}}}{K_L^2(p)} \exp(i\mathbf{p}\cdot\mathbf{ r})
    \nonumber \\
    \, &=& \frac{\delta(\mathbf{r})}{3 K_L^2(\infty)}  + \frac{1-3\mathbf{\hat{r}}\mathbf{\hat{r}}}{4 \pi K_L^2(\infty) r^3} + \mathbf{D}(\mathbf{r})
 \end{eqnarray}
where we have split off the singularity of the integral at large wave numbers, leaving the remaining term $\mathbf{D}(\mathbf{r})$ as a
 contribution
to the traceless dipole-dipole coupling described by the second term. Since $\mathbf{D}(\mathbf{r})$ is, by construction,  the Fourier transform of a function that decays to zero for large $p$,
it is free from a Dirac distribution, and even non-singular as $\mathbf{r}\rightarrow 0$. We will show this explicitly in Sec.~\ref{sec2dipoles} for the recurrent scattering from two dipoles.  As a result,
the first two terms in Eq.~(\ref{GL}) dominate on small scales.
The first, subtle Lorentz contact term is  a genuine Dirac distribution and vanishes for $\mathbf{r}\neq 0$, but for $\mathbf{r}=0$ makes a genuine contribution to DOS.
Since the transverse field $\mathbf{G}_T(\mathbf{r}) \sim 1/r$
for $kr < 1 $ is much less singular, we conclude that
 \begin{eqnarray}\label{Gr0}
    \mathbf{G}(k,\mathbf{r}\rightarrow 0) \rightarrow  \mathbf{G}_{0,L}(K_L(\infty),\mathbf{r})
 \end{eqnarray}
This takes the same form as the familiar dipole-dipole regime of the bare Green's function $\mathbf{G}_0(\mathbf{r})$, with however the wave number $k= \omega/c_0$ in vacuum replaced by a complex-valued and frequency-dependent wave-vector $K_L(\infty)$.
For finite-size dielectric scatterers one may argue that at small
scales described by $p\rightarrow \infty$ the effective medium is homogeneous and $ K_L(\infty)$ must be some real-valued wave number.
For atomic atomic dipolar scatterers however, we will see that  the complex value of
$K_L(p)$   extends up to infinity.
The complex value of $ K_L(\infty)$ indicates that the dipole-dipole coupling, dominating in the near field,
is not static but depends on frequency and contributes to the DOS. In Sec.~\ref{sectionDOS} we will calculate $K_L(\infty)$ numerically in all orders of the density for a model of randomly positioned electric dipoles.

At long distances $kr \rightarrow \infty$, small wave numbers prevail in Eq.~(\ref{GL}) so that
 \begin{equation}\label{DD}
    \mathbf{G}_L(k,\mathbf{r}\rightarrow \infty) = \frac{1-3\mathbf{\hat{r}}\mathbf{\hat{r}}}{4 \pi K_L^2(0) r^3}
 \end{equation}
with $K_L(p)$ now evaluated at $p=0$. If this propagator would not be compensated, the far field would contain an algebraically small longitudinal term
which would severely affect the random-walk picture of transverse electromagnetic wave transport.
However, it is compensated very generally by a part of the  transverse
 propagator $\mathbf{G}_T(k,\mathbf{p})$.
 For  $kr \gg 1$ it is useful to make the following decomposition,
  \begin{eqnarray}\label{GT}
    \mathbf{G}_T(k,\mathbf{r}) &=& \sum_\mathbf{p} \frac{\mathbf{\Delta}_p}{K^2_T(p) -p^2} \exp(i\mathbf{p}\cdot\mathbf{ r}) \nonumber \\
   &=&\frac{1}{2\pi^2}\left(-\bm{\nabla}^2 + \bm{\nabla} \bm{\nabla}\right)
    \frac{1}{2ir}\int_\Gamma dp \frac{e^{ipr}}{p}
     \frac{1}{K^2_T(p) -p^2} \nonumber \\
    &+& \left(-\bm{\nabla}^2 + \bm{\nabla} \bm{\nabla}\right)  \frac{1}{4\pi K_T^2(0)r}
 \end{eqnarray}
 Here $\Gamma$ denotes the line $(-\infty, +\infty)$ that avoids the origin $p=0$ via a small contour in the upper complex $p$-plane, and
 which generates the last term.
 In the far field, since necessarily $K_T(0)= K_L(0)$, the last term of Eq.~(\ref{GT})  cancels \emph{exactly }against the longitudinal far field  in Eq.~(\ref{DD}).
 The Green's function $\mathbf{G}(k, \mathbf{r})$ as a whole is  therefore determined by the denominator of the first
 term and
  \begin{eqnarray}\label{GTL}
    \mathbf{G}(k,\mathbf{r}\rightarrow \infty) =
    \frac{\mathbf{\Delta}_r }{4\pi^2 ir}\int_{-\infty}^{\infty} dp \,
     \frac{p\, e^{ipr}}{K^2_T(p) -p^2}
 \end{eqnarray}
This indicates that the electric field is asymptotically dominated by transverse modes and also transverse to the direction of propagation $\mathbf{r}$.
If $K_{T}(p)$ has an analytical extension at least over a small sheet $\mathrm{Im}\, p < K_T''$ in the upper
complex $p$-plane, $\mathbf{G}(k,\mathbf{r})$  will
 decay at least as $\exp (-K_T''r)/r$. Different ``effective medium'' approaches exist to calculate $\mathbf{G}(k,\mathbf{r})$ for various models \cite{sheng}.
 The easiest method is to assume the presence of a simple pole
 $K_{T}(p) = k_T + i/2\ell $, in
 which case normal exponential behavior emerges with
 the decay length equal to (twice) the  elastic scattering  mean free path $\ell$.

 We conclude that the Green's tensor of the effective medium has a true longitudinal component
 ($\partial_i G_{ij}(k,\mathbf{r}) \neq 0$) that affects wave propagation at small scales $r < 1/k$.
 In the far field, the electric field is always transverse to propagation ($\hat{r}_i G_{ij}(k,\mathbf{r}) = 0$).
 Decay is exponential under broad conditions with a decay length $\ell$. This implies that radiative transfer should still be compatible with a
 random walk with step length $\ell$, though with possibly new
 mechanisms for energy transport
 in the near field provided by the presence of longitudinal fields, that can become dominant when $k\ell \approx 1$.
 This idea will be worked out concretely in the next subsections for an ensemble of randomly distributed dipolar electric scatterers (``dipoles'' for short).

 \subsection{Independent electric dipole scattering}\label{section1D}

 In the independent scattering approximation (ISA)  applied to point-like electric dipole scatterers with number density $n$ and $T$-matrix $t(k)$,
 $\mathbf{\Sigma}_{\mathrm{ISA}}(k, \mathbf{p}) = nt(k)$. In this work we assume each dipole to be impenetrable for
light outside, and to have only longitudinal excitations in its vicinity, at scales much smaller than the wavelength. This conveniently labels material energy as longitudinal
 states that take part in the scattering process. By definition, the $T$-operator of a general polarizable scatterer perturbs wave propagation in
 free space according to $\mathbf{G}(k) = \mathbf{G}_0(k) + \mathbf{G}_0(k) \cdot \mathbf{T}(k)\cdot
 \mathbf{ G}_0(k)$. If we set $\mathbf{T}(k)=  | \mathbf{r}_d \rangle \mathbf{t}(k)\langle  \mathbf{r}_d | $ to describe an a electric
 dipole at position $\mathbf{r}_d$, and impose
 $\langle \mathbf{r} | \mathbf{G}({k}) |\mathbf{r}_d \rangle =0 $ for any $\mathbf{r }\neq \mathbf{r}_d$ for it to be ``impenetrable'', then it follows that
 \begin{equation}\label{born}
    \mathbf{t}(k) = \frac{-1}{\langle \mathbf{r}_d | \mathbf{G}_0({k}) |\mathbf{r}_d \rangle} = - \left[\sum_\mathbf{p}
    \left( \frac{\mathbf{\hat{p}}\mathbf{\hat{p}} }{k^2} + \frac{\mathbf{ \Delta}_p}{k^2 -p^2 + i0}\right)\right]^{-1} \
 \end{equation}
 This model can be refined to acknowledge finite penetration of light into the dipoles \cite{theo},
 but the present choice highlights the role of longitudinal waves and is arguably
 the best description of elastic scattering from an atom without going into the details of atomic physics. Both the longitudinal and the
 transverse integral diverge, the first essentially due to the Lorentz contact term.
 We will regularize the first as
 $ \sum_\mathbf{p} \mathbf{\hat{p}}\mathbf{\hat{p}} = 1/3u$ and the transverse part as $  \sum_\mathbf{p} \mathbf{ \Delta}_p/p^2 = 1/ 6\pi \Gamma $.
 It follows that
 \begin{equation}\label{tED}
    \mathbf{t}(k)=  \frac{-6\pi \Gamma k^2}{k_0^2 -k^2 - i k^3 \Gamma }
 \end{equation}
 Both $\Gamma$ (with dimension of length) and $u$ (a volume) are genuine properties of the dipole, independent of frequency or
 polarization of the light. In particular $k_0^2 = 2\pi \Gamma /u$ determines the resonant frequency of the dipole. For $k=k_0$ longitudinal and transverse
 singularities, opposite in sign, cancel each other.

 For small $k$, the static
 polarizability $\alpha(0)$ is related to the $t$-matrix as  $t =-\alpha(0) k^2 $ \cite{PR}, and we can identify $\alpha(0) = 3u $. This relation can be understood from classical electrodynamics.
 We recall the Lorentz relation $\mathbf{E}(0) = \mathbf{E}-
 \frac{1}{3} \mathbf{P}$ for the homogeneous electric field inside the dipole, assumed spherical. Since we have imposed  $\mathbf{E}(0) = 0$, the polarization density must
 equal $3$ times the local electric field
 $\mathbf{E}$. The dipole moment is thus $ u \mathbf{P} \equiv \alpha(0)\mathbf{ E} = 3 u \mathbf{E}$ with $u$  the volume of the dipole, and hence $\alpha(0) = 3u$.
 The line width in frequency near the resonance is related to $\Gamma$  according to
 $\gamma = k_0^2c_0\Gamma  = \alpha(0) k_0^4 c_0/6\pi $, a known relation
 for the radiative decay rate of a semi-classical two-level atom in the electric-dipole approximation \cite{AllanEberly}. We can identify the quality factor
 $Q_0= \omega_0/\gamma = 6\pi /\alpha(0) k_0^3$. Near the resonance, we can thus write
  \begin{equation}\label{tEDreso}
    \mathbf{t}(k= \omega/c_0)= -\frac{6\pi}{k_0} \frac{\gamma/2 }{\omega_0 - \omega - i\gamma/2 }
 \end{equation}
  The $t$-matrix satisfies the optical theorem,
 \begin{equation*}
    -\mathrm{Im}\, t  = \sum_{\mathbf{p}'} {|t(k)|^2} \cdot \mathbf{ \Delta}_p \, \pi \delta (k^2-p^2)=  \frac{|t(k)|^2 k }{6\pi}
 \end{equation*}
 This expression is consistent  with Eq.~(\ref{ward}), worked out linearly in the dipole density $n$ on both sides,
 with $U^{\mathrm{ISA}}_{\mathbf{pp}'}= n|t(k)|^2$ the ISA collision operator and
 $\mathbf{\Sigma}_{\mathrm{ISA}}(\mathbf{p}) = nt(k) $. For its relative simplicity, many exact
 numerical simulations have been carried out with media filled randomly with electric dipoles \cite{felipe, sergey0, pool, remi},
 and many theoretical treatments exist already \cite{jpc, bartpre, Dalibard}, not only because one can go far without making further approximations
 but also because they constitute a good and complete model for multiple scattering of light from simple atoms. We notice that the $t$-matrix of a single dipole is independent
 of both polarization, $\mathbf{p}$ and $\mathbf{p}'$. As a result, a single dipole can scatter microstates with arbitrary state of polarization, and with arbitrary $\mathbf{p}$ towards arbitrarily large $\mathbf{p}'$.

\begin{figure}
\includegraphics[width=7cm]{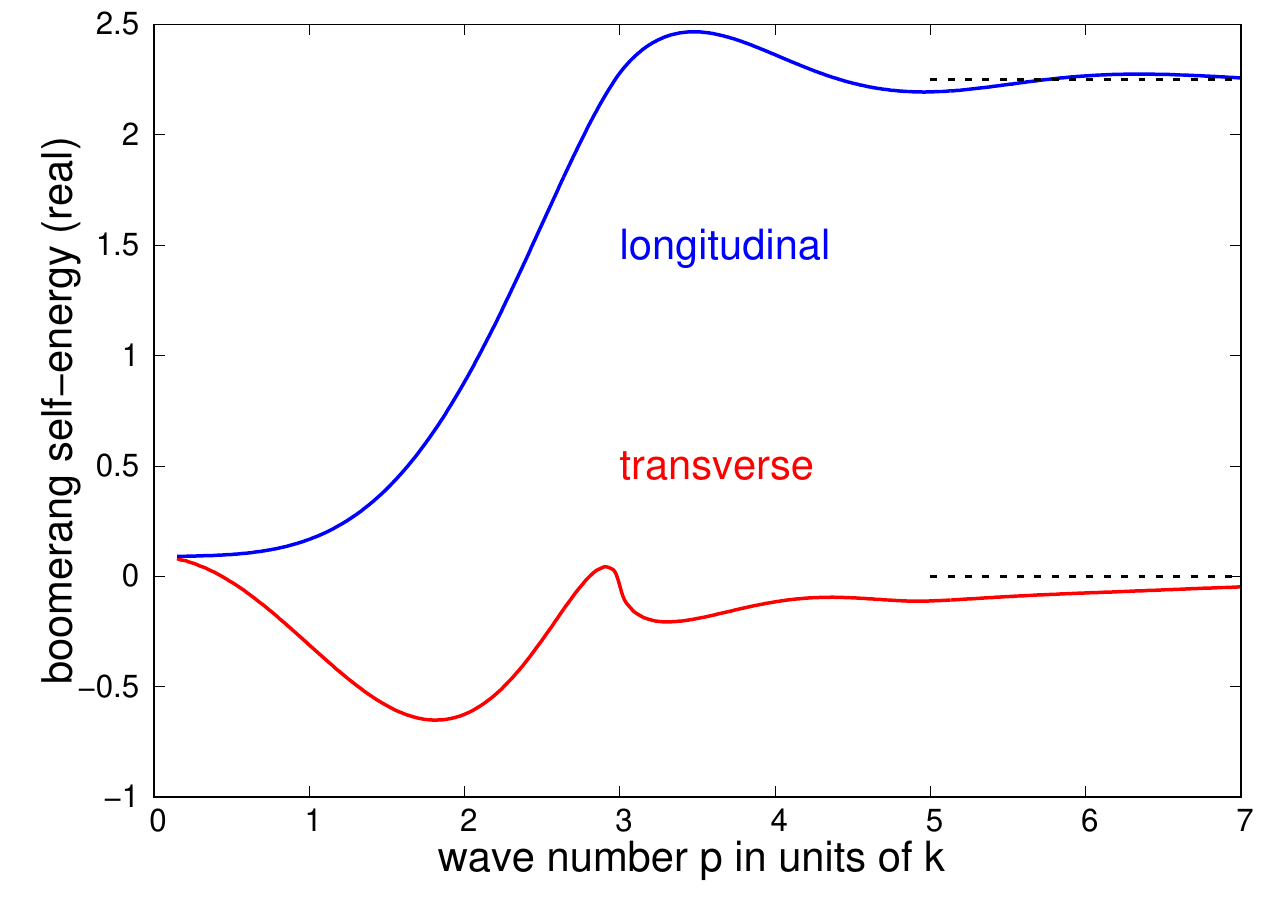}\\
\includegraphics[width=7cm]{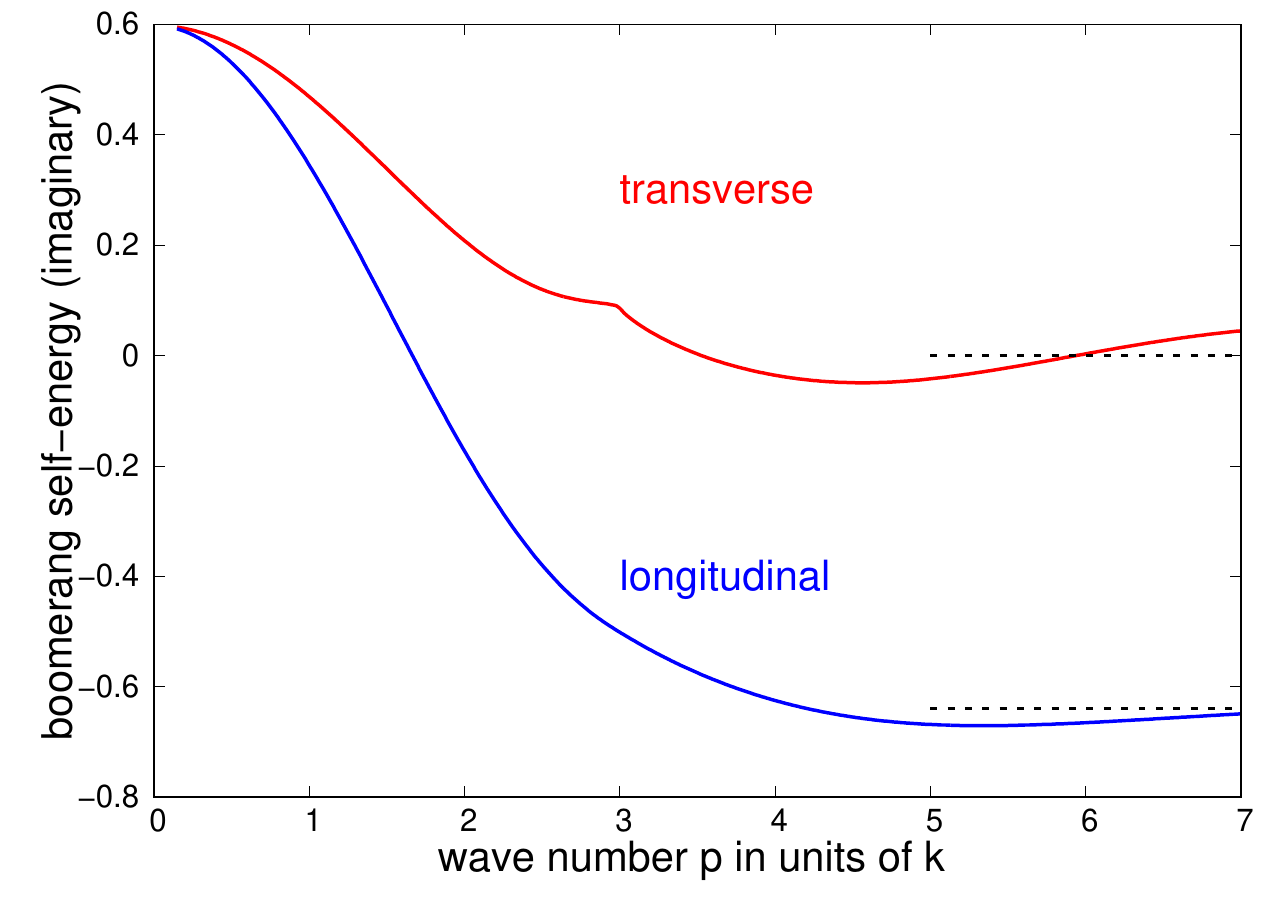}
  \caption{Real part (top, on resonance $\omega = \omega_0$) and imaginary part (bottom, for a detuning {$\delta = (\omega-\omega_0)/\gamma = -0.75$}) of the ``boomerang'' self-energy
  associated with two dipoles as a function of wave number $p$.
  The wave number latter  is expressed in units of $k=\omega/c_0$ in free space, the self-energy is expressed  in units of $(4\pi n/k^3)^2 \times k^2$.
  The transverse self-energy $\Sigma_T(k,p)$ converges asymptotically to zero (dashed line) for all detunings,
  meaning that the Lorentz local field term
  $\Sigma_{LL} (k,p) = - \frac{1}{3} n^2 t^2 $, part of the boomerang diagrams but independent of $p$, is canceled. The longitudinal self-energy $\Sigma_L(k,p)$,
   on the other hand, converges asymptotically to
  $-n^2t^2$
  (dashed line), as expressed by Eq.~(\ref{sigma2}). The characteristic wave number to reach the asymptotic constant value is  $p = 3k$ and is associated with the triple round
  trip of light between two dipoles in the
  boomerang diagrams, as described by
  Eq.~(\ref{self}).}\label{boom}
\end{figure}

\subsection{Extinction involving two electric dipoles}
\label{sec2dipoles}

The extinction caused by recurrent scattering from two dipoles was  discussed in Ref. \cite{bartpre} for scalar waves, in Ref. \cite{wellens} for low-energy electrons,   and  in Refs. \cite{cherro,kwong,jpc} for electromagnetic waves.  The last two works mainly focused on diffusion of transverse light, but used the full Green's tensor~(\ref{dyson}) to describe recurrent scattering.
In Ref.~\cite{kwong} correlations between dipoles were included and
compared successfully to numerical simulations.
Despite the singular Green's tensor $\mathbf{G}_0(k, \mathbf{r})$ in the near field,
 no new divergencies were encountered provided the whole series of recurrent scattering is summed.
 In the following section we will explicitly include the longitudinal field in the transport. To that end, we need to understand the behavior of the self-energy tensor $ \mathbf{\Sigma}(k,\mathbf{p})$
at large $p$. The self-energy involving one or two different dipoles is given by \cite{jpc}
\begin{eqnarray}\label{self}
   \mathbf{\Sigma}(k, \mathbf{p}) &=& nt \mathbf{1}  + n^2 \int d^3 \mathbf{r} \frac{t^3 \mathbf{G}_0^2(\mathbf{r})}{\mathbf{1}-t^2 \mathbf{G}_0^2(\mathbf{r})} \nonumber \\
   &+& n^2 \int d^3 \mathbf{r} \frac{t^4\mathbf{G}_0^3(\mathbf{r})}{\mathbf{1}-t^2 \mathbf{G}_0^2(\mathbf{r})}
e^{i\mathbf{p}\cdot \mathbf{r}} + \mathcal{O}\left(n^3 \ln n\right)\;\;\;\;\;\;\;
\end{eqnarray}
We have dropped the explicit reference to $k= \omega/c_0$ in $t(k)$ and in $\mathbf{G}_0(k+i\epsilon,\mathbf{r})$. The first term is the ISA, the second term involves recurrent loops between two dipoles.
They are both independent of $p$ and necessarily isotropic tensors. We will show in Sec.~\ref{sectionDOS} that, in our model, loop diagrams of arbitrary order
rigorously determine the energy stored in longitudinal modes, and exploit this notion numerically.
The third term, summing up the so-called boomerang diagrams $\mathbf{\Sigma}_B$ (see Fig. \ref{boom}) provides the first $p$-dependent contribution and causes $ \Sigma_T(p) \neq \Sigma_L(p)$.
Higher orders in number density involve 3 different dipoles or more.
The boomerang diagrams generate a subtle contribution via the Lorentz contact term $\delta (\mathbf{r}) /3 k^2$ in
$\mathbf{G}_0(\mathbf{r})$ \cite{Dalibard, nienhuis}, which gives rise to the well-known Lorenz-Lorentz correction $ - n^2 t^2/3k^4$ to both the
longitudinal and transverse dielectric functions, and that is independent of $p$.  {Adding a small anti-correlation between the dipoles to avoid
their physical overlap does not eliminate this
 correction but rather transfers it to a new correlation diagram} \cite{nienhuis}.
Nevertheless, as $p\rightarrow \infty$, this term is compensated, again subtly, in the transverse self-energy and reappears as a  purely
 longitudinal self-energy.
This can be seen by subtracting  the transverse photon field
$\mathbf{G}_{0,T} (\mathbf{r})$ in free space, derived in Eq.~(\ref{GT}) for the effective medium, which is free from the Lorentz contact term. The
boomerangs become,
\begin{eqnarray}\label{boom2}
   \mathbf{\Sigma}_B(\mathbf{p}) &=& n^2 t^2 \int d^3 \mathbf{r}
    \, \left[\frac{\mathbf{G}_0(\mathbf{r})}{1-t^2\mathbf{G}_0^2(\mathbf{r})} - \mathbf{G}_{0,T}(\mathbf{r}) \right] \exp(i\mathbf{p}\cdot \mathbf{r}) \nonumber \\
   & +&  n^2 t^2 \int d^3 \mathbf{r}
    \, \left[\mathbf{G}_{0,T}(\mathbf{r}) - \mathbf{G}_0(\mathbf{r}) \right] \exp(i\mathbf{p}\cdot \mathbf{r})
 \end{eqnarray}
In the first term, the Lorentz contact term at $\mathbf{r}=0$  no longer contributes and the integral vanishes for large $p$.
The second integral is just equal to (minus) the longitudinal Green's tensor $\mathbf{G}_{L,0}(k,\mathbf{p})$
in Fourier space. Hence we find the somewhat surprising relation for infinite $p$,
\begin{equation}\label{sigma2}
    \lim_{p\rightarrow\infty}  \mathbf{\Sigma} (\mathbf{p})  = {\Sigma}_{\mathrm{ISA}} + {\Sigma}_{\mathrm{Loop}} -\frac{n^2t^2}{k^2}\mathbf{\hat{p}}\mathbf{\hat{p}}
\end{equation}
Figure~\ref{boom}   illustrates
 by numerical integration of Eq.~(\ref{self}) that for large wave vectors, the Lorentz contact term is canceled in the transverse (boomerang) self-energy.
 It converges always to zero, whereas the longitudinal boomerang self-energy converges asymptotically to $\Sigma_L(p) = -n^2t^2/k^2$. Neither one of them
converges to  $ -n^2t^2/3k^2$, associated with the Lorentz contact term.
 The asymptotic limit established in Eq.~(\ref{sigma2}) is important
since it  demonstrates that  $K_L(\infty) \neq K_T(\infty)$, the first introduced earlier in Eq.~(\ref{GL}) describing the
dynamic dipole-dipole coupling in the near field.

It is instructive to calculate the longitudinal Green's function~(\ref{GL}) associated with the self-energy in Eq.~(\ref{self}). Only the boomerang
diagrams $\Sigma_B$ depend on wave number $p$. Hence, up to order $n^2$,
\begin{eqnarray*}
    \mathbf{G}_L(k, \mathbf{r}) &=& \sum_\mathbf{p}  \hat{\mathbf{p}}\hat{\mathbf{p}} \frac{1}{k^2 - \Sigma_L (p)}\exp(i\mathbf{p}\cdot \mathbf{r})\\
     = &-&\bm{\nabla} \bm{\nabla} \cdot  \sum_\mathbf{p} \left[ \frac{\mathbf{1}}{k^2- \Sigma_0} + \frac{1}{k^4}
     \mathbf{\Sigma}_B(\mathbf{p}) \right]  \frac{\exp(i\mathbf{p}\cdot \mathbf{r}) }{p^2}
\end{eqnarray*}
with $\Sigma_0= \Sigma_{\mathrm{ISA}} + \Sigma_{\mathrm{Loop}}$. Upon inserting the boomerang diagrams and using
$-\bm{\nabla} \bm{\nabla} (1/4\pi r)  = \delta(\mathbf{r})/3+ (1- 3\hat{\mathbf{r}}\hat{\mathbf{r}})/4\pi r^3 = k^2 \mathbf{G}_{0,L}(\mathbf{r})$, one obtains,
\begin{eqnarray*}
    \mathbf{G}_L(k, \mathbf{r}) &= &  \mathbf{G}_{0,L}( (k^2 -\Sigma_0)^{1/2},\mathbf{r} )  \\
     &+&   \frac{n^2}{k^2} \int d^3\mathbf{r}' \, \mathbf{G}_{0,L}(\mathbf{r}-\mathbf{r}')\cdot
     \frac{  t^4 \mathbf{G}^3_0(\mathbf{r}')}{1-t^2\mathbf{G}_0^2(\mathbf{r}')}
\end{eqnarray*}
This determines the longitudinal Green's tensor at all distances, and also depends on frequency for all distances. The first term stands for ordinary dipole-dipole coupling of the type $1/r^3$
with a modified prefactor from the effective medium that arises because we consider the electromagnetic Green's tensor and not the potential energy of the dipoles.
The second term really changes the propagator  from $\mathbf{r}=0$ to $\mathbf{r}'$, because a dipole can be situated at $\mathbf{r}=0$, that first couples via a high-order
dipole interaction to
a dipole at $\mathbf{r}'$ (a single coupling is already counted in the effective medium) before finally arriving at $\mathbf{r}$.
In the following we show 1) that this coupling fully disappears at large distance (contrary to Ref.~\cite{theoL})
and 2) that for small distances we recover the dipole-dipole coupling found earlier in Eq.~(\ref{GL}), with the complex wave number $K_L(\infty)$.

For $kr\gg 1 $, we can take $\mathbf{G}_{0,L}(\mathbf{r})$ out of the integral, and recognize the remainder as the boomerang self-energy at $p=0$. Hence,
\begin{eqnarray}\label{DrL}
    \mathbf{G}_L(k, \mathbf{r} \rightarrow \infty ) =   \mathbf{G}_{0,L}(K_L(0),\mathbf{r})
\end{eqnarray}
This result agrees with Eq.~(\ref{DD}) and was seen to cancel against a similar term in the transverse part of the Dyson Green's function.   For  $kr \ll 1$ we can write,
\begin{eqnarray*}
    \mathbf{G}_L(k ,\mathbf{r}\rightarrow 0) &= &  \mathbf{G}_{0,L}((k^2 -\Sigma_0)^{1/2} ,\mathbf{r})   \\
     &-&\frac{n^2 t^2 }{k^2} \int d^3\mathbf{r}' \, \mathbf{G}_{0,L}(\mathbf{r}-\mathbf{r}')\cdot
     \mathbf{G}_0(\mathbf{r}')     \\
     &+&   \frac{n^2}{k^2} \int d^3\mathbf{r}' \, \mathbf{G}_{0,L}(\mathbf{r}-\mathbf{r}')\cdot
     \frac{  t^2 \mathbf{G}_0(\mathbf{r}')}{1-t^2\mathbf{G}_0^2(\mathbf{r}')}
\end{eqnarray*}
The last term is regular and $\mathbf{r}=0$ can be inserted. The second term is  equal to $-(n^2t^2/k^4)  \mathbf{G}_{0,L}(\mathbf{r})$ and adds up to
the first term. Since by Eq.~(\ref{sigma2}) we have
$K_L^2(\infty) = k^2 -\Sigma_0 + n^2t^2/k^2$,
\begin{eqnarray}\label{DrS}
    \mathbf{G}_L(k,\mathbf{r} \rightarrow 0 ) &= &  \mathbf{G}_{0,L}(K_L(\infty),\mathbf{r}) +\mathbf{D}(0)\nonumber
\end{eqnarray}
This agrees with Eq.~(\ref{GL}) and attributes a finite complex, frequency-dependent value to $\mathbf{D}(\mathbf{r}=0)$,
\begin{equation}
   \mathbf{ D}(0)= \frac{n^2t^2}{k^2} \int d^3\mathbf{r}' \, \frac{  \mathbf{G}_{0,L}(\mathbf{r}')\cdot
       \mathbf{G}_0(\mathbf{r}')}{1-t^2\mathbf{G}_0^2(\mathbf{r}')}
\end{equation}
We note that $\mathbf{D}(0)$ is  negligible compared to the dipolar coupling $G_L \sim 1/r^3$.

\subsection{Density of states}\label{sectionDOS}

In this section we derive the density of states (DOS) of electromagnetic waves in disordered media, express it in terms of the effective medium,
identify its longitudinal part (DOLS) and calculate it for our model of randomly positioned
electric dipoles with volume number density $n$.
The total electromagnetic spectral density at frequency $\omega = kc_0$  in a polarizable medium is defined by
\begin{equation*}
    N_{tot}(k)  = \frac{|k|}{c_0} \mathrm{TR}\, \delta\left(k^2 - \mathcal{H}\right)
\end{equation*}
with $\mathcal{H} = \varepsilon(\mathbf{r})^{-1/2} (\mathbf{p}^2-\mathbf{pp})\varepsilon(\mathbf{r})^{-1/2}$ the Helmholtz operator and $\mathrm{TR}$ the trace
in the Hilbert space spanned by all
eigenfunctions, including a strongly degenerate longitudinal eigenspace  with eigenvalue $0$. Written in this way, the spectral density is defined (and equal) for positive and negative frequencies and
normalized to the dimension of the Hilbert space,
\begin{equation*}
    \int_{-\infty}^\infty d\omega \,  N_{tot} (k) = \mathrm{TR}
\end{equation*}
independent of $\varepsilon (\mathbf{r})$, and formally infinite. We can work out the trace in real space as
\begin{equation*}
    N_{tot}(k)  = \int d^3\mathbf{r}\, \frac{|k|}{c_0} \langle \mathbf{r}| \mathrm{Tr}\, \delta\left(k^2 - \mathcal{H}\right)|\mathbf{r}\rangle
\end{equation*}
with $\mathrm{Tr}$ the trace over 3 polarizations only, and identify the integrand as the local density of states,
\begin{equation*}
    N(k,\mathbf{r})= -\frac{k}{c_0} \frac{1}{\pi }  \mathrm{Im} \mathrm{Tr}\, \mathbf{G}_\mathcal{H}( k+i\epsilon, \mathbf{r}, \mathbf{r})
\end{equation*}
with $\mathbf{G}_\mathcal{H} = [(k+i\epsilon)^2 - \mathcal{H})]^{-1}$ . After
ensemble-averaging it becomes independent of $\mathbf{r}$, and we can express it in terms of the Dyson Green's function (\ref{dyson}),
\begin{eqnarray}\label{DOS1}
    \langle N(k,\mathbf{r})\rangle &=& \left\langle -\frac{k}{c_0} \frac{1}{\pi }
    \right. \nonumber \\
    &\times& \left.
    \mathrm{Im} \mathrm{Tr}\,
    \langle \mathbf{r}| \, \varepsilon^{1/2}(\mathbf{r})\cdot  \mathbf{G}(k+i\epsilon)\cdot\varepsilon^{1/2}(\mathbf{r})
    |\mathbf{r}\rangle
    \vphantom{\frac{1}{1}}
    \right\rangle\nonumber \\
    &=& -\frac{k}{c_0} \frac{1}{\pi } \sum_\mathbf{p}  \mathrm{Im} \, \mathrm{Tr }\, \frac{p^2 \mathbf{\Delta}_p}{k^2} \cdot \mathbf{G} (k+i\epsilon, \mathbf{p})
\end{eqnarray}
Both lines in this expression count, by construction, all states but, quite surprisingly, the second line projects on the transverse states only with however a
large weight on large wave numbers $p \gg k$. The reason is that the first line counts electrical energy, including the longitudinal modes, whereas the second
line counts magnetic energy, which has only transverse modes. Equation~(\ref{DOS1}) states that the density of states can be calculated from either the magnetic or electrical energy, provided
the latter includes also the longitudinal states.

For our model of electric dipoles we expect that the DOS is the sum of transverse traveling waves and stocked longitudinal waves.
To show this   we go back
to the first line of Eq.~(\ref{DOS1}). For $\varepsilon (\mathbf{r}) = 1 + {\delta\varepsilon}(\mathbf{r})$, we identify $\mathbf{V}=-{\delta\varepsilon}(\mathbf{r})k^2 $
as the  interaction operator in the Born series of light scattering \cite{PR}.
Before doing the configurational average, we can consider $M$ dipoles in a finite volume $V$ (see also Appendix \ref{appA}). Rigorous scattering theory imposes the operator identity
 $\mathbf{V}\cdot \mathbf{G}(k) = \mathbf{T} \cdot \mathbf{G}_0(k)$. Hence
\begin{eqnarray*}
    N(k,\mathbf{r}) &=& -\frac{k}{c_0} \frac{1}{\pi } \mathrm{Im} \mathrm{Tr}
    \, \mathbf{G}(k+i\epsilon, \mathbf{r},\mathbf{r})  \\
    &+& \frac{k}{c_0} \frac{1}{\pi } \mathrm{Im} \mathrm{Tr} \langle \mathbf{r}| \frac{\mathbf{T}}{k^2} \cdot
     \mathbf{G}_0(k+i\epsilon)
    |\mathbf{r}\rangle
\end{eqnarray*}
This equation is still exact and depends on the position $\mathbf{r}$. Since the polarizability density ${\delta\varepsilon}(\mathbf{r})$ has disappeared explicitly we can
consider the special case of scattering from identical, impenetrable  electric dipoles, associated with a dielectric susceptibility ${\delta\varepsilon}(\mathbf{r}) \rightarrow \infty$, and described
 by Eq.~(\ref{born}). For  $M$ such dipoles,
\begin{equation}\label{Tmma}
    \mathbf{T}(k) = \sum_{mm'}^M  \mathbf{T}_{mm'}(k)|\mathbf{r}_m\rangle  \langle \mathbf{r}_{m'}|
\end{equation}
with, for $m,m'$ fixed,  the $3 \times 3 $ matrix $\mathbf{T}_{mm'}(k)$.
To have $\mathbf{G}(\mathbf{r}_m, \mathbf{r})=0$ inside all dipoles at $\mathbf{r}_m$ and for arbitrary $\mathbf{r}$ outside imposes that  $\mathbf{T}_{mm'}(k)$  be
given by the inverse of the $3M \times 3M $ matrix $-\mathbf{G}_0 (k, \mathbf{r}_m, \mathbf{r}_m')$.
It easily follows that
 \begin{equation*}
 \langle \mathbf{r}| \mathbf{T}\cdot  \mathbf{G}_0(k+i\epsilon)
    |\mathbf{r}\rangle = -\mathbf{1} \sum_{m=1}^M \delta(\mathbf{r}-\mathbf{r}_m) = -n(\mathbf{r})
\end{equation*}
Since this is purely real-valued, it cancels in the expression above for $N(k,\mathbf{r})$. Upon averaging and letting $M,V \rightarrow \infty$ at constant number density,
the remaining term yields
 \begin{equation}\label{DOS3}
     \langle N(k) \rangle = -\frac{k}{c_0} \frac{1}{\pi } \sum_\mathbf{p} \mathrm{Im} \mathrm{Tr}\,
    \mathbf{G}(k+i\epsilon, \mathbf{p})
 \end{equation}
 in terms of the Dyson Green's function~(\ref{dyson}). This is recognized as
 $\langle |\mathbf{E}(\mathbf{r})|^2\rangle$, proportional to the  energy density $\langle \mathbf{E}(\mathbf{r})^2\rangle /8\pi $,
 averaged over disorder and cycles, and having
 both longitudinal $N_L(k)$ and transverse $N_T(k)$ parts.
 We emphasize that Eq.~(\ref{DOS3}) only applies for our model that excludes any light inside the scatterer. As a result
 no stored energy density $\langle \mathbf{E}(\mathbf{r})\cdot\mathbf{ P}(\mathbf{r})\rangle$ exists as e.g. in Mie scattering \cite{PR}. In this model,
  the stocked dipole-dipole energy is entirely described by longitudinal (electric) waves.

\begin{figure}
  \includegraphics[width=7cm]{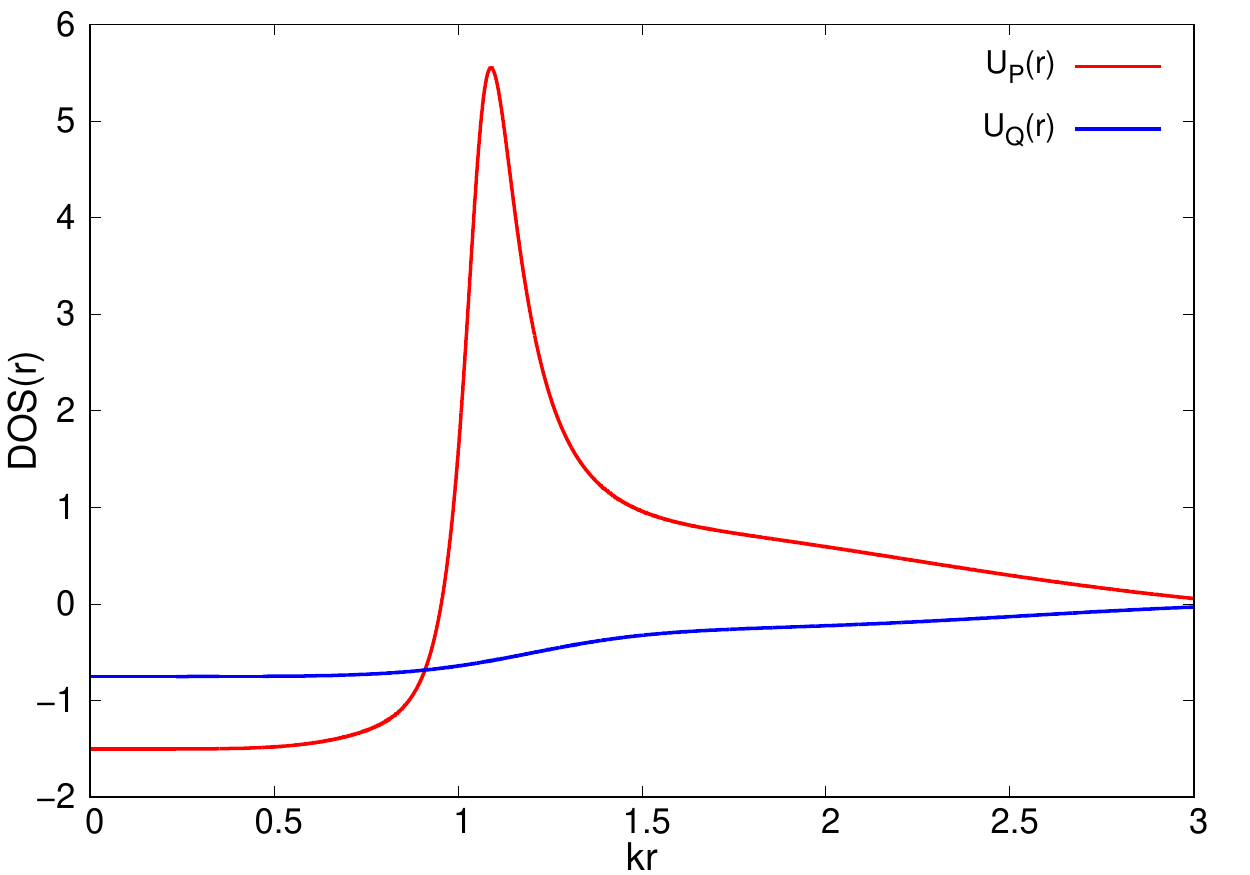}
  \includegraphics[width=7cm]{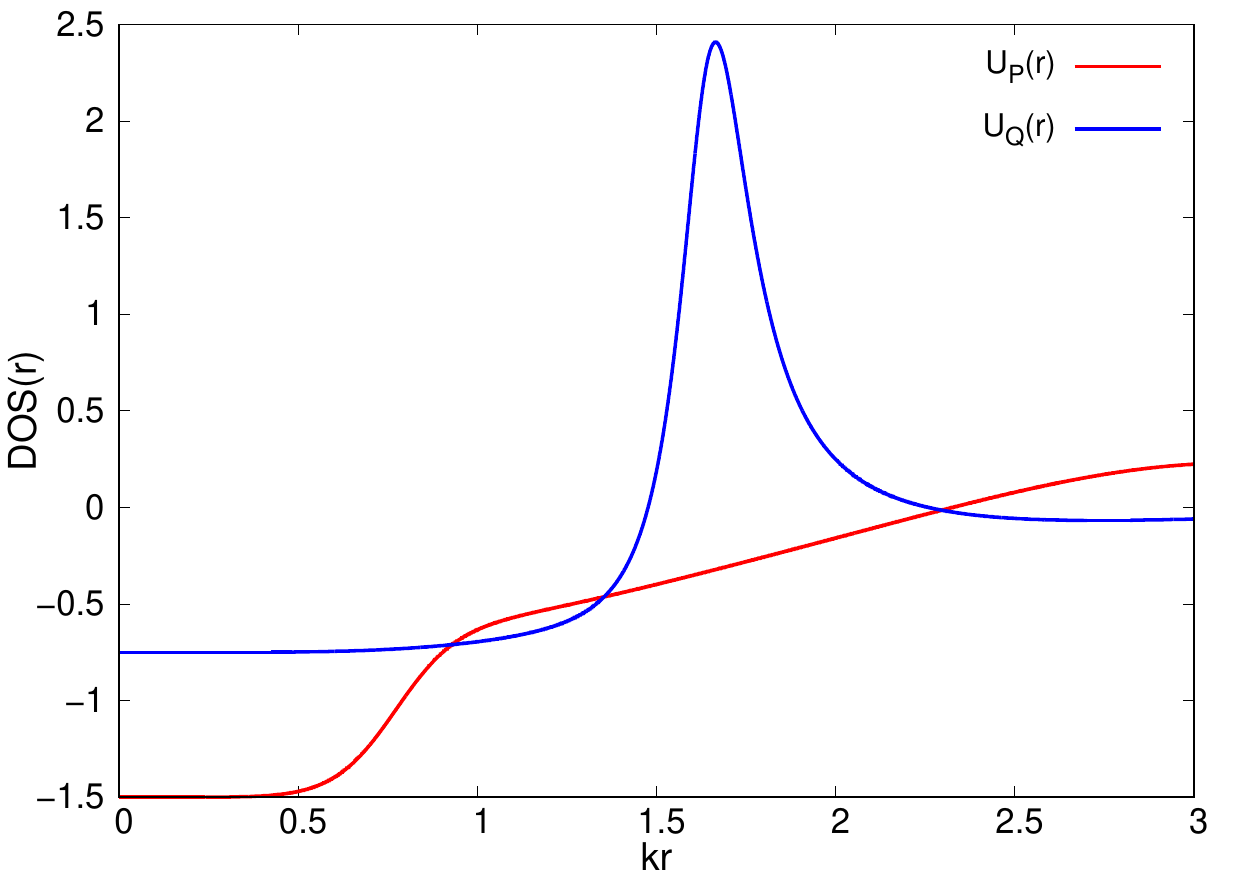}
  \caption{The contribution of dipole-dipole coupling  to the DOS as a function of distance between the dipoles.
  The volume integral of the functions shown produces  the second term of Eq.~(\ref{DOSISA}). $U_P(r)$ is associated with the electric field
  perpendicular to $\mathbf{r}$, $U_Q(r)$ with the electric field directed along $\mathbf{r}$. Top: $\delta = (\omega-\omega_0)/\gamma = -0.5$ (redshift). Bottom: $\delta = 0.5$
  (blueshift). $U_P(r)$ has a subradiant peak only for positive detuning whereas $U_Q(r)$ only for negative detunings.
  }\label{PQfig}
\end{figure}

We can insert the Dyson Green's function obtained in Sec.~\ref{greensection} into Eq.~(\ref{DOS3}),
\begin{eqnarray*}
    \langle N(k) \rangle &=& - \frac{k}{\pi c_0} \mathrm{Im} \left[\sum_\mathbf{p} \frac{1}{K^2_L(\infty)}
    \right. \\
    && \left.+ \sum_\mathbf{p} \left( \frac{1}{K^2_L(p)}
    - \frac{1}{K^2_L(\infty)} \right)\right. \\ &&  \left. +2 \sum_\mathbf{p}  \frac{1}{K^2_T(p) -p^2} \right]
\end{eqnarray*}
The first  {diverging }term  { expresses the singular Lorentz cavity }$ \sum_\mathbf{p} = \delta(\mathbf{r}=0)$.
It is entirely governed by longitudinal excitations, and is regularized using
$\sum_\mathbf{p} = 3/\alpha(0)= k_0^3 Q_0/2\pi $ consistent with  Eq.~(\ref{born}). The second term, proportional to $\mathbf{D}(\mathbf{r}=0)$ in Eq.~(\ref{GL}),
is non-zero but a factor $Q_0$ smaller and shall be neglected. Finally the last term is just the density of states of transverse waves. We shall assume the existence of a
well-defined complex pole $K_T= k_T + i /2\ell$. This gives
 \begin{eqnarray}\label{DOSLT}
    \langle N(k) \rangle = \frac{k}{2\pi^2 c_0}  \left[ - Q_0 \mathrm{Im} \frac{k_0^3}{K^2_L(\infty)}  +  {k_T}\right]
\end{eqnarray}
The ratio of longitudinal and transverse DOS is thus
\begin{equation}\label{ratioDOSLT}
    \frac{ \langle N_L(k) \rangle }{ \langle N_T(k) \rangle } = - Q_0 \frac{k_0^3 }{k_T} \mathrm{Im} \frac{1}{K^2_L(\infty)}
\end{equation}
In view of the factor $Q_0$ this can be a large number, proportional to the density of the dipoles. For low density is
$K_L(\infty) \approx  K_T \approx k + i /2\ell$ so that $ \langle N_L \rangle /\langle N_T \rangle = Q_0/k\ell$. This ratio will be discussed in the next section

\begin{figure}
  \includegraphics[width=8cm]{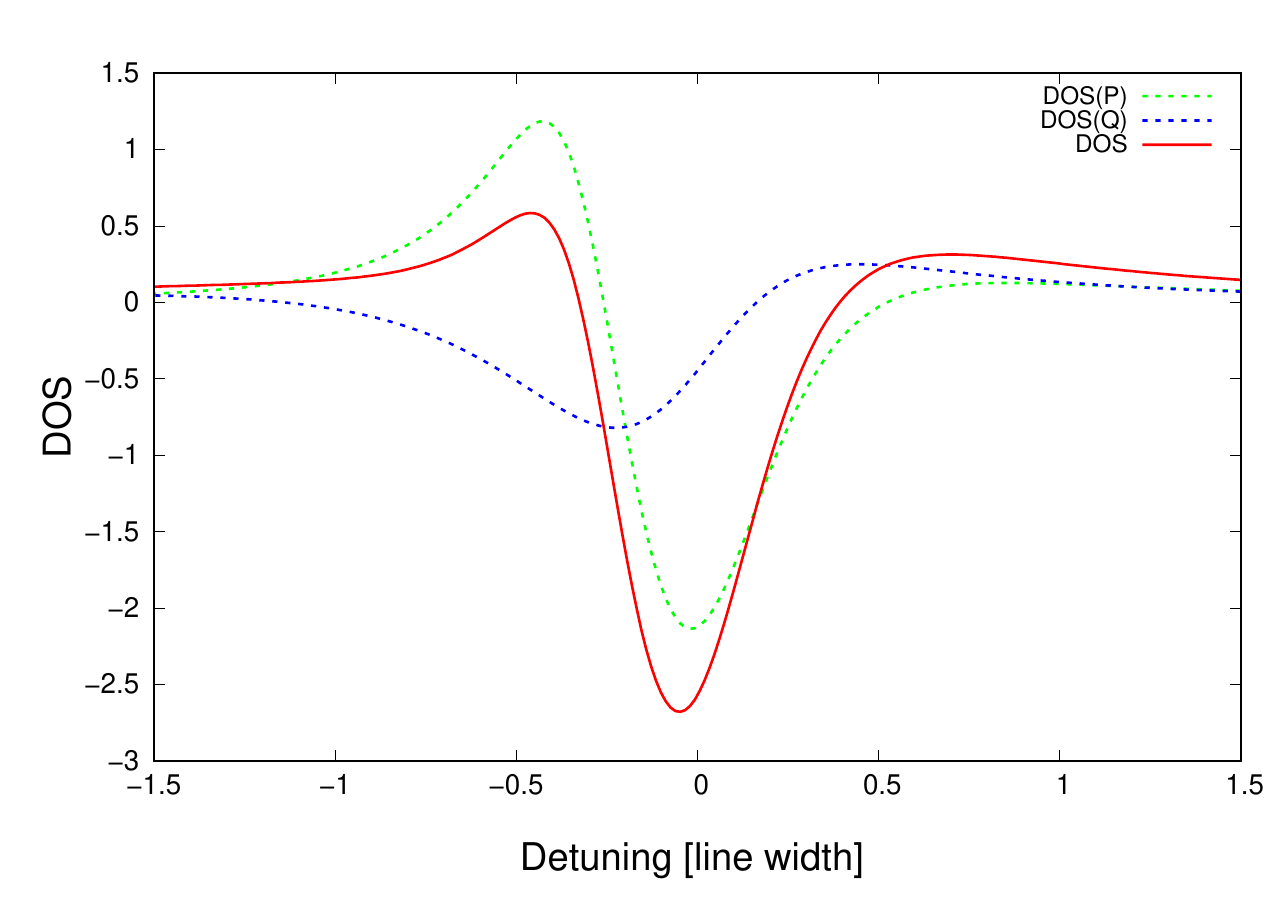}
  \caption{The contribution of dipole-dipole coupling  to the DOS (in units of $N_0 \times Q_0 \times (4\pi n/k_0^3)^2$) as a function of
  detuning $\delta = (\omega-\omega_0)/\gamma$.
  The dashed lines show the separate contributions of the modes with electric field perpendicular ($P$) and parallel $(Q)$ to the separation vector
  $\mathbf{r}$.}\label{DOSDD}
\end{figure}

A rigorous expression can be derived for DOS without relying on the existence of a complex pole by taking into account the $p$-dependence of the
self-energy $\Sigma_T(k,p)$ associated with the scattering from two electric dipoles. A direct expansion in dipole density yields
\begin{eqnarray*}
   \langle  N(k&&) \rangle = -\frac{k}{\pi c_0} \mathrm{Im} \mathrm{Tr}\,   \sum_\mathbf{p}  \left[ \vphantom{\frac{\, }{\, }}
    \mathbf{G}_0(k, \mathbf{p}) \right. \\
    &&+ \mathbf{G}_0(k, \mathbf{p}) \cdot \mathbf{\Sigma}(k, \mathbf{p})\cdot \mathbf{G}_0(k, \mathbf{p})\\
    &&+ \left.
    \mathbf{G}_0(k, \mathbf{p}) \cdot \mathbf{\Sigma}(k, \mathbf{p})\cdot \mathbf{G}_0(k, \mathbf{p})\cdot \mathbf{\Sigma}(k, \mathbf{p})\cdot
    \mathbf{G}_0(k, \mathbf{p})  \vphantom{\frac{\, }{\, }} \right] \\
    &&+ \mathcal{O}(n^3)
\end{eqnarray*}
Several singular longitudinal terms, stemming from the Lorentz cavity  can be seen to cancel.
The first term describes the free electromagnetic field and the longitudinal field drops out trivially.
The longitudinal component of the second term contains a  singular Lorentz cavity $(nt + \Sigma_{\mathrm{Loop}} -n^2t^2/k^2)
\sum_\mathbf{p} \mathbf{G}_L^2(\mathbf{p})$ stemming from Eq.~(\ref{sigma2}).
Similarly, the third  term generates a singular longitudinal contribution $n^2t^2 \sum_\mathbf{p} \mathbf{G}_L^3(\mathbf{p})$ that cancels exactly against the local field $-n^2t^2/k^2$  generated by the previous term.
We can work out the wave number integral in the expression for $\langle N(k) \rangle$ exactly by
  inserting Eq.~(\ref{self}), and  use the cyclic property of the trace,
  \begin{eqnarray*}
    \langle N(k) \rangle  &=& \frac{k^2}{2\pi^2 c_0} + \frac{k}{\pi c_0}\mathrm{Im} \mathrm{Tr}\, \left[
  nt \frac{\partial}{\partial k^2} \mathbf{G}_0 (k,0) \right. \\
  & +&
   n^2 t^3 \int d^3 \mathbf{r} \frac{\mathbf{G}_0^2(\mathbf{r})}{\mathbf{1}-t^2 \mathbf{G}_0^2(\mathbf{r})} \cdot \frac{\partial}{\partial k^2} \mathbf{G}_0 (k,0)
   \\
   &+& n^2 t^4 \int d^3 \mathbf{r} \frac{\mathbf{G}_0^3(\mathbf{r})}{\mathbf{1}-t^2 \mathbf{G}_0^2(\mathbf{r})} \cdot
  \frac{\partial}{\partial k^2}  \mathbf{G}_0 (k, \mathbf{r}) \\
   &+& \left. n^2t^2 \int d^3\mathbf{ r} \, \mathbf{G}_0 (k, \mathbf{r}) \cdot
   \frac{\partial}{\partial k^2}  \mathbf{G}_0 (k, -\mathbf{r}) \right]
    \end{eqnarray*}
We have transformed the integral over wave vectors $\mathbf{p}$ of the last term $ \mathbf{G}_0 \cdot \mathbf{\Sigma} \cdot \mathbf{G}_0 \cdot \mathbf{\Sigma} \cdot
    \mathbf{G}_0$  in $\langle N(k) \rangle$  to real space. Using again the relation $\mathbf{1}/t(k) = -  \mathbf{G}_0 (k, \mathbf{r}=0) $ this can be rearranged to
 \begin{eqnarray}\label{friedel}
   \langle N(k) \rangle  &-& N_0(k) = - \frac{3n}{2\pi} \frac{d}{dk} \mathrm{Im}\, \ln t \nonumber \\
   &-& \frac{n^2}{4\pi c_0} \frac{d}{d k} \mathrm{Im} \mathrm{Tr} \int d^3 \mathbf{r} \ln \left[\mathbf{1}-t^2 \mathbf{G}_0^2(\mathbf{r}) \right]
 \end{eqnarray}
 with $N_0 =  {k^2}/{2\pi^2 c_0}$ the LDOS of transverse waves in free space. The appearance of a full frequency derivative in the DOS
 is a manifestation of Friedel's theorem
 \cite{mahan}.
 The second term is recognized as the
 dipole-dipole energy
 expressed as the ``return trip operator'', widely used in the theory of Casimir energy in matter \cite{miloni}, and involves loop paths only. The integral is well-defined at both $r=0$ and $r\rightarrow \infty$. Since the dominating frequency dependence comes from $dt/dk \approx  -2Q_0t^2/6\pi$,
  \begin{eqnarray}\label{DOSISA}
 \frac{ \langle \Delta  N(k ) \rangle  }{N_0(k)}&=&  -\frac{Q_0}{3k^2}
\mathrm{Im }\mathrm{Tr} \left[ {nt }\mathbf{1} + n^2\int d^3 \mathbf{r}  \frac{t^3\mathbf{G}_0^2(\mathbf{r})}{1-t^2\mathbf{G}_0^2(\mathbf{r})}  \right]
\nonumber
 \\ &=&  -\frac{Q_0}{k^2} \mathrm{Im } \left( \Sigma_{\mathrm{ISA}}  +  {\Sigma}_{\mathbf{\mathrm{Loop}}}  \right)
\end{eqnarray}
This expression suggests that in general the modification of DOS is dominated by ISA $+$ loop diagrams, describing longitudinal excitations, even if mediated by
transverse, propagating waves.
This, in turn, implies  that
the complex longitudinal wave number $K_L(\infty)$ is governed by loop diagrams only. In Appendix \ref{appA} we demonstrate that this statement holds
rigorously. More precisely, if
we recall the $T$-matrix~(\ref{Tmma})  of $M$ electric dipoles randomly distributed in a volume $V$, then
\begin{equation}\label{allloop}
    \frac{1}{K_L^2(\infty)} = \frac{1}{k^2} +  \frac{1}{3k^4} n \left\langle  \mathrm{Tr}\,\mathbf{T}_{mm}(k) \right\rangle
\end{equation}
for $M\rightarrow \infty$ at constant  $M/V=n$. The first two terms in the density expansion clearly coincide with Eq.~(\ref{DOSISA}). All higher order terms are rigorously loop diagrams
and
the ensemble-average of the diagonal element $\mathbf{T}_{mm}$ over all other $M-1$ dipoles must make it proportional to the identity matrix.

\begin{figure}
\includegraphics[width=0.7\columnwidth, angle=-90]{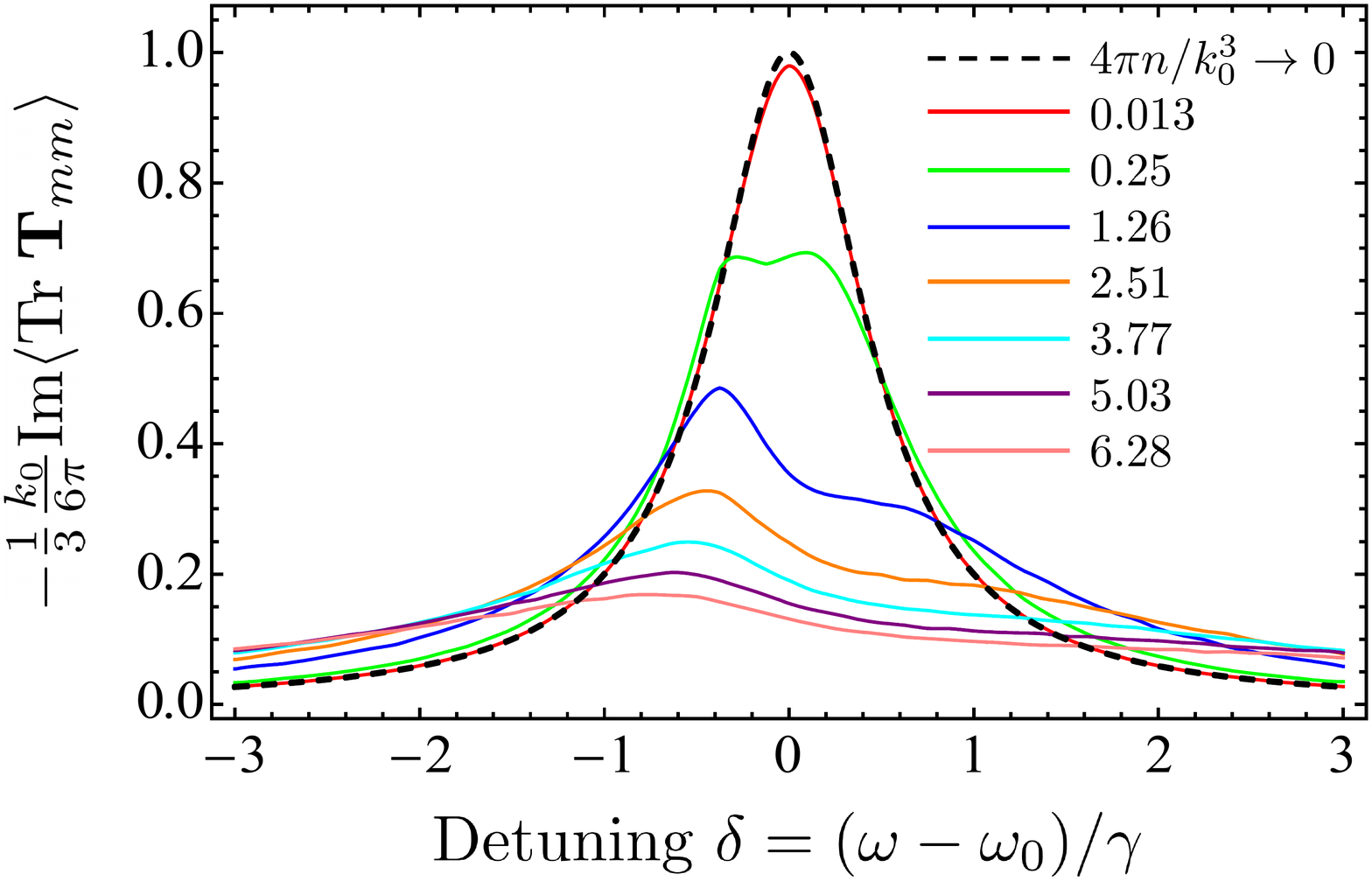}\\
\vspace*{-5mm}
\includegraphics[width=0.7\columnwidth, angle=-90]{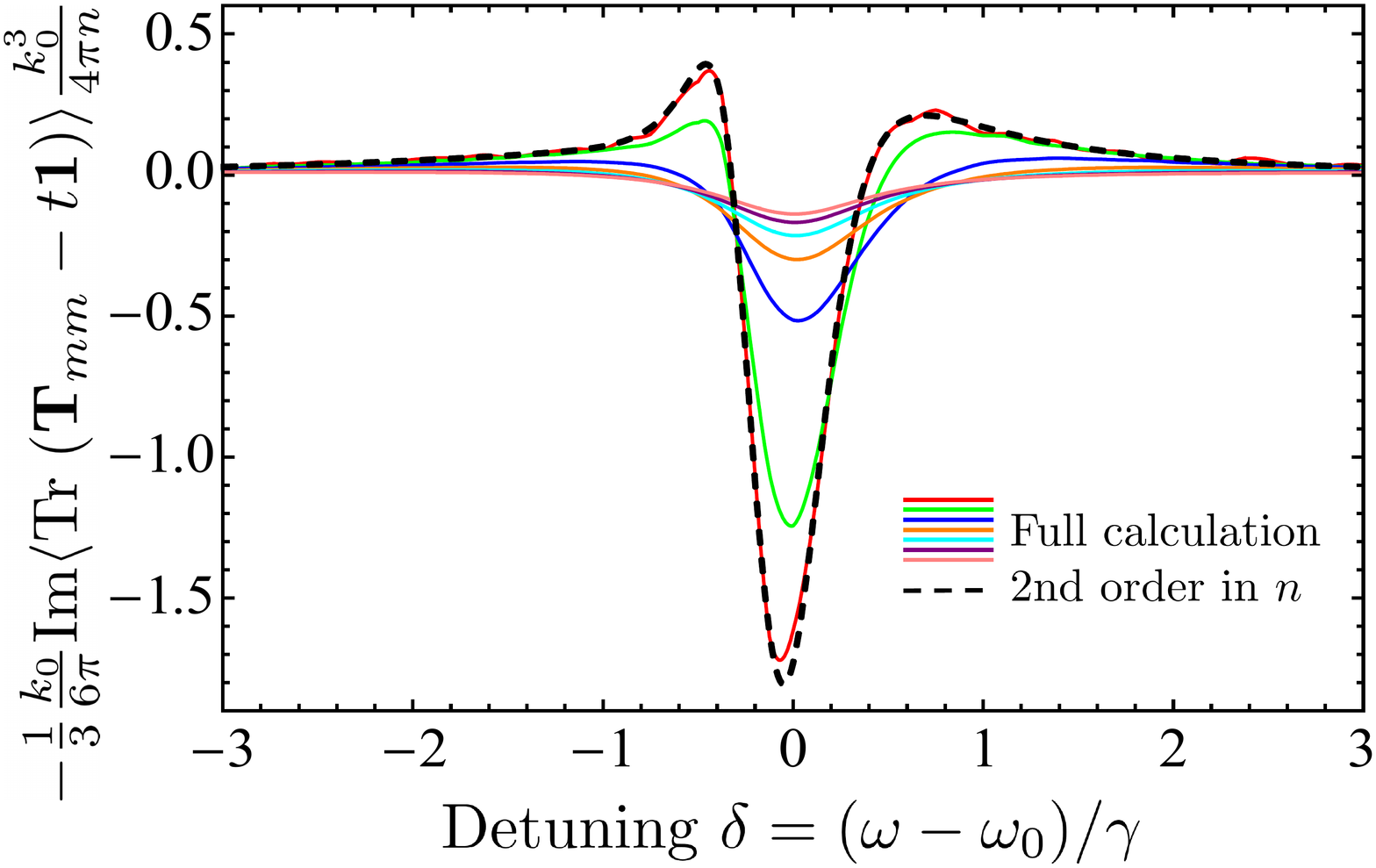}
\vspace*{-7mm}
\caption{\label{fig_dosl}
Top: Numerical simulation of the averaged imaginary part of the diagonal elements of the $T$-matrix as a function of detuning, for several densities $n$ of the dipoles.
Via Eqs.~(\ref{DOSLT}) and (\ref{allloop}) this quantity determines the DOLS. Bottom: Same but with the ISA approximation subtracted and normalized by
$4 \pi n/k_0^3$.
The dashed line shows the second-order in density term in Eq.~(\ref{DOSISA}) as also shown in Fig.~\ref{DOSDD}.
}
\end{figure}

Near resonance  the first ISA term of $\langle \Delta N\rangle/N_0$ has a Lorentzian profile with a large peak
height inversely proportional to $Q_0$, to be associated with the excitation of a single dipole.
The second term of Eq.~(\ref{DOSISA}) becomes important when $4\pi n/k_0^3 \approx 1$
and constitutes an
inhomogeneous contribution to the line-profile. Using
{$\mathbf{G}_0(k, \mathbf{r}) = -\exp(i k r)/(4 \pi r) [P(kr) \mathbf{\Delta}_r + Q(kr) \hat{\mathbf{r}}\hat{\mathbf{r}}]$ for $\mathbf{r} \ne 0$}
\cite{PR} the integrand
can be split up
into two interactions $U_Q(r)$ and $U_P(r)$, that govern the near-field coupling of two dipoles in real space. Both are shown in Fig.~\ref{PQfig}.
The total dipole-dipole coupling, shown in Fig.\ \ref{DOSDD}, is negative around the resonance.
 Because local field singularities cancel in the DOS,  the solid curve in Fig.\ \ref{DOSDD} is the same as  found in
Ref.~\cite{cherro}.

In Fig.~\ref{fig_dosl} we have calculated numerically
the diagonal elements of the $3M\times 3M$ matrix $\mathbf{T}_{mm}$ for $M = 10^4$ dipoles homogeneously distributed in a sphere at density $n=M/V$,
thereby averaging over all $3M$ diagonal elements as well as
over 10 independent random configurations of the dipoles. This calculation confirms that $K_L(\infty)$ is a genuine complex quantity and in general different from the complex wave
 vector $K_T = k_T + i/2\ell$ associated with the transverse modes (see also Figs.~\ref{fig_keff} and \ref{fig_tr} in Appendix B).  For low dipole densities, the calculation agrees accurately with the analytical calculation of the loops between two dipoles in Eq.~(\ref{DOSISA}). The line profile of the longitudinal
DOS broadens significantly well beyond the single-dipole line profile as the dipole density increases. Nevertheless,
the total surface underneath remains constant. This is to be expected since  each dipole contributes exactly $3$ microstates to the DOS and this number
cannot be affected by dependent scattering (see Appendix A).


\subsection{Equipartition between longitudinal and transverse waves}

 {In the following we show that transverse and longitudinal excitations are mutually converted both  by independent scattering from a single dipole and by recurrent scattering from two dipoles.
The efficiency of this process is proportional to the average density of longitudinal (DOLS) or transverse (DOTS) states  of the disordered medium, and leads eventually to a steady ratio of longitudinal and
transverse energies.  We identify \emph{exactly two } events where the longitudinal field creates a singularity and which therefore dominate the dynamics of this process. }

Let $\tau_{i} $  {be the life time of the excitation} $i$   due to scattering to all other modes. { Since scattering is assumed elastic all involved modes have the same frequency. }
The scattering implies a change in  wave number $\mathbf{p}$  {and/or in polarization, characterized by either the longitudinal polarization} $\hat{\mathbf{p}}$  {or by one of the  two transverse polarizations}
$\hat{\mathbf{g}}_T$. Let  $\rho_i$ be the density of states of excitation $i$,  and $n_i$ its average occupation number.
If $U_{ji}$ is the matrix element  converting $j$ to $i$,   {the  transport equation takes the generic form} ,

\begin{equation}\label{transportgen}
 \frac{d n_i }{dt} = - \frac{ n_i}{\tau_{i} }+   \sum_j \rho_j n_j U_{ji}
\end{equation}
 {The total energy $\sum_i \rho_i n_i(t)$ is conserved in time provided that $\rho_i /\tau_{i} =  \sum_j \rho_j U_{ij}  $.
This is {
akin to the Ward identity}~(\ref{ward}), and we can identify - apart from  a constant factor with the dimension of a velocity} -
$U_{ij}$ with the polarization matrix elements of the irreducible vertex and  $ \rho_i/\tau_i$ with the imaginary part of the self energy $- \mathrm{Im }\Sigma_i$. If $U_{ij} = U_{ji} $ the transport equation has
the solution $n_i(t) \equiv n$ to which it finally converges and which corresponds to  equipartition of energy in phase space.
 {The conversion rate} $1/\tau(T \rightarrow L' )$ of one transverse excitation $T=(\mathbf{\hat{g}}_T,\mathbf{p})$ to all longitudinal
excitations $L'= (\mathbf{\hat{p}}',\mathbf{p}')$ of
arbitrary wave number $\mathbf{p}'$ can thus be identified as,

 \begin{equation*}
 \frac{ \rho_T(\mathbf{p})}{\tau(T \rightarrow L') }=   \sum_{\mathbf{p}'} \rho_L(\mathbf{p}') \times \mathbf{ \hat{p'}\hat{p'}}\cdot U_{\mathbf{p}\mathbf{p}' }
 \cdot \hat{\mathbf{g}}_T \hat{\mathbf{g}}_T
\end{equation*}
with  $\rho_L=  -\mathrm{Im} \, G_L(k,p)\approx  -n\mathrm{ Im }\, t(k) /k^4$ the DOLS, independent of $\mathbf{p}$. { Independent scattering from an electric dipole }
gives rise to a  conversion rate from
a transverse excitation to a longitudinal  excitation of arbitrary wave number, proportional to the
  vertex   $U^{\mathrm{ISA}}= n |t|^2$,

\begin{eqnarray}\label{div1}
 \frac{ \rho_T(\mathbf{p})}{\tau(T \rightarrow L') } =  \sum_{\mathbf{p}'} U^{\mathrm{ISA}}  \rho_L(\mathbf{p}')  =
    -Q_0\frac{n^2 |t|^2 \mathrm{Im}\, t }{6\pi  k}
   \end{eqnarray}
Since the  integral diverges at large $p'$,
 we have used the same regularization $\sum_\mathbf{p } = Q_0 k^3/2\pi$ as the one employed earlier for the $T$-matrix of one dipole. The resulting scattering rate is large and positive.

\begin{figure}
  \includegraphics[width=0.95\columnwidth]{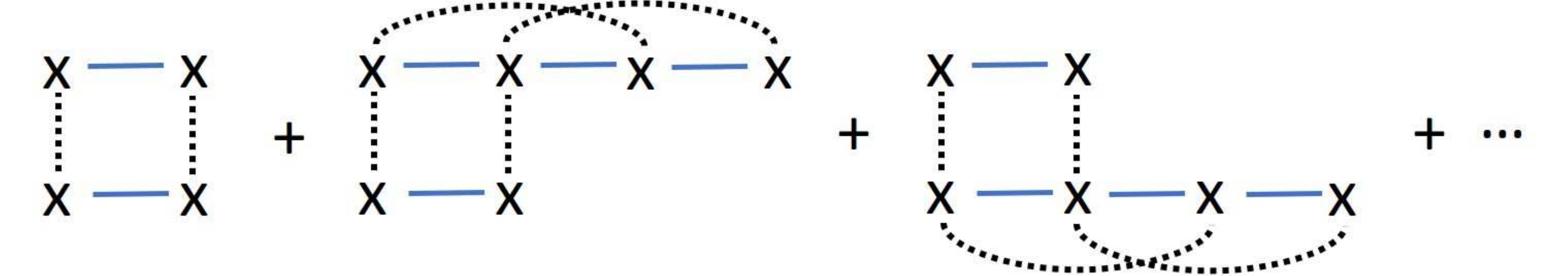}
  \caption{Diagrammatic presentation of the ladder series (without external lines)  involving two different electric dipoles. Dashed lines connect identical dipoles, solid lines denote the Green's tensor $\mathbf{G}_0(\mathbf{r})$,
  crosses denote transition matrix $t(k)$, bottom line denotes Hermitian conjugation. The first diagram on the left is reducible (a simple product) and is not part of the
  collision operator $U_{\mathbf{pp'}}$.}\label{figIL}
\end{figure}

  { In independent scattering  the conversion from longitudinal excitations back to transverse waves does not have the same singularity and has a rate -- in the same units --
  equal to the inverse mean  free path.}
 The matrix element  $U^{(2)}_{\mathbf{pp}'}$ involving all recurrent scattering from two dipoles \cite{cherro} can mode-convert any initial state to transverse states,
 with a rate proportional to $n^2$, of same order as the process in Eq.~(\ref{div1}). { We can establish that the  vertex} $U^{(2)}_{\mathbf{pp}'}$,  {together with
 the self-energy}~(\ref{self})  {associated with two dipoles, satisfies the Ward identity}~(\ref{ward}),  {but the detailed proof is beyond the scope of this work.}
 A close look identifies \emph{only one }event part of  $U^{(2)}_{\mathbf{pp}'}$ that gives rise to a singular scattering rate, displayed in Fig.~\ref{figIL}.
 These so-called irreducible ladder diagrams $U^{(\mathrm{LAD})}_{\mathbf{pp}'}$   add up to
 \begin{eqnarray}\label{UL2}
    U^{(\mathrm{LAD})}_{\mathbf{pp}'} &=& n^2 |t|^4 \int d^3 \mathbf{r} \,  \nonumber \\
      && \left[ \frac{\mathbf{G}_0}{1-t^2\mathbf{G}_0^2}\left( \frac{\mathbf{G}_0}{1-t^2\mathbf{G}_0^2}\right)^*
    - \mathbf{G}_0\mathbf{G}^*_0  \right]
 \end{eqnarray}

Our tensor notation is $  (\mathbf{AB})_{ij|kl} = A_{ik}B_{lj}$, equivalent to $(\mathbf{AB})\cdot \mathbf{S}= \mathbf{A}\cdot \mathbf{S}\cdot \mathbf{B}$.
 This vertex is independent of $\mathbf{p}$ and $\mathbf{p}'$. The second term must be subtracted since it stands for a reducible event that is not
 part of the collision operator $U$. However, this subtraction creates a diverging contribution at $\mathbf{r}=0$ in the integral due to the singular longitudinal field.
 To repair this in a way consistent with previous sections, we extract the transverse photon propagator $\mathbf{G}_{0,T}(\mathbf{r})$,
and write
 \begin{eqnarray*}
    U^{(\mathrm{LAD})}_{\mathbf{pp}'} &=& n^2 |t|^4 \int d^3 \mathbf{r}\,  \left[ \frac{\mathbf{G}_0}{1-t^2\mathbf{G}_0^2}\left( \frac{\mathbf{G}_0}{1-t^2\mathbf{G}_0^2}\right)^*
    \right. \nonumber \\
    &-& \left. \mathbf{G}_{0,T}\mathbf{G}^*_{0,T}  \vphantom{ \frac{\,}{\, }}      \right] \\
    &+& n^2 |t|^4 \sum_{\mathbf{p}''}\,  \left[ \mathbf{G}_{0,T}(\mathbf{p}'')\mathbf{G}^*_{0,T}(\mathbf{p}'') - \mathbf{G}_0(\mathbf{p}'')\mathbf{G}^*_0(\mathbf{p}'')
 \right]\end{eqnarray*}
We have used Parseval's identity to convert the second integral into an integral over wave vectors. The first term of $U^{(\mathrm{LAD})}_{\mathbf{pp}'}$ now converges at small $r$ and shall be ignored,
the second term can be dealt with as
before, giving
 \begin{eqnarray}
    U^{(\mathrm{LAD},s)}_{\mathbf{pp}'} = -Q_0\frac{ 3 n^2 |t|^4 }{6\pi k} {S}
    + Q_0\frac{ n^2 |t|^4 }{6\pi k_0} \left(\frac{1}{3}\mathbf{ 11} - {S}
 \right) \,\,\,  \, \end{eqnarray}
with $S \equiv \langle \mathbf{\hat{p}} \mathbf{\hat{p}} \mathbf{\hat{p}} \mathbf{\hat{p}}\rangle$ the fully symmetric four-rank tensor.
The first term of this collision operator can convert longitudinal waves $L= (\hat{\mathbf{p}},\mathbf{p})$ to all available transverse waves $T' =(\mathbf{g}'_T, \mathbf{p}')$. Since the sum over the two transverse polarizations
$\sum  \hat{\mathbf{g}}_T'\hat{\mathbf{g}}_T' =\mathbf{ \Delta}_{p'}$, the  rate is given by
   \begin{eqnarray}\label{div2}
&&    \frac{\rho_L(\mathbf{p})}{\tau(L \rightarrow T')} = \sum_{\mathbf{p}'}\rho_T(\mathbf{p}') \times \mathbf{ \hat{p}\hat{p}}\cdot U^{(\mathrm{LAD},s)}_{\mathbf{pp}'}\cdot \mathbf{\Delta}_{p'} \nonumber \\
   && = - Q_0 \frac{ n^2 |t|^4 }{(6\pi)^2 }
   \end{eqnarray}
Since $t$ satisfies the optical theorem,  the two secular scattering rates (\ref{div1}) and (\ref{div2}) are equal but of opposite sign.  This is a manifestation of
the Ward identity~(\ref{ward}). Since the life-times of longitudinal and transverse modes are not singular and easily found from Eq.~(\ref{self}),
its righthand side $\sum_j U_{ij} \rho_j$ must be free of  divergencies.

 {However, the two singular scattering events do} \emph{not}  {cancel in the transport equation}~(\ref{transportgen})  {as long as } $n_L \neq n_T$,
 {and therefore govern the dynamics of the equipartition
process. }The time found in Eq.~(\ref{div1}) must be the characteristic time for equipartition to set in. We can compare it to the characteristic time for mode conversion between transverse waves,
given by $ \rho_T /\tau(T\rightarrow T') = k/\ell$ in the same units. Hence
\begin{equation}\label{tau}
    \frac{\tau^{-1} (T\rightarrow L')}{\tau^{-1}(T\rightarrow T')} = \frac{1}{3}\frac{Q_0}{ k\ell}
\end{equation}
Once equipartition is established,
 the ratio of average longitudinal and transverse energy densities is constant and given by
\begin{eqnarray}\label{EP}
  \frac{E_L(k)}{E_T(k)} =  \frac{\rho_L}{\rho_T} = \frac{Q_0}{k\ell }
\end{eqnarray}
If $k\ell  /Q_0  \ll 1$ we see that $ E_L \gg   E_T $ and $1/\tau_{TL} \gg 1/\tau_S$.  These inequalities imply that
longitudinal states dominate in energy, and equilibrate in phase space as fast as the transverse waves. In fact, when $k\ell  /Q_0  < 1$,
the intermediate scattering to a longitudinal wave becomes more efficient for
 transverse waves to equilibrate among themselves than accomplished  by ISA single
scattering. The atomic quality factor $Q_0$ is large, and experiments \cite{nice} and numerical simulations \cite{pool,sergey0,remi}
exist where $Q_0 \gg k\ell$.

\section{Kubo formalism}\label{kubosection}

In this section we use the rigorous Kubo formalism for the DC conductivity, adapted from electron conduction \cite{mahan} to scalar classical
 waves \cite{bara1} and electromagnetic waves \cite{bara2}. We investigate how photon diffusion is affected by the existence of longitudinal waves.

\begin{figure}
  \includegraphics[width=7cm]{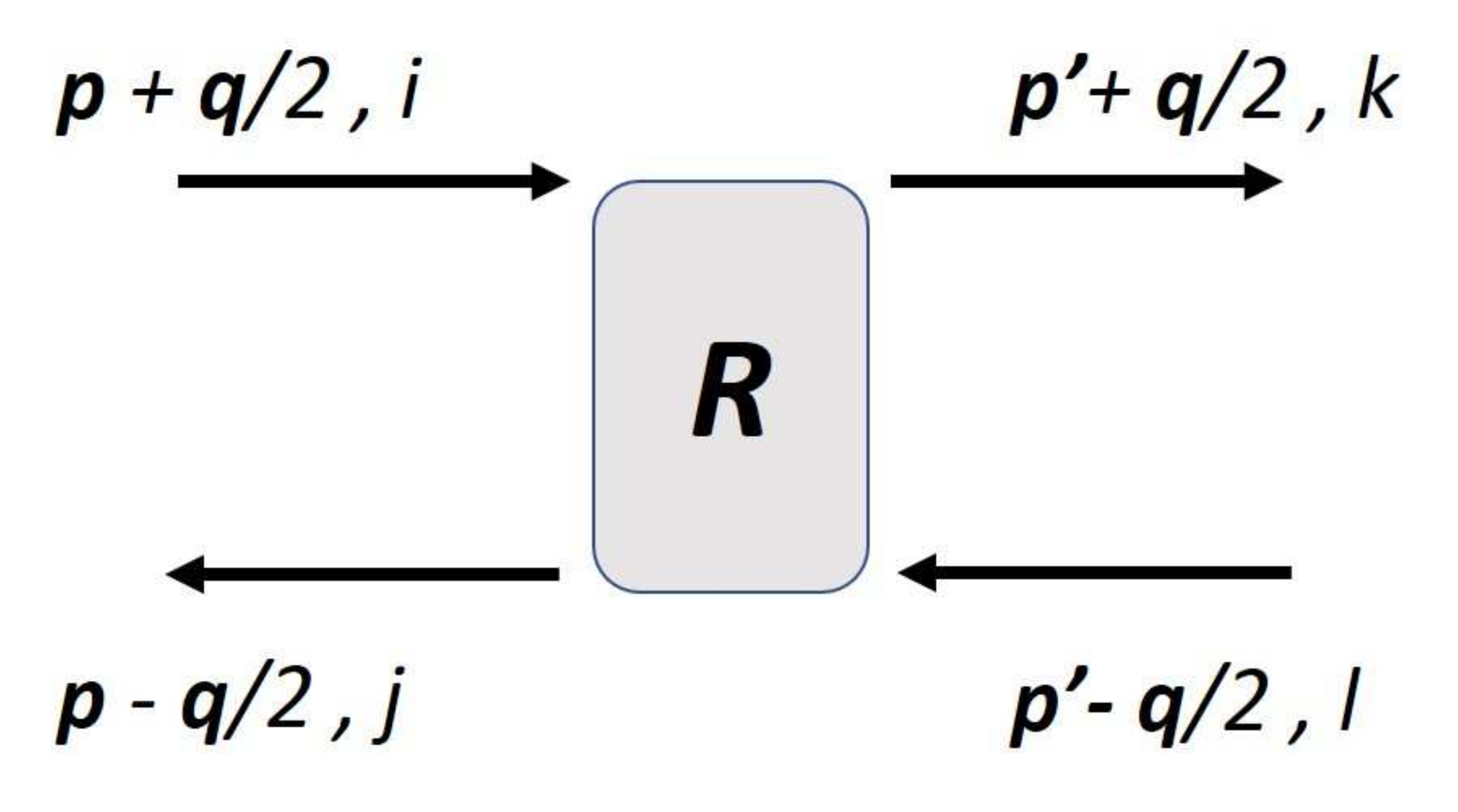}
  \caption{The diagrammatic convention associated with the reducible vertex $R_{ij|kl,\mathbf{ pp'}}(\mathbf{q})$, with external lines. Top line denotes
 retarded Green's function $\mathbf{G}(k + i0)$, bottom line $\mathbf{G}^\dag(k + i0) =\mathbf{G}(k - i0)$ is the advanced
 Green's function and travel in the opposite direction. The polarization labels are $ij$ on the left hand side (``observer'') and $kl$ on the right
 hand side (``source'').The sum of incoming and outgoing wave numbers is conserved.}\label{Reps}
\end{figure}

Before averaging over the disorder, the electric field at frequency $\omega = kc_0$ is given formally
by the operator identity $\mathbf{E}(k) = \mathbf{G}(k) \otimes \mathbf{s}(k)$, with $\mathbf{s}(k)$ a source ($\otimes$ stands for the matrix product in full Hilbert space, whereas $\cdot$
stands for matrix product in $3 \times 3$ polarization space). Transport theory describes the
correlation function $\phi_{ij} = \langle E_i\bar{E}_j \rangle $ of the electric field at two different frequencies and for
two different positions. We can formally
relate it to the source correlation function $\mathbf{S}$ according to  $\phi_{ij} = R_{ij|kl} \otimes S_{kl}$, which introduces the reducible
four-rank vertex ${R}$. It
satisfies the Bethe-Salpeter equation,
\begin{equation}\label{BS}
   {R} = \mathbf{GG}^\dag + \mathbf{GG}^\dag \otimes{U}\otimes  {R}
\end{equation}
This equation identifies the irreducible vertex ${U}$ as the scattering operator, and $\mathbf{G} \mathbf{G}^\dag$ as the transport between scattering events.
(We use $ \dag$ for Hermitian conjugate in full Hilbert space as opposed to $*$ for Hermitian conjugate of a $3\times 3$ matrix with polarization components; a bar denotes complex conjugation of a scalar.)
 The Green's function of the
effective medium was introduced in Eq.~(\ref{dyson}) and has transverse and longitudinal parts. We recall the tensor convention
$\mathbf{AB} \cdot \mathbf{S}= \mathbf{A}\cdot \mathbf{S}\cdot\mathbf{B}$, or equivalently $(\mathbf{A} \mathbf{B})_{ij|kl} =A_{ik}B_{lj}$, with the matrix
$\mathbf{B}$ displayed as the bottom line of a Feynman diagram, propagating backwards in time. Similarly in Hilbert space,
$(\mathbf{G} \mathbf{G}^\dag)_{\alpha \beta|\kappa \gamma} =G_{\alpha \kappa }G^\dag_{\gamma \beta} = G_{\alpha \kappa }\bar{G}_{\beta \gamma} $.
After averaging, translational symmetry can be exploited so that the vertex in Fourier space (Fig.~\ref{Reps})
can be written as
${R}_{\mathbf{pp}'}(\mathbf{q})$, with $\mathbf{p}'$ and $\mathbf{p}$ interpreted as incident and outgoing wave numbers,
and $\mathbf{q}$ conjugate to
 distance between source and observer. Thus, the electromagnetic ``Wigner function'' takes the form
 \begin{equation}\label{FFSS}
    \phi_{ij}(\mathbf{p},\mathbf{q}) \equiv \langle E_i(\mathbf{p}^+)\bar{E}_j(\mathbf{p}^-)\rangle =
    \sum_{\mathbf{p}'} R_{\mathbf{pp'}; ij|kl}(\mathbf{q}) S_{kl}
    (\mathbf{p}',\mathbf{q})
 \end{equation}
 and
 \begin{eqnarray}\label{BS2}
   {R}_{\mathbf{pp}'}(\mathbf{q}) &\,&=  \mathbf{G}(\mathbf{p}^+)\mathbf{G}^*(\mathbf{p}^-)  \delta_{\mathbf{pp}'} \nonumber \\
    + &\,& \mathbf{G}(\mathbf{p}^+)\mathbf{G}^*(\mathbf{p}^-)  \cdot \sum_{\mathbf{p}''} U_{\mathbf{pp}''}(\mathbf{q})
   \cdot {R}_{\mathbf{p}''\mathbf{p}'}(\mathbf{q})
 \end{eqnarray}
with $\mathbf{p}^\pm =\mathbf{p} \pm \mathbf{q}/2$ (see Fig.~\ref{Reps}) and $ \delta_{\mathbf{pp}'} \equiv (2\pi)^3 \delta(\mathbf{p}-\mathbf{p}')$. The two terms describe direct propagation with extinction of the mode $\mathbf{p}$ and
scattering from $\mathbf{p}'$ towards $\mathbf{p}$,
respectively. One important property is reciprocity \cite{bara2, rogerboek}. Since without external magnetic fields, the (unaveraged) Green's function satisfies
$\langle \mathbf{p}, i | \mathbf{G}(k+i0)| k, \mathbf{p}'\rangle = \langle -\mathbf{p}', k | \mathbf{G}(k+i0)| i, -\mathbf{p} \rangle$
we easily check that
\begin{equation}\label{reci}
  {R}_{ij|kl,{\mathbf{pp}'}}(\mathbf{q})   = {R}_{kl|ij,-\mathbf{p}'-\mathbf{p}}(-\mathbf{q})
\end{equation}

A second property follows from complex conjugation, equivalent to switching bottom and top lines of the diagram,
\begin{equation}\label{CC}
    {R}_{ij|kl,{\mathbf{pp}'}}(\mathbf{q})   = \bar{{R}}_{ji|lk,\mathbf{p}\mathbf{p}'}(-\mathbf{q})
\end{equation}
If Eq.~(\ref{ward}) is satisfied, $R$ is known to exhibit long-range diffusion ($q\rightarrow 0$), as its equivalent in electron-impurity scattering \cite{mahan}, that decouples input and output, and takes the form
\begin{equation}\label{diffR}
    {R}_{ij|kl,{\mathbf{pp}'}}(\mathbf{q})  = \frac{d_{ij}(\mathbf{p},\mathbf{q}) d_{kl}(\mathbf{p}',\mathbf{q})}{\pi N(k) D(k) q^2}
\end{equation}
where $N(k)$ is the DOS given in Eq.~(\ref{DOS1}) and the eigenfunction associated with long-range diffusion is written as
\begin{equation}\label{dij}
    \mathbf{d}(\mathbf{p},\mathbf{q}) = -\mathrm{Im}\, \mathbf{G}(\mathbf{p}) -\frac{i}{2}\mathbf{J}(\mathbf{p},\mathbf{q}) +\mathcal{ O}(q^2)
\end{equation}
The first term $-\mathrm{Im}\, \mathbf{G} (\mathbf{p}) \equiv -[\mathbf{G}(\mathbf{p}) -\mathbf{G}^*(\mathbf{p})] /2i $ is proportional to the
spectral function and implies perfect equipartition of the electromagnetic energy in phase space.
The second term is linear in $\mathbf{q}$ and describes a small perturbation
due to gradients of $\Phi_{ij} (\mathbf{r}) $ in real space that trigger diffuse energy flow.
For it to be small for all momenta  $\mathbf{p}$ imposes constraints
to be discussed later. Because $\mathbf{d}(\mathbf{p},\mathbf{q})$ describes
an electric field correlation function, it must satisfy $d_{ij}(\mathbf{p},\mathbf{q}) =\bar{d}_{ji}(\mathbf{p},-\mathbf{q}) $, consistent with Eqs.~(\ref{CC}) and (\ref{reci}).
Thus,  $ \mathbf{J}(\mathbf{p},\mathbf{q})= - \mathbf{J}^*(\mathbf{p},-\mathbf{q})$ and, being
linear in $\mathbf{q}$ by construction, we conclude that the tensor $\mathbf{J}(\mathbf{p},\mathbf{q})$ is Hermitian.

Following common treatments in radiative
transfer, many microscopic approaches interpret the expansion~(\ref{dij}) as one in the angular anisotropy of scattered radiation with wave numbers in equipartition and imposed
 near the frequency shell, as described by the first term. If we ignore electromagnetic polarization and without any kind of explicit anisotropy in space,
 the only possible choice of this expansion is,
\begin{equation}\label{dijscal}
    \mathbf{d}(\mathbf{p},\mathbf{q}) = -\mathrm{Im}\, \mathbf{G}(\mathbf{p}) \left[ 1 - iJ_0(p)\mathbf{ p}\cdot \mathbf{q} + \cdots \right]
\end{equation}
For diffusion of cold atoms, $ \mathbf{J}(\mathbf{p},\mathbf{q})$ was obtained by solving numerically the Bethe-Salpeter equation \cite{marie}.
Alternatively, the unknown function $J_0(p)$ can be chosen such
that the first angular moment $\sum_\mathbf{p} \mathbf{p}\mathbf{ d}(\mathbf{p},\mathbf{q}) $ matches the divergence  $-i\mathbf{q}\cdot \mathbf{K}$ of the
energy current density. This leads to $J_0(p)= 1/p^2$ \cite{sheng} and makes the vector $\mathbf{K}$ the only unknown.
This choice conveniently circumvents divergencies
that occur in rigorous theory for large $p$.  For vector waves, $ \mathbf{J}(\mathbf{p},\mathbf{q}) $ is a tensor containing longitudinal and transverse components, and even their
 {interferences.
It obeys the
Bethe-Salpeter equation}~(\ref{BS2})   {linearized in the gradient vector} $\mathbf{q}$  {that can be obtained straightforwardly as for scalar waves} \cite{mahan}
\begin{eqnarray}\label{BSJ}
  \mathbf{J}(\mathbf{p},\mathbf{q}) &=& \mathbf{J}^D(\mathbf{p},\mathbf{q}) \nonumber \\
  &+& \mathbf{ G}(\mathbf{p})\mathbf{G}^*(\mathbf{p}) \cdot \sum_{\mathbf{p}'} {U}_{\mathbf{pp}'}\cdot \delta_\mathbf{q}
   \mathbf{G}(\mathbf{p}',\mathbf{q})\nonumber \\
   &+& \mathbf{ G}(\mathbf{p})\mathbf{G}^*(\mathbf{p}) \cdot \sum_{\mathbf{p}'} {U}_{\mathbf{pp}'}\cdot \mathbf{J}(\mathbf{p}',\mathbf{q})
\end{eqnarray}
The first term is often referred to as the Drude contribution to diffusion and depends only on the effective medium properties. It reads
\begin{equation}\label{Drude}
    \mathbf{J}^D(\mathbf{p},\mathbf{q})  = \mathbf{G} (\mathbf{p})\cdot\mathbf{ L}(\mathbf{p},\mathbf{q})  \cdot \mathbf{G}^* (\mathbf{p})
    - \delta_\mathbf{q}
  \mathrm{ Re}\, \mathbf{G}(\mathbf{p},\mathbf{q})
\end{equation}
in terms of the bilinear Hermitian tensor $\mathbf{ L}_{ij}(\mathbf{p},\mathbf{q}) = 2(\mathbf{p}\cdot \mathbf{q}) \delta_{ij} -p_iq_j-q_ip_j $ and the notation is
$\delta_\mathbf{q}
 \mathrm{  Re}\, \mathbf{G}(\mathbf{p},\mathbf{q}) = (\mathbf{q}\cdot \partial_\mathbf{p}) \mathrm{Re}\, \mathbf{G}(\mathbf{p}) $.
 The second and third terms in Eq.~(\ref{BSJ})  are
    genuine contributions from
    scattering. They vanish only for isotropic events in $U_{\mathbf{pp}'}$ but not in general.
    It is straightforward to demonstrate that the (cycle-averaged) Poynting vector $\mathbf{K} = c_0 \mathrm{Re}\, (\mathbf{E}\times \bar{\mathbf{B}})/8\pi $
   is related to the
    correlation function of the electric field according to
  \begin{eqnarray}\label{Poyn}
    K_n(k,\mathbf{q}) &=& \frac{c_0}{8\pi k} \sum_\mathbf{p} \left( p_n \delta_{ik}- \frac{1}{2} p_k\delta_{in} -
    \frac{1}{2} p_i\delta_{kn} \right) \phi_{ki}(\mathbf{p},\mathbf{q}) \nonumber \\
    &+& \frac{c_0}{8\pi k} \sum_\mathbf{p} q_k \frac{1}{2}( \phi_{kn}(\mathbf{p},\mathbf{q}) - \phi_{nk}(\mathbf{p},\mathbf{q}) )
  \end{eqnarray}
In the absence of external magnetic fields,
 $ \phi_{ki}(\mathbf{p},\mathbf{0}) = \phi_{ik}(\mathbf{p},\mathbf{0})$ so that the second term vanishes in linear order of $\mathbf{q}$. Upon inserting Eq.~(\ref{dij}),
  the first term involves only the diffusion current tensor $\mathbf{J}$.
  Some  manipulations lead to
\begin{eqnarray}\label{K1}
  iq_nK_n  &=& \frac{1}{4\pi  N(k)} \sum_\mathbf{p}  L_{ik}(\mathbf{p},\mathbf{q}) J_{ki}(\mathbf{p},\mathbf{q}) \nonumber \\
   &\times&  \frac{1}{D(k) q^2}\frac{c_0}{8\pi k}\sum_{\mathbf{p}'}  -\mathrm{Im}\, {G}_{lj}(\mathbf{p}') \cdot {S}_{lj} (\mathbf{p}')
\end{eqnarray}
The factor that has been split off on the right hand side can be identified as the (cycle-)averaged energy density $\rho(\mathbf{r})=
 \langle |\mathbf{E}|^2 + |\mathbf{B}|^2 \rangle /16\pi$ released by the source and diffusing out.
 This can be established by noting that the tensor
$\mathbf{J}$, being odd in $\mathbf{p}$,
does not
contribute to the energy density. Since the first term  in Eq.~(\ref{dij}) obeys equipartition,
$\langle|\mathbf{E}(\mathbf{r})|^2 \rangle= \langle|\mathbf{B}(\mathbf{r})|^2\rangle$,
the energy density is
\begin{eqnarray}\label{E}
    \langle\rho(\mathbf{q}) \rangle &=& \frac{1}{\pi N(k) } \frac{k}{c_0}\mathrm{Tr}\, \sum_\mathbf{p} \frac{p^2}{k^2}
    \mathbf{\Delta}_p \cdot -\mathrm{Im}\,  \mathbf{G}(k,\mathbf{p}) \nonumber \\
   &\times& \frac{1}{D(k)q^2} \frac{c_0}{8\pi k}\mathrm{Tr }\sum_{\mathbf{p}'}  -\mathrm{Im}\, \mathbf{G}(\mathbf{p}') \cdot \mathbf{S}(\mathbf{p}')
\end{eqnarray}

If we recall that the electromagnetic DOS is given by Eq.~(\ref{DOS1}),
the first factor in Eq.~(\ref{E}) equals one. In real space Eq.~(\ref{K1}) thus becomes
$\bm{\nabla} \cdot \mathbf{K} = - D \bm{\nabla}^2 \rho(\mathbf{r})$ with
\begin{equation}\label{Kubo}
    \pi N(k) D(k) = \frac{1}{4}\mathrm{Tr} \, \sum_\mathbf{p}
    \mathbf{L} (\mathbf{p},\hat{\mathbf{q}}) \cdot \mathbf{J}(\mathbf{p},\hat{\mathbf{q}})
\end{equation}
This is the Kubo formula for the electromagnetic diffusion constant. Since $D$ is a scalar in this work,
the right hand side does depend on the direction of $\mathbf{q}$. The left hand side can be identified as the electromagnetic DC conductivity $\sigma(k)$ (here in units of $1/$m)
expressed as the (Einstein) product of DOS and
diffusion constant. With this definition, that we prefer in view of the presence of $\pi ND$ in Eq.~(\ref{diffR}), the "electromagnetic conductance" of a slab with surface $A$ and
length $L$ takes the form of a Landauer formula $\langle
\sum_{ab}T_{ab} \rangle = 4\sigma A/L$ \cite{LB}.  In terms of the energy density $\rho(\mathbf{q})$ the electric field correlation function is
\begin{eqnarray}\label{DAn}
    \phi_{ij}(\mathbf{p},\mathbf{q}) = \frac{d_{ij}(\mathbf{p},\mathbf{q})}{\pi N(k) } \times \frac{8\pi k}{c_0} \rho(\mathbf{q})
\end{eqnarray}

\subsection{Diffusion current tensor}
The diffusion current tensor $\mathbf{J}(\mathbf{p},\mathbf{q})$ must be a parity-even, Hermitian tensor, linear in the gradient vector $\mathbf{q}$. For our problem, with no
explicit anisotropy present,
this
leaves
us with the following general form
\begin{eqnarray}\label{Jgeneral}
\mathbf{J}(\mathbf{p},\mathbf{q}) &=& J_0 (p) ({\mathbf{p}}\cdot \mathbf{q}) \Delta_\mathbf{p}
+ J_1(p)   ({\mathbf{p}}\cdot \mathbf{q})
\hat{\mathbf{p}} \hat{\mathbf{p}}
\nonumber
 \\ && + J_{2} (p) ({\mathbf{p}} \mathbf{q}+{\mathbf{q}}{ \mathbf{p}} ) +
J_{3} (p) i ({\mathbf{p}} \mathbf{q}- {\mathbf{q}} {\mathbf{p}} )
\end{eqnarray}
with four real-valued functions $J_i(p)$ to be determined. A fifth term $i\epsilon_{ijk}q_k$ is in principle allowed but is
excluded for scattering that respects parity symmetry. Alternatively, we could have defined the mode $J_2(p)$ in terms of the tensor
${\mathbf{p}}\mathbf{q}+\mathbf{q}{\mathbf{p}} - 2({\mathbf{p}}\cdot \hat{\mathbf{q}}) \hat{\mathbf{p}}\hat{\mathbf{p}}$ in which case all 4 modes would be mutually
orthogonal.

The four functions can be associated with four different aspects in diffuse transport. By restricting only to the first, the transport problem reduces to the common approximation made in Eq.~(\ref{dijscal}).
The modes $J_1$, $J_2$ and $J_3$ are clearly genuine vector effects, absent in
a scalar theory. However, only $J_0$ and $J_2$ carry a Poynting vector, with $J_0$
associated with the transport of transverse waves in the far field, and $J_2$ associated with a novel process that involves
the interference of longitudinal and transverse waves.
By restricting to the purely transverse term $J_0(p)$, transport theory almost reduces to a scalar theory. The term $J_1$
describes how the longitudinal energy density $ |E_L(\mathbf{p})|^2 $ achieves an anisotropy in phase space
due to the spatial gradient of energy, but without inducing an energy current.
The presence of $J_3$ is more subtle and can be associated with the \emph{imaginary} part of the complex Poynting vector, discussed for instance in
Ref.~\cite{jackson}.
Let us call $\mathrm{Im} \, \mathbf{K} = c_0 \mathrm{Im}\, ({\mathbf{E}}\times \bar{\mathbf{B}})/8\pi $.
We readily find, similar to the derivation of its real part in
Eq.~(\ref{Poyn}), that in terms of the field correlation function $\phi_{ik} (\mathbf{p},\mathbf{q})$,
 \begin{eqnarray}\label{imPoyn}
    \mathrm{Im}\, K_n(k,\mathbf{q}) &=& \frac{-ic_0}{8\pi k} \sum_\mathbf{p} \left( q_n \delta_{ik}- \frac{1}{2} q_k\delta_{in} -
    \frac{1}{2} q_i\delta_{kn} \right) \phi_{ki} \nonumber \\
    &-& \frac{ic_0}{8\pi k} \sum_\mathbf{p} p_k ( \phi_{ki} - \phi_{ik} )
  \end{eqnarray}
The first term is independent of $\mathbf{J}$ and can be evaluated without any approximation. The integral over wave numbers is proportional to the total
DOS $N(k)$ and cancels this same factor in
 the denominator of Eq.~(\ref{DAn}). As a result it is completely
independent of the presence of the dipoles.
The second term requires anti-symmetry
in the diffusion tensor $J_{ij}$, described only by $J_3(p)$. We obtain
\begin{equation}\label{DIM}
    \mathrm{Im}\, \mathbf{K}(k,\mathbf{q}) = \frac{2}{3} \left(\frac{c_0}{k} + \frac{1}{\pi N(k)} \sum_\mathbf{p} p^2 J_3(p) \right) (-i\mathbf{q}) \rho(\mathbf{q})
\end{equation}
Like the real part, the ``current density'' $ \mathrm{Im}\, \mathbf{K}$ is proportional to minus the gradient in energy density, with however
a very small ``fictitious'' diffusion constant $D_I = \frac{2}{3} c_0/k $ associated with
$ \mathrm{Im}\, \mathbf{K}$, and a correction from $J_3$ calculated in the next section.

Even if  $J_1$ and $J_3$ do not carry current themselves, they cannot be ignored because the  Bethe-Salpeter equation~(\ref{BSJ}) couples in principle all $J_i$
 through scattering. From Eq.~(\ref{BSJ}) we can identify four different contributions to  $\mathbf{J}(\mathbf{p},\mathbf{q})$,
written as
\begin{equation}\label{JDall}
    \mathbf{J}(\mathbf{p},\mathbf{q}) = \mathbf{J}^D  + \mathbf{J}^{\delta \Sigma} + \mathbf{J}^{\delta G} +\mathbf{J}^{S}
\end{equation}
In this expression, the Drude diffusion current in Eq.~(\ref{Drude}) has been further split up into the first two terms above.  The first is given by,
\begin{equation}\label{drudeagain}
    \mathbf{J}^D(\mathbf{p},\mathbf{q})  = \mathbf{G} (\mathbf{p})\cdot\mathbf{ L}(\mathbf{p},\mathbf{q})  \cdot \mathbf{G}^* (\mathbf{p})
    -\mathbf{G} (\mathbf{p})\cdot\mathbf{ L}(\mathbf{p},\mathbf{q})  \cdot \mathbf{G} (\mathbf{p})
\end{equation}
The second term is generated by the  explicit dependence of the self-energy on wave number,
\begin{equation}\label{drudesigma}
    \mathbf{J}^{\delta\Sigma}(\mathbf{p},\mathbf{q})  =  - \mathrm{Re} \, \mathbf{G}(\mathbf{p})\cdot (\mathbf{q}\cdot
    \partial_\mathbf{p})\mathbf{ \Sigma }(\mathbf{p})  \cdot  \mathbf{G}(\mathbf{p})
\end{equation}
with the convention that $\mathrm{Re} \, \mathbf{A} = (\mathbf{A}+\mathbf{A}^*)/2$. The final
two terms $\mathbf{J}^{\delta G}$ and $\mathbf{J}^{S}$ are defined as the two last scattering terms involving $U_{\mathbf{pp}'}$ in Eq.~(\ref{BSJ}).

 {To summarize the above analysis, the diffusion tensor} $\mathbf{J}(\mathbf{p},\mathbf{q}) $  {can have 4 different symmetries, denoted by} $J_i$.  {Each term can originate from 4 different parts
of the Bethe-Salpeter equation}~(\ref{BSJ}).
The mode  $J_2$ { implies a new mechanism of long-range diffusion stemming in all 4 cases from the mixture of  longitudinal and transverse  fields.}
One peculiarity is the direction of the Poynting vector associated with the diffuse mode expressed by Eq.~(\ref{dij}).
The  mode $J_0$ of the pure transverse field generates a Poynting vector
 whose component along the gradient varies as  $\cos^2\theta$ in phase space, with $\theta$ the angle between wave vector $\mathbf{p}$ and gradient vector $\mathbf{q}$, and is
 thus largest \emph{along} the gradient vector.
For the mode $J_2$  this component varies as $\sin^2\theta$, which is largest \emph{orthogonal }to the gradient vector.


In  the following subsections \ref{drudesection}--\ref{scattdifsection} we discuss these 4 contributions to $\mathbf{J}$ separately, and show that the scattering from
two electric dipoles
generates all four channels in Eq.~(\ref{Jgeneral}). The results are summarized in subsection~\ref{secsummary} and in Table~(\ref{tabeldiff}).

\subsubsection{Drude current tensor}
\label{drudesection}

 {The Drude current tensor} $\mathbf{J}^D(\mathbf{p},\mathbf{q})$ { is defined as the contribution of the effective medium, as expressed by } Eq.~(\ref{drudeagain})  {and  is thus by definition
independent of the collision operator} $U_{\mathbf{pp}'}$ { and not subject to interference}. It is therefore the easiest to calculate.
We will split   $\mathbf{J}^D(\mathbf{p},\mathbf{q}) $ further up into
a pure transverse part and an interference term and write
\begin{equation}\label{JDall2}
    \mathbf{J}^D(\mathbf{p},\mathbf{q}) = \mathbf{J}^D_{TT}(\mathbf{p},\mathbf{q})  + \mathbf{J}^D_{TL}(\mathbf{p},\mathbf{q})
\end{equation}
The first part stems from purely transverse propagation and contributes only to the $J_0$-channel in Eq.~(\ref{Jgeneral}). The
second term is produced
by a mixture of longitudinal and transverse propagation and contributes to the channels $J_1$, $J_2$ and $J_3$.
Since $\mathbf{p}\cdot \mathbf{L}(\mathbf{p},\mathbf{q}) \cdot \mathbf{p} =0$, the Drude current tensor features no purely longitudinal mode $\mathbf{J}^D_{LL}(\mathbf{p},\mathbf{q})$, proportional to $|G_L(p)|^2$.

The transverse Green's function $\mathbf{G}_T(\mathbf{p})$ is given by the second term in Eq.~(\ref{dyson}). It follows
\begin{eqnarray}\label{JT}
    \mathbf{J}^D_{TT}(\mathbf{p},\mathbf{q}) &=& 2 (\mathbf{p}\cdot \mathbf{q})\mathbf{ \Delta}_p \left( |G_T(p)|^2 - \mathrm{Re }\, G_T(p)^2 \right)
    \nonumber \\
    &=&4 (\mathbf{p}\cdot \mathbf{q})\mathbf{ \Delta}_p \mathrm{Im }^2\, G_T(p)
\end{eqnarray}
This function is heavily peaked near the frequency shell of the effective medium. We can ignore any $p$-dependence in
$\Sigma_T(p)$ and approximate it by $\Sigma_T(p=k)$. For $G_T= (K_T^2-p^2)^{-1}$ and $K_T= k_e + i/2\ell$ a complex wave vector independent of $p$,
we can use,
\begin{equation*}
    \frac{\sum_\mathbf{p} 2 p^2 \mathrm{Im}^2\, G_T(p)}{\sum_\mathbf{p} -\mathrm{Im}\, G_T(p) } = k_e\ell
\end{equation*}
In terms of the density of transverse states (DOTS) this produces the classical Drude diffusion constant in the $J_0$-channel,
\begin{equation}\label{DDrude}
    D^D_0(k) = \frac{1}{3}\times \left( c_0\frac{k_e}{k} \frac{N_T(k)}{N(k)} \right) \times \ell(k) \equiv \frac{1}{3} v_E \ell
\end{equation}
It is customary to write $k_e/k = c_0/v_p$ in terms of the phase velocity $v_p$. The ratio $N_T/N = N_T/(N_L + N_T)$ is a factor that can be very
small near
the resonance $\omega_0$. We recall that for our electric dipole scatterers all stored energy resides in the longitudinal field. In the Drude approximation for the transverse field
we recover the familiar picture of  light
diffusion, with the extinction length as the mean free path, and $v_E$ as energy transport velocity \cite{PR}.

The perturbation expansion in $\mathbf{q}$ is valid for the transverse waves as long as $2pq  |\mathrm{Im} \,G_T(p)|^2 < |\mathrm{Im} \,G_T(p)| $. This is most stringent near the frequency shell
$p=k_e$ where the spectral function $|\mathrm{Im} \,G_T(p)|$ is maximal and not stringent at all for large momenta. This gives $q < |\mathrm{Im} \,\Sigma_T(k_e)|/2k_e = 1/2\ell$.
This could have been an intuitive estimate.

The diffusion tensor $\mathbf{J}^D_{TL}$ is given by
\begin{eqnarray}\label{LT}
    \mathbf{J}^D_{TL} (\mathbf{p},\mathbf{q}) &=& 2\mathrm{Im}\, G_T \mathrm{Im}\, G_L \, ( 2 \hat{\mathbf{p}} \hat{\mathbf{p}} (\mathbf{p}\cdot \mathbf{q})  -\mathbf{pq} -
    \mathbf{qp} ) \nonumber \\
    &+& i \mathrm{Im}\,  [\bar{G}_L {G}_T] \,  (\mathbf{pq} - \mathbf{qp})
\end{eqnarray}
with contributions to the channels  $J_1$, $J_2$ and $J_3$ in Eq.~(\ref{Jgeneral}). We focus first on the Poynting vector for which only the channel
$J_2(p)$ is relevant. Inserting the first term
into Eq.~(\ref{Kubo})  gives
\begin{equation}\label{DDsing}
    \pi N(k) D^{D}_{2}(k) = \sum_\mathbf{p}\mathrm{Im}\, G_T(p) \mathrm{Im}\, G_L(p) \, [p^2- (\mathbf{p}\cdot \hat{\mathbf{q}})^2]
\end{equation}
This diffusion is clearly determined by the overlap of transverse and longitudinal modes in phase space. Because the longitudinal spectral function is essentially independent of $p$,
this overlap is significant and the integral even diverges as $\sum_\mathbf{p } 1/p^2$. We can extract and regularize it as earlier by $Q_0k_0/4\pi$,
 \begin{equation}\label{DTL}
 \pi N(k) D^{D}_{2}(k) = \frac{Q_0}{6\pi }\frac{\left(\mathrm{Im }\Sigma_{\mathrm{ISA}}\right)^2 }{k^3}
\end{equation}
This singular term is proportional to the square of the density of the dipoles and  will later be seen to cancel. As a result, the leading  current tensor is,
\begin{eqnarray}
  \mathbf{J}^D_{TL}  = \frac{2\pi }{k\ell} \delta(k^2-p^2) \frac{1}{p^2}&\,&( 2 \hat{\mathbf{p}} \hat{\mathbf{p}} ({\mathbf{p}}\cdot \mathbf{q})  -
  \mathbf{{p}q} -
    \mathbf{q{p}} ) \nonumber \\
    + && i \mathrm{Im}\,  [\bar{G}_L {G}_T] \, \,
   (\mathbf{{p}q} - \mathbf{q{p}})
\end{eqnarray}
The use of the Dirac distribution implies here implicitly that a typical Kubo integral of the kind
$\sum_\mathbf{p} \mathbf{p }\mathbf{J}(\mathbf{p},\mathbf{q})$
converges for large $p$, with no need for regularization.
 This expression will turn out to be leading for $J_2$ and dominating for $J_3$. The Drude  diffusion constant associated
 with the mixture of transverse and longitudinal waves is thus given by
 \begin{equation}\label{Dlong}
    D^{D}_2 = \frac{1}{3\pi N(k) }\sum_\mathbf{p} \frac{2\pi }{k\ell} \delta(k^2-p^2) \approx \frac{1}{3}v_E \frac{1}{k^2\ell}
 \end{equation}
This diffusion constant can be considered as the ISA of electromagnetic diffusion in the $J_2$-channel. Its value is \emph{positive} and,
apart from the universal pre-factor $v_E$ in diffusion, depends linearly on the density of the dipoles.
In Ref.~\cite{theoL} one finds a correction induced by dipole-dipole coupling that can be written as $\Delta D = \frac{1}{3} v_E  F(\delta) /k_0^3\ell^2$,
with the function $F$ varying over the resonance. Like the diffusion found in Eq.~(\ref{Dlong}),
it is  positive and proportional to $v_E$, but unlike Eq.~(\ref{Dlong}) it scales as $n^2$. The interference of longitudinal and transverse waves is excluded
in Ref.~\cite{theoL} which explains why this
leading term~(\ref{Dlong}) is not found.

For the hydrodynamic expansion made in Eq.~(\ref{dij}) to hold for the transport channel  $J_2$, we must have $|\mathrm{Im} \,G_T(p)| > |J^D_{TL}(p)| pq/2$, or equivalently,
  $pq < 1/|\mathrm{Im}\, G_L| \approx|k^2/ |\mathrm{Im}\, \Sigma |$.
Since transverse waves already impose    $q < 1/\ell$ we conclude that  $p < k^3\ell^2 $. This becomes restrictive once $k\ell$
approaches unity.

 It is straightforward to obtain the Drude approximation for the fictitious diffusion constant in Eq.~(\ref{DIM})
 associated with $\mathrm{Im}\,\mathbf{ K}$, and which was seen to be governed by $J_3$. Since $J_3 =  \mathrm{Im}\, (\bar{G_L} {G}_T)$ we can write
  \begin{eqnarray*}
   \frac{1}{\pi N } \mathrm{Im} \, \sum_\mathbf{p} p^2\, (\bar{G_L} {G}_T) &=& \frac{1}{\pi N } \mathrm{Im} \sum_\mathbf{p} \left[  p^2\bar{G}_L \frac{1}{-p^2}  +  \frac{z^2}{\bar{z}^2} G_T \right] \\
&\approx& -\frac{c_0}{k} \frac{N_L + \frac{1}{2}N_T }{N_L + N_T}
\end{eqnarray*}
where we used the expression~(\ref{DOS3})  of the DOS split up in its longitudinal part $N_L$ and its transverse part $N_T$. With $v_E= c_0 N_T/(N_L+N_T)$ we find from Eq.~(\ref{DIM}),
\begin{equation}\label{DIM2}
       D^D_{I}=  \frac{1}{3} v_E k^{-1}
\end{equation}
The singular longitudinal DOS,  proportional to $Q_0$, cancels. In the Drude approximation the fictitious diffusion constant $D_I$ of the mode $J_3$ is
a factor $k\ell$ \emph{larger }than the   diffusion constant $D_2$ of the channel $J_2$, and a factor $k\ell$ \emph{smaller} than the transverse
diffusion of mode $J_0$.
This suggests that they all become of the same order near
$k\ell =1$.

\subsubsection{Self-energy dependent on wave number}

Any dependence on $p$ of the self-energy contributes to the diffusion current via the term  $\mathbf{J}^{\delta\Sigma}$  derived in  Eq.~(\ref{drudesigma}).
For electric dipoles such dependence on wave number comes in via the boomerang diagrams discussed in Eq.~(\ref{self}) with the subtle local field correction at large momenta derived in
Eq.~(\ref{sigma2}), on which we shall focus.
If we insert this term into  Eq.~(\ref{drudesigma}) we find an interference term between longitudinal and transverse propagation,
\begin{eqnarray*}
  \mathbf{J}^{\delta\Sigma}(\mathbf{p},\mathbf{q}) &=& -\mathrm{Re }\, \left(\frac{ n^2t^2}{k^2} G_T(p) G_L(p) \right) \frac{1}{p}\\
\times &&  ( 2 \hat{\mathbf{p}} \hat{\mathbf{p}} (\hat{\mathbf{p}}\cdot
   \mathbf{q}) -\hat{\mathbf{p}}  \mathbf{q} -
    \mathbf{q} \hat{\mathbf{p}}  )
\end{eqnarray*}
Its contribution to the Poynting vector in the $J_2$-channel diverges again as $\sum_\mathbf{p}1/p^2$. Restricting to large wave vectors,
\begin{equation}\label{DTL2}
 \pi N(k) D^{\delta\Sigma}_2(k) = \frac{Q_0}{12 \pi }\frac{\mathrm{Re}\left(n^2t^2\right) }{k^3}
\end{equation}

The remainder of $\mathbf{\Sigma}(p)$ in Eq.~(\ref{self}) provides contributions to $J_0$, $J_1$ and $J_2$, and is proportional to $n^2$ once
 the
divergency has been removed. Some formula manipulation gives the following closed expression for the diffusion constant caused by the dependence on wave number of the
self-energy of two electric dipoles,
\begin{eqnarray}\label{DTL4}
 \pi N(k) D^{\delta \Sigma}(k) =  \frac{1}{4} n^2 \mathrm{Re}\, \mathrm{Tr}\, \int d^3\mathbf{r } (\mathbf{r}\cdot\hat{\mathbf{ q}})^2 \nonumber \\
 \left(
 \frac{t^2\mathbf{G}_0^2}{1-t^2\mathbf{G}_0^2}
 - t^2 \mathbf{G}_0 \cdot \mathbf{G}_{0,T} \right)
\end{eqnarray}
This expression is free from any singularity but is beyond the scope of this work, being a factor $1/k\ell$ smaller than
what was found in Eq.~(\ref{Dlong}) for the $J_2$-channel, and even a factor $1/(k\ell)^3$ smaller than the leading contribution in the $J_0$-channel.

\subsubsection{Scattering diffusion current tensor}
\label{scattdifsection}

The scattering diffusion current tensor $\mathbf{J}^{\delta G}(\mathbf{p},\mathbf{q})$ is given by the second term in Eq.~(\ref{BSJ}).
It vanishes for
any isotropic scattering in $\mathbf{U}_{\mathbf{pp}'}$, among which (here) single scattering.
 Among the different scattering events generated by two
electric dipoles, only the most-crossed diagrams and the forward-crossed diagrams induce an anisotropy in scattering. They are given by
\begin{equation}\label{MC2}
    {U}^{\mathrm{MC}}_{\mathbf{pp}'}= n^2 |t|^2 \int d^3\mathbf{r} \,\frac{ t\mathbf{G}_0}{1-t^2 \mathbf{G}_0^2 }
    \left( \frac{ t\mathbf{G}_0}{1-t^2 \mathbf{G}_0^2}\right)^* \mathrm{e}^{i(\mathbf{p}+\mathbf{p}')\cdot \mathbf{r}}
\end{equation}
and
\begin{eqnarray}\label{FC}
&&  {U}^{\mathrm{FC}}_{\mathbf{pp}'} = n^2 |t|^2 \int d^3\mathbf{r} \nonumber\\    &&  \  \    \left[\frac{ \mathbf{1} }{1-t^2 \mathbf{G}_0^2 }
    \left( \frac{\mathbf{1}}{1-t^2 \mathbf{G}_0^2}\right)^*  - \mathbf{11} \right]\mathrm{e}^{i(\mathbf{p}- \mathbf{p'})\cdot \mathbf{r}}
\end{eqnarray}
The most-crossed diagrams generate a  contribution to $ \mathbf{J}^{\delta G}(\mathbf{p},\mathbf{q})$ of the type $J_2$ leading to
a diffusion constant free from any singularity at large $p$, and of the same order as was found in Eq.~(\ref{DTL4}). We will ignore them for the same reason and
focus on the forward-crossed diagrams. We can write
\begin{eqnarray*}
    &&\sum_\mathbf{p' } {U}^{\mathrm{FC}}_{\mathbf{pp}'} \cdot \delta_\mathbf{q} \mathbf{G}_0(\mathbf{p}')  = n^2 |t|^2 \int d^3\mathbf{r}
    \, (i\mathbf{q}\cdot \mathbf{r})
    \\
    &&  \left[\frac{ \mathbf{1} }{1-t^2 \mathbf{G}_0^2 } \cdot \mathrm{Re}\, \mathbf{G}_0
    \cdot \left( \frac{\mathbf{1}}{1-t^2 \mathbf{G}_0^2}\right)^*  - \mathrm{Re}\, \mathbf{G}_0 \right]\mathrm{e}^{i\mathbf{p}\cdot \mathbf{r}}
\end{eqnarray*}
This integral is regular for all $p$, but does not decay fast enough with $p$ to prevent singularities in the channels $J_2$ and $J_3$.  To see this,
the factor between brackets is written as $ \mathbf{F}= F_0(r) \mathbf{1} + F_1(r)
\hat{\mathbf{r}}\hat{\mathbf{r}} $.
The space integral above can be done to get,
\begin{eqnarray*}
  \mathbf{J}^{\delta G}(\mathbf{p},\mathbf{q}) =  n^2|t|^2 \mathbf{ G}(\mathbf{p})\mathbf{G}(\mathbf{p})^* &\cdot &  \\
   (\hat{\mathbf{p}}\cdot \mathbf{q}) f_0(p)+ \hat{\mathbf{p}}\hat{\mathbf{p}} f_1(p)
  &+& (\hat{\mathbf{p}}{\mathbf{q}}+{\mathbf{q}}\hat{\mathbf{p}}) f_2(p)
\end{eqnarray*}
with $3$ known functions related to $F_i(r)$. The first term with $f_0(p)$ is part of $J_0$, and
constitutes a high-order correction to transverse diffusion, of no interest here.
The  longitudinal term with $f_1$ produces no Poynting vector.
We concentrate on the  term with $f_2$, given by
\begin{equation*}
    pf_2(p) =-\int d^3 \mathbf{r} \,  F_1(r) {j_2(pr)}
\end{equation*}
This integral is finite for all $p$ but does not decay with $p$ since
  \begin{eqnarray*}
    \lim_{p\rightarrow \infty } pf_2(p) &=& \lim_{p\rightarrow \infty } \frac{-1}{p^3}\int d^3 \mathbf{y} \,  F_2(y/p) {j_2(y)} \\
    &=& -\frac{3}{4\pi k^2}\int d^3 \mathbf{y} \, \frac{{j_2(y)}}{y^3}= -\frac{1}{k^2}
\end{eqnarray*}
 We have used that for small $r$,
$\mathbf{F}(\mathbf{r})= -\delta(\mathbf{r})/3k^2 -(1-3\hat{\mathbf{r}}\hat{\mathbf{r}} ) /4\pi k^2 r^3$. The local contact term
does not
contribute.
The diffusion constant in the $J_2$-channel is given by,
\begin{equation}\label{S1}
    \pi N(k)D^{\delta G}_2(k)=-\frac{1}{3}  n^2|t|^2  \sum_\mathbf{p} p f_2(p) \mathrm{Re}\, G_{0,L}(p)\bar{G}_{0,T}(p)
\end{equation}
This equation thus suffers from a divergence. Upon splitting it off and regularizing  $\sum_\mathbf{p} 1/p^2 = Q_0 k/4\pi $ we find
\begin{equation}\label{S2a}
    \pi N(k)D^{\delta G}_2(k)= -  Q_0  \frac{n^2|t|^2 }{12\pi k^3} + \mathcal{O}(n^2)
\end{equation}
This is the third diverging term that will cancel against the two already found earlier.  The term proportional to $f_2(p)$ also produces a
contribution to the $J_3$-channel,
\begin{equation}\label{J3S}
J_3^{\delta G}(p)= -\frac{ n^2|t|^2}{k^2} \frac{f_2(p)}{p} \mathrm{Im}\, G_T(p)
\end{equation}
This function decays rapidly as $1/p^6$ for large $p$. It is easily checked that the integral $\sum_\mathbf{p} pJ_3(p)$ is not singular at large $p$ and produces a correction of order $n^2$ in
Eq.~(\ref{imPoyn}) that will not be further discussed.

\subsubsection{Weak localization}

The last term in Eq.~(\ref{BSJ}), defined as $\mathbf{J}^S(\mathbf{p},\mathbf{q})$,  mixes in principle all four transport  mechanisms $J_i$.
For our model of electric dipoles, the ISA makes no contribution since isotropic, but  the
diagrams~(\ref{MC2}) and (\ref{FC}) do. The leading order is obtained by inserting on the right hand the Drude expression for the transverse
diffusion current tensor $\mathbf{J}^{TT}$ found in Eq.~(\ref{JT}). Since this current is strongly peaked near $p=k$  we can approximate
$ \mathbf{J}^{TT}(\mathbf{p},\mathbf{q}) = 2\pi \ell
(\hat{\mathbf{p}}\cdot\mathbf{ q})\mathbf{ \Delta}_p \delta(k^2 -p^2)$ so that,
\begin{equation}\label{WL1}
    \mathbf{J}^S(\mathbf{p},\mathbf{q})= \frac{k\ell}{2\pi } \mathbf{G}(\mathbf{p})\mathbf{G}(\mathbf{p})^* \cdot
    \int \frac{d\hat{\mathbf{ k}} }{4\pi}U_{\mathbf{pk}}\cdot
   \mathbf{ \Delta}_k (\hat{\mathbf{k}}\cdot \mathbf{q})
\end{equation}
Only the angle-dependent scattering $U^{MC}$ and $U^{FC}$ survive this integral. For convenience  we can summarize Eqs.~(\ref{MC2}) and (\ref{FC}) by
\begin{equation}\label{CminF}
    U_{\mathbf{pp}'} = \int d^3 \mathbf{r}\, \left[ U^{MC}(\mathbf{r}) e^{i(\mathbf{p}+\mathbf{p}')\cdot \mathbf{r}} +
    U^{FC}(\mathbf{r}) e^{i(\mathbf{p}-\mathbf{p}')\cdot \mathbf{r}}\right]
\end{equation}
The angular integral over $\hat{\mathbf{k}}$ can be performed to get
\begin{eqnarray*}
  && \mathbf{J}^{S}(\mathbf{p},\mathbf{q}) =  \frac{k\ell}{2\pi i} \mathbf{G}(\mathbf{p})\mathbf{G}(\mathbf{p})^* \cdot
\int d^3 \mathbf{r} \, e^{i\mathbf{p}\cdot \mathbf{r}}  \cdot \\
 &&\ \ \ \  \left[U^{MC}(\mathbf{r}) -
    U^{FC}(\mathbf{r}) \right] \cdot\\
&& \left[\frac{j_2(kr)}{kr} ( \hat{\mathbf{r}}\mathbf{q}+ {\mathbf{q}} \hat{\mathbf{r}}) -(\hat{\mathbf{r}}\cdot \mathbf{q}) \left( j_1(kr) -\frac{j_2(kr)}{kr}  + j_3(kr)\hat{\mathbf{r}} \hat{\mathbf{r}} \right)\right]
 \end{eqnarray*}
The integrand of this equation for $\mathbf{J}^S(\mathbf{p},\mathbf{q})$ involves the difference $U=U^{MC} - U^{FC}$ between most-crossed and forward-crossed diagrams.
They both contain sub-radiant poles (where $t^2\mathbf{G}_0^2 \approx 1$), and quite remarkably, this singularity cancels significantly in this subtraction.
The equation  generates all 4 transport modes,
\begin{eqnarray*}
    \mathbf{J}^{S}(\mathbf{p},\mathbf{q}) &=&   J_0^{S}(p)  (\hat{\mathbf{p}}\cdot \mathbf{q})
    +   J_1^{S}(p) \hat{\mathbf{p}}  \hat{\mathbf{p }} (\hat{\mathbf{p}}\cdot \mathbf{q}) \\
    &+& J_2^{S}(p) (\hat{\mathbf{p}} {\mathbf{q }} +{\mathbf{q}}\hat{\mathbf{p}})  + J_3^{S}(p) i(\hat{\mathbf{p}} {\mathbf{q }} - {\mathbf{q}}
      \hat{\mathbf{p }} )
\end{eqnarray*}
We will show that the mode ${J}^S_0$ exhibits the standard weak localization correction, of relative order $1/k\ell $ and
negative in diffusion constant. Also the mode ${J}^S_2$ is subject to a  weak localization correction,
of order $1/(k\ell)^2$ and\emph{ positive}, showing that
not all modes are affected similarly by interference.

We first focus on $J_0^S$. Contrary to $U^{FC}$, $U^{MC}$ associated with two dipoles induces a singular
angular dependence
of the kind $1/|\mathbf{p}+\mathbf{p}'|$, and therefore dominates ${J}_0^S$.
The space integral is dominated by large $r$ so we insert $U^{MC} = (6\pi /\ell)^2 \mathbf{C}(\mathbf{r})\mathbf{ C}(\mathbf{r})^*$ with
$\mathbf{C}\approx - \mathbf{\Delta}_r (\exp(ikr)/4\pi r)$.
The angular integral over $\hat{\mathbf{r}} $ can be done. The end result is written as
\begin{eqnarray*}
 J_0^{MC}(p)  &=& - \frac{9}{2}\frac{k}{ \ell }|G_T(p)|^2  \\
 \times   \int_0^\infty &dr&  \left( \frac{4}{5} j_1(kr)  - \frac{1}{5}j_3(kr) \right)
 \left( \frac{4}{5} j_1(pr)  - \frac{1}{5}j_3(pr) \right) \\
 &=& - \frac{3\pi }{20}\frac{1}{\ell}|G_T(p)|^2\left( 2 \frac{k^2}{p^2} + \frac{9}{7}\frac{k^4}{p^4} \right)
\end{eqnarray*}
The last equality holds only for $p \ge k$; $J_0^{MC}(p)$ decays rapidly as $1/p^6$ and has most of it weight near $p=k$.
The weak localization correction  can be obtained from Eq.~(\ref{Kubo}),
\begin{eqnarray}\label{WL0}
    \Delta D^{WL}_0 &=& \frac{1}{4\pi N(k)} \mathrm{Tr} \sum_\mathbf{p} \mathbf{L}(\mathbf{p},\hat{\mathbf{q}}) \cdot \mathbf{\Delta}_p
    (\hat{\mathbf{p}}\cdot \hat{\mathbf{q}}) \nonumber \\
    &\times& \frac{-3\pi^2 }{20 k }\frac{23}{7} \delta(k^2-p^2) =  - \frac{1}{3}\frac{v_E }{k} \frac{69\pi  }{280}
\end{eqnarray}
or equivalently $ \Delta D^{WL}_0/D^T = - 0.774/k\ell$. The numerical factor is actually larger than the leading one ($\pi/6=0.523 $) obtained for scalar
waves \cite{wellens}.
We can
compare this weak localization correction to the positive diffusion constant~(\ref{Dlong})  found for the $\mathbf{J}_2$ channel. If we extrapolate
to small values for $k\ell$, we conclude that the diffusion in the $J_2$-channel compensates the first weak localization correction
in the $J_0$-channel  for $k\ell < 1.3$.

The channel $J_2^S$ is more complicated. It is instructive split the Green's function up into $\mathbf{G}_0(\mathbf{r})\sim P(r) \mathbf{\Delta}_r +
Q(r)\hat{\mathbf{ r}} \hat{\mathbf{r}}$, and to express the tensor $U(\mathbf{r}) = U^{MC} - U^{FC}$  as,
\begin{eqnarray*}
    U(&&\mathbf{r}) = U^{TT} (r) \mathbf{\Delta}_r \mathbf{\Delta}_r  + U^{LL} \hat{\mathbf{r}}\hat{\mathbf{r}}\hat{\mathbf{r}}\hat{\mathbf{r}}
    \\
    &&+ \mathrm{ Re }\, U^{TL} ( \hat{\mathbf{r}}\hat{\mathbf{r}} \mathbf{\Delta}_r
    + \mathbf{\Delta}_r \hat{\mathbf{r}}\hat{\mathbf{r}} )
    + i \mathrm{Im}\, U^{TL} ( \mathbf{\Delta}_r \hat{\mathbf{r}}\hat{\mathbf{r}} -  \hat{\mathbf{r}}\hat{\mathbf{r}}\mathbf{\Delta}_r )
\end{eqnarray*}
This corresponds to 4 different scattering events  involving two dipoles at distance $\mathbf{r}$ with  the electric field vector
either along or perpendicular to $\mathbf{r}$,
as well as  their interferences. With the angular integral
of $\hat{\mathbf{r}}$ performed analytically, they give each the following contribution to $J_2^S$,
\begin{eqnarray*}
    J_2^{TT}(p) &=& \frac{2\ell}{k} \mathrm{Re}\, G_T(p) \int_0^\infty dr\, r^2 \,  U^{TT}(r) \\
     &\times& \left(j_1(kr)-\frac{j_2(kr)}{kr}\right)
    \frac{j_2(pr)}{pr}
\end{eqnarray*}
\begin{equation*}
    J_2^{LL}(p)= -\frac{4\ell}{k} \mathrm{Re}\, G_T(p) \int dr_0^\infty \, r^2 \, U^{LL}(r) \frac{j_2(kr)}{kr} \frac{j_2(pr)}{pr}
\end{equation*}
\begin{eqnarray*}
    J_2^{TL1}(p) &=&  \frac{2\ell}{k} \mathrm{Re}\, G_T(p) \int_0^\infty dr\,r^2\,  \mathrm{Re}\, U^{TL}(r) \\
   &\times& \left( j_1(pr)-2\frac{j_2(pr)}{pr} \right) \frac{j_2(kr)}{kr}
\end{eqnarray*}
and finally
\begin{eqnarray*}
    J_2^{TL2}(p) =  \frac{2\ell}{k} \mathrm{Im}\, G_T(p) \int_0^\infty dr\, r^2\, \mathrm{Im}\, U^{TL}(r)j_1(pr)
\frac{j_2(kr)}{kr}
\end{eqnarray*}
The weak localization correction is found by
\begin{equation}\label{D2J2}
    \Delta D_2 (k) = -\frac{1}{3} \frac{1}{\pi N(k) }\sum_\mathbf{p} p^2 (J_2^{TT} +  J_2^{LL}+  J_2^{TL1}+ J_2^{TL2})
\end{equation}
To perform the integral over the wave vector $\mathbf{p}$ we use that
\begin{equation*}
 \sum_\mathbf{p} \frac{p}{(k+i\epsilon)^2 - p^2} j_1(pr) = -\frac{ik^2 }{4\pi }h_1^{(1)}(kr)
\end{equation*}
and
\begin{equation*}
 \sum_\mathbf{p} \frac{p}{(k+i\epsilon)^2 - p^2} \frac{j_2(pr)}{pr} = -\frac{ik}{4\pi r } \left[h_2^{(1)}(kr) + \frac{3i}{(kr)^3}\right]
\end{equation*}
Consequently, only the radial integrals $\int dr$ remain to be done numerically. The weak-localization correction $\Delta D_2 (k)$ is proportional to the density of the electric dipoles.

Figure~{\ref{DTLfig} shows the total diffusion constant $D_2^{D}+\Delta D_2$ in the $J_2$ channel around the resonance frequency,
as well as the contributions stemming from the 4 individual  terms in Eq.~(\ref{D2J2}).
The weak localization correction $\Delta D_2$ is dominated by the purely transverse and
longitudinal channels $TT$ and $LL$ between the two dipoles, who have competing signs. For negative detuning  the purely longitudinal mode $LL$ dominates, for positive
detunings the $TT$ channel dominates and is more than twice as large as the Drude contribution $D_2^D$.
We note that the weak localization correction to the diffusion constant $D_2$  of the channel $\mathbf{J}_2(\mathbf{p},\mathbf{q})$ is of the same order as the Drude
approximation in Eq.~(\ref{Dlong}).
The sum of the 4 weak localization terms and the Drude approximation is strictly positive. At fixed density, positive detuning has the largest diffusion constants in the
$J_2$ channel. Note that the ratio $D_2/D_0$ is of same  order $1/(k\ell)^2$, but of opposite sign compared to the standard (Cooperon) weak localization correction $-1/(k\ell)^2$.
This will be discussed more in detail in the next section, for which it will turn out useful to define a function $F(\delta) = (k \ell)^2 D_2/D_0$.

\begin{figure}
  \includegraphics[width=0.95\columnwidth]{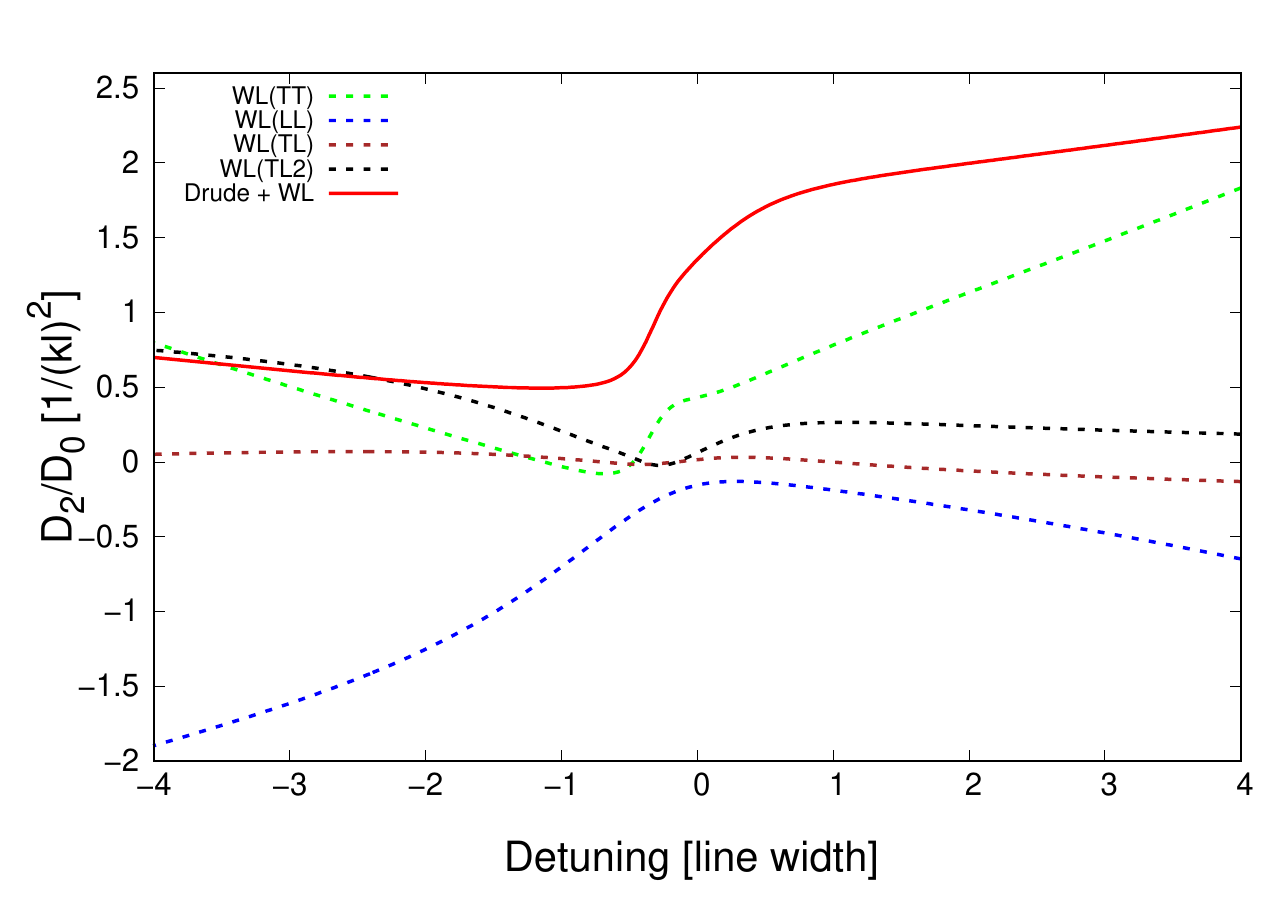}
  \caption{The diffusion constant in the $J_2$ channel (solid line), being the sum of the Drude approximation $D_2^D$ plus the weak-localization correction $ \Delta D_2 (k) $ in Eq.~(\ref{D2J2}) from two dipoles, as a function of the detuning $\delta =
  (\omega-\omega_0)/\gamma$.  It is normalized by the diffusion constant in the $J_0$ channel. The 4 weak localization corrections
  discussed in this section are shown separately as dashed lines.}\label{DTLfig}
\end{figure}

\subsubsection{Summary of previous subsections}
\label{secsummary}

We have identified four mechanisms in the transport of electromagnetic waves, expressed by the diffusion current tensor~(\ref{Jgeneral}).
The results have been summarized in Table~\ref{tabeldiff}.
The mechanism described by $J_0(p)$ is the familiar picture of transverse wave diffusion near the shell $p\approx k$
and results in the diffusion constant (\ref{DDrude}). It is inversely proportional to the density of the electric dipoles and contains
an energy velocity that can be small since  the impenetrable electric dipole scatterers contain temporarily stored, longitudinal energy.
The mechanism associated with $J_2$ is caused by interference of longitudinal and transverse fields, necessary condition to carry a Poynting vector.
The leading term~(\ref{Dlong}), linear in the dipole density, comes from the Drude approximation.
Upon considering all scattering events involving two dipoles, we have been able to identify  three singular terms. After regularization, they are expressed by
Eqs.~(\ref{DTL}), (\ref{DTL2}) and (\ref{S2a}) and  proportional to the large quality factor $Q_0$ and the density squared.
They add up to
\begin{eqnarray}\label{DQ}
    \pi N(k) \Delta D_2 (k) &=&  \frac{Q_0}{6\pi k^3} n^2 \left( (\mathrm{Im}\, t)^2 + \frac{1}{2}\mathrm{Re} \, t^2 - \frac{1}{2}|t|^2 \right)\nonumber \\  &=&0
\end{eqnarray}
This explicit cancelation in the $J_2$-channel is very important and not entirely obvious since the 3 terms stem from entirely different parts in
transport theory (Drude diffusion, Lorentz local field and enhanced forward scattering). Without cancelation they would have given an electromagnetic  conductivity  $Q_0/k\ell^2$,
and not small at all with respect to the traditional transverse conductivity, of order  $k^2\ell$ since  $Q_0$ is large for an atomic oscillator.
Their cancelation also supports  the general renormalizability of electromagnetic transport theory with point-like dipoles.
It is highly plausible that this cancelation happens in all orders of perturbation theory, but this is currently impossible to prove in general.
We will use this hypothesis in the next session.

\begin{table}

\begin{tabular}{|c||c|c|c|c|c|}
  \hline
    & Drude $\mathbf{J}^D$ & $\mathbf{J}^{\delta \mathbf{\Sigma}}$ & $\mathbf{J}^{\delta \mathbf{G}}$  & WL $\mathbf{J}^{S}$ \\ \hline\hline
  $J_0$ & $+\ell$ &  $ 1/k^3\ell^2$ &  $1/k^3\ell^2$  & $ -0.774/k $  \\
   & (\ref{DDrude}) & (\ref{DTL4})  &  NC   & (\ref{WL0}) \\ \hline
  $J_2$ & $+1/k^2\ell $ & $-Q(\delta)/2k^3\ell^2$ & $-{Q(\delta)}/2k^3\ell^2$ & $+{F(\delta)}/{k^2\ell}$ \\
  & $+ Q(\delta)/k^3\ell^2$   (\ref{Dlong}) & (\ref{DTL2})  &  (\ref{S2a})  & (\ref{D2J2}) Fig.~\ref{DTLfig} \\   \hline
  $J_3$  & $+1/k$  &  $ 1/k^3\ell^2$  & $ 1/k^3\ell^2$  &   ${1}/{k^2\ell}$    \\
   & (\ref{DIM2}) &  (\ref{DTL4})   &   (\ref{J3S})  & NC  \\ \hline
  \hline
\end{tabular}
\caption{Contributions to the transport mean free path for various transport channels $J_i$ and the 3 different diagrammatic classes identified in
 the Bethe-Salpeter equation~(\ref{JDall}). When an explicit sign is found, it is indicated.  Most values depend also explicitly on detuning, not indicated if not calculated.
 Numbers refer to the corresponding equations. The channel $J_1$ does not contribute to transport mean free path,  $J_3$ only contributes to the transport mean free path associated with the
 imaginary part of the Poynting vector. NC stands for ``not calculated'', WL for ``weak localization". The terms proportional to $Q(\delta)$ are regularized singularities depending
 on detuning $\delta$ that cancel in the transport mean free path.  }\label{tabeldiff}
\end{table}

The  $J_2$ channel, in the leading order modified by the weak localization from 2 electric dipoles, exhibits a positive diffusion constant, linear in the dipole density.
Although usually small compared to standard transverse diffusion, it must
be realized that this diffusion stems from a sofar unexplored mechanism for electromagnetic wave diffusion,
involving the interference of longitudinal and transverse waves. In this transport channel, the first weak localization correction induced
 by two electric dipoles is
actually of the same order as the Drude value and again \emph{positive}, showing that in the channel $J_2$ interferences behave differently
in comparison to the traditional transverse channel.
 {In early experiments on light scattering by cold Rubidium atoms \cite{nice}, values of $k \ell$ are of the order $1000$
and hence the $J_2$ channel should be irrelevant for optical transport. However, atomic Ytterbium clouds with very large densities
can be created  \cite{japan} with  $k \ell < 1$. Understanding light transport in such clouds will certainly require taking into
account $J_2$ channel involving  longitudinal modes.
}

\section{Radiative force density}\label{sectionrad}

A well-known relation exists between diffuse flow and radiative forces. In radiative transfer,
the energy flux is driven by the spatial  gradient of the total energy density,  and automatically carries momentum. In the presence of an induced polarization density $\mathbf{P}$, the electromagnetic force density $\mathbf{f}$ is caused by
the Lorentz force acting on the induced Coulomb charge
density $\rho (\mathbf{r})= -\bm{\nabla} \cdot \mathbf{P}$ and on the induced current density $\mathbf{j}=\partial_t \mathbf{P}$.
Maxwell's equations allow the formulation of a momentum conservation law that is  very generally valid. It takes the form (before cycle averaging) \cite{loudonmom,abrahamik},
 \begin{eqnarray}\label{momentumcon}
 \partial_t \mathbf{\mathcal{G} }&+& \mathbf{f}  = \bm{ \nabla} \cdot  \mathbf{T}
 \end{eqnarray}
 with $\mathbf{\mathcal{G}}= (\mathbf{E} \times  \mathbf{B})/4\pi c_0$ the electromagnetic momentum density, and $\mathbf{T}$ the momentum stress tensor,
 \begin{equation}
 \mathbf{T} = \frac{1}{4\pi }\left[ \mathbf{EE }+\mathbf{ BB}
 -\frac{1}{2} \left( {\mathbf{E}}^2 + \mathbf{B}^2\right) + \mathbf{X}        \right]
 \end{equation}
 The tensor $\mathbf{X}$ is related to internal angular momentum inside the particle
 that we shall ignore here.

In the regime of multiple scattering, and after cycle averaging, Eq.~(\ref{dij}) expresses that $\langle E_i(\mathbf{r})\bar{E}_j(\mathbf{r}) \rangle
 = \frac{1}{3} \langle |\mathbf{E}(\mathbf{r})|^2 \rangle  \delta_{ij}$, and idem for the magnetic field. The stress-tensor $\mathbf{T}$ is thus diagonal on average,
 meaning that the $i^{\mathrm{th}}$ component of the electromagnetic momentum only flows in the direction $i$. For stationary flow, we thus obtain
 \begin{equation}\label{radiativeforce}
  \langle  \mathbf{f}(\mathbf{r})\rangle = -\frac{1}{3}\bm{\nabla} \left< \frac{|\mathbf{E}(\mathbf{r})|^2 +|\mathbf{B}(\mathbf{r})|^2 }{16\pi} \right>
 \end{equation}
For a medium filled with impenetrable electric dipoles we have shown in  Eq.~(\ref{DOS3}) that
 $|\mathbf{E}|^2/16\pi$ is the total electric energy density having both longitudinal and transverse components, and equal to the magnetic energy density.
 In the diffusion approximation, we write the averaged Poynting vector as
 $\langle \mathbf{K}\rangle  = -D \bm{\nabla}  \left< {\left[ |\mathbf{E}(\mathbf{r})|^2 +|\mathbf{B}(\mathbf{r})|^2 \right] }/{16\pi} \right>$.  This leads to a simple relation
 \begin{equation}\label{KisP}
    \langle \mathbf{f} \rangle = \frac{1}{3D}\langle \mathbf{ K} \rangle
 \end{equation}
 between
Poynting vector and radiative force density.
 In the ISA, $D= \frac{1}{3} v_E \ell/(1 - \langle \cos\theta \rangle )$, and this reduces to the almost intuitive expression
 $\langle \mathbf{f} \rangle= n\sigma (1- \langle \cos \theta\rangle )  \langle \mathbf{K} \rangle /v_E$ involving
 the product of particle density and pressure cross-section of one scatterer.
The second factor accounts for transfer of momentum  from the light to a single  scatterer, and
 of course for independent electric dipoles $\langle \cos\theta \rangle=0$.

 The factor $1/v_E$ is less intuitive in this model. For one isolated scatterer this would clearly be $1/c_0$,
 since for a plane wave with arbitrary direction in vacuum, momentum current density and energy current density (the Poynting vector) differ by a factor $1/c_0$.
 In a medium filled with resonant dipoles, stocked, longitudinal energy contributes to the momentum current density $\mathbf{T}$ but not to the
 energy current density $\mathbf{K}$. Put otherwise, scattering of a transverse state with wave number $p\approx k$ to a longitudinal mode with large wave number induces a significant recoil,
 but does
 not generate an energy current. For the medium filled with dipoles, the ratio $f/K$ thus achieves a factor $(N_L + N_T)/N_Tc_0 \approx 1/v_E$.

\section{No Anderson localization of light? }\label{sectionAL}

In the following we will make a first attempt to include the 4 transport mechanisms, introduced in the previous section, into the self-consistent transport
theory for localization of light.
This theory is celebrated by some for its surprisingly simple description of the transition from long-range diffusion to localization. Others criticize the theory for its oversimplified  nature, neglecting many scattering events in the collision operator
$U_{\mathbf{pp}'}(\mathbf{q})$ introduced in Eq.~(\ref{BS2}).
The self-consistent theory  predicts the Ioffe-Regel
criterion $k_e\ell \approx 1$ for the mobility edge in 3D, produces the universal finite-size scaling in arbitrary dimension, and can easily be engineered with. However, the theory predicts a wrong critical exponent of the localization transition and fails in the presence of an external magnetic field.

In the standard theory, adapted from electron localization \cite{vw},
the most-crossed diagrams are included into the diffusion constant of the light. These diagrams involve the interference of time-reversed waves and are part of the
scattering vertex  $U_{\mathbf{pp}'}(\mathbf{q})$. By reciprocity, the most-crossed diagrams also contain a
hydrodynamic pole, featuring the same diffusion constant. This immediately turns the calculation of $D$ into a self-consistent problem because the most-crossed diagrams,
modify the diffusion current $J(\mathbf{p},\mathbf{q})$ in Eq.~(\ref{BSJ}). We recall that in the case of
electromagnetic waves the diffusion current is a tensor with 4 independent parts.  No Anderson
localization was seen to occur in recent numerical simulations with electric dipoles \cite{sergey0}. The intention of this section is to discover
 what exactly breaks down in this theory  when taking into account longitudinal waves.


We here summarize the various approximations made, which are basically equivalent to the ones made in previous works, even if often adopted implicitly \cite{sheng, akkermans, vw}.
\begin{itemize}
  \item The most-crossed diagrams, involving scattering events associated with many dipoles, are the only angle-dependent scattering events that influence the
  diffusion current tensor $\mathbf{J}(\mathbf{\mathbf{p}},\mathbf{q})$ when going beyond the Drude
  approach. The existence of other diagrams is only acknowledged implicitly to guarantee flux conservation.
  Weak localization effects caused by low-order scattering events, such as those described by Eqs.~(\ref{WL0}) and (\ref{D2J2}) are not included either,
  although this could be done without dramatic changes in the theory. The
  implicit existence of other diagrams is also necessary to justify the cancelation of UV-singularities  in transport theory. For scattering events involving only two electric dipoles,
  UV-divergencies were seen to cancel \emph{explicitly} earlier in this work, but no general demonstration is known.

  \item The diffuse regime of the most-crossed diagrams, only valid on spatial scales well beyond the mean free path, is assumed to hold on
  scales up to the mean free path.
  On this scale we may expect the diffusion kernel to be of the type $D(q)q^2$ which is disregarded in the standard version of the self-consistent theory.

  \item The electromagnetic self-energies $\Sigma_{T/L}(k,p)$ are assumed not to depend on $p$.
  In particular this means that $\Sigma_T(k) = \Sigma_{L}(k) $. This is definitely an approximation, even for point-like dipoles, that needs
  more study, but in general, such wave number dependence is not believed to be essential for Anderson localization.

\end{itemize}

The contribution of the most-crossed diagrams to the scattering vertex $U_{\mathbf{pp'}} (\mathbf{q})$ can be obtained from the reducible vertex
$R_{\mathbf{pp'}} (\mathbf{q})$ introduced in Eq.~(\ref{BS}) by removing the four external Dyson propagators, and time-reversing the bottom line.
This gives
\begin{equation}\label{MC}
    U^{MC}_{\mathbf{pp}'; ij|kl}(\mathbf{q}) = \frac{\tilde{d}_{il}(\mathbf{f}+\mathbf{q},\mathbf{Q})\tilde{ d}_{kj}(-\mathbf{f}+\mathbf{q},\mathbf{Q}) + \mathcal{O}(Q^2)}
    {\pi N D\mathbf{ Q}^2}
\end{equation}
with the notation
$\mathbf{Q}=\mathbf{p}+\mathbf{p}'$ and $\mathbf{f}=(\mathbf{p}-\mathbf{p}')/2$.
In this expression the tensor $\tilde{\mathbf{d}}(\mathbf{p},\mathbf{Q})$  is the diffuse eigenfunction defined in Eq.~(\ref{dij}) stripped
from  the 4 external lines in Fig.~\ref{Reps} (transforming $-\mathrm{Im}\,\mathbf{ G} + i\mathbf{J}/2 $ into $-\mathrm{Im}\,\mathbf{ \Sigma} + i\mathbf{j}/2 $ with $\mathbf{j}$ again a Hermitian bilinear form). This leads to  $\tilde{\mathbf{d}}(\pm \mathbf{f}+\mathbf{q},\mathbf{Q}) = -\mathrm{Im}\, \Sigma (\pm \mathbf{f}+\mathbf{q}) + \mathbf{j}(\pm \mathbf{f}+\mathbf{q},
 \mathbf{Q})$.

A  generalized Ward identity,
\begin{eqnarray}\label{ward2}
    (\mathbf{q}\cdot \partial_\mathbf{p})  \mathrm{Re}\, \mathbf{\Sigma}(\mathbf{p}) &=& \sum_{\mathbf{p}'} U_{\mathbf{pp}'} (0)
    \cdot  (\mathbf{q}\cdot \partial_{\mathbf{p}'})  \mathrm{Re}\, \mathbf{G}(\mathbf{p}') \nonumber \\
    &&  + \sum_{\mathbf{p}'}  \delta_\mathbf{q} U_{\mathbf{pp}'} (\mathbf{q}) \cdot
    \mathrm{Im}\, \mathbf{G}(\mathbf{p}')
\end{eqnarray}
can be used to eliminate the second term in
Eq.~(\ref{BSJ}), which then transforms into
\begin{eqnarray}\label{BSJ2}
  \mathbf{J}(\mathbf{p},\mathbf{q}) &=& \mathbf{J}^D(\mathbf{p},\mathbf{q}) \nonumber \\
  &+& \mathbf{ G}(\mathbf{p})\cdot (\mathbf{q}\cdot \partial_\mathbf{p})  \mathrm{Re}\, \mathbf{\Sigma}(\mathbf{p}) \cdot \mathbf{G}^*(\mathbf{p})
  \nonumber \\
  &-& \mathbf{ G}(\mathbf{p})\mathbf{G}^*(\mathbf{p}) \cdot  \sum_{\mathbf{p}'}  \delta_\mathbf{q} U_{\mathbf{pp}'} (\mathbf{q}) \cdot
    \mathrm{Im}\, \mathbf{G}(\mathbf{p}') \nonumber \\
   &+& \mathbf{ G}(\mathbf{p})\mathbf{G}^*(\mathbf{p}) \cdot \sum_{\mathbf{p}'} {U}_{\mathbf{pp}'}\cdot \mathbf{J}(\mathbf{p}',\mathbf{q})
\end{eqnarray}
The first and second terms cannot depend on diffusion constant. Because  $U^{MC}$ depends on
both diffusion constant $D$ and the entire diffusion tensor $\mathbf{J}(\mathbf{p},\mathbf{q})$, the self-consistent theory would, in its most advanced version, be a non-linear
integral equation for the second-rank tensor $\mathbf{J}$.

In the following we apply the approximations specified above. The above hydrodynamic limit of $U^{MC}$ is assumed  valid when
$|\mathbf{p}+\mathbf{p}'| \ll 1/\ell$. In the standard approach  of the self-consistent theory
one focusses on its diffuse pole near $\mathbf{p}\approx - \mathbf{p}'$, and neglects all other dependence on $\mathbf{p}'$.
Secondly, wave number  dependence of the self-energy is ignored.
In that case the self-consistent problem simplifies  to the following equation,
\begin{eqnarray}\label{SC1}
   \mathbf{ J}(\mathbf{p},\mathbf{q}) &\approx& \mathbf{J}^D(\mathbf{p},\mathbf{q}) + \mathbf{G}(\mathbf{p})\mathbf{G}(\mathbf{p})^* \cdot
  \nonumber \\  &\, & \sum_{|\mathbf{Q}| < q_m} U^{MC}_{\mathbf{Q}=\mathbf{p}+\mathbf{p}'}(0) \cdot\mathbf{ J}(-\mathbf{p} ,\mathbf{q})
\end{eqnarray}
with the Drude current tensor $\mathbf{J}^D$ given in Eq.~(\ref{drudeagain}). In particular, the third term in Eq.~(\ref{BSJ2}) becomes proportional to
$\sum_\mathbf{Q} \mathbf{Q} /D\mathbf{Q}^2$ and drops out. The sum over $\mathbf{Q}$ that remains in Eq.~(\ref{SC1}) is recognized as the
return Green's function of the diffusion
equation in real space \cite{akkermans}, though with short, non-diffusive paths eliminated by the condition $Q < q_m$.

Diffusion constant and diffusion current tensor are related by
the Kubo formula
\begin{eqnarray}\label{SC2}
\sigma \equiv \pi N D  &=&   \frac{1}{4}\mathrm{Tr} \, \sum_\mathbf{p}
    \mathbf{L} (\mathbf{p},\hat{\mathbf{q}}) \cdot \mathbf{J}(\mathbf{p},\hat{\mathbf{q}})
\nonumber \\ &=& \frac{1}{6\pi^2}\int_0^\infty dp \,  p^4 \left( J_0(p) -J_2(p) \right)
\end{eqnarray}
It can readily be seen that Eq.~(\ref{SC1}), despite its simplicity,  couples  the four diffusion current tensors
identified in Eq.~(\ref{Jgeneral}), among which $J_0(p)$ and $J_2(p)$ are relevant in Eq.~(\ref{SC2}). A mobility edge is characterized by $D=0$.
The large weight of large wave numbers ($p \gg k$) in the Kubo formula is evident and UV-divergences will occur that will be regularized with
the argument that other diagrams exist that compensate.

\subsection{Transverse approximation}
In most applications of the self-consistent theory for localization of light one ignores polarization and focusses on the transverse channel $J_0(p)$ and, not unrelated,
assumes  this channel to be governed by
excitations near the frequency shell $p\approx k_e$ of the effective medium where their DOS is largest. In this approximation weak localization of light becomes
essentially equivalent to the one of scalar waves. As a matter of fact, this approximation applies to localization of elastic waves with all
polarizations modes  propagating with the same velocity everywhere. We refer to the work of Zhang and Sheng \cite{zhang} where the self-consistent theory for
localization of scalar waves is derived and discussed in great detail.

We will first neglect the weak localization found in Eq.~(\ref{WL0}) associated with two dipoles and incorporate it in the next section when dealing with the mode $J_2(p)$.
Upon putting $\mathbf{J}(\mathbf{p},\mathbf{q}) = J_0(p) (\hat{\mathbf{p}}\cdot \mathbf{q}) \mathbf{\Delta}_p$ into Eq.~(\ref{SC1}), and by assuming that $\mathrm{Im}\, \Sigma(p)$ is independent of $p$,
the explicit solution is just
\begin{equation}\label{SC3}
    J_0(k,p) = J_0^D(k,p)\left[1 + \frac{\sigma_c}{\sigma} A(k,p) \right]^{-1}
\end{equation}
with the dimensionless function $A(k,p) = |G_T(k,p)|^2 \times \mathrm{Im}^2 \Sigma_T(k,p) $, and
a critical conductivity defined as  $\sigma_c \equiv \sum_Q 1/Q^2 = q_m/2\pi^2 $. Note that $A(k,p) \leq 1$ is a bounded function of $p$. Equation~(\ref{SC3}) says that the amount of weak
localization varies in phase space,
and is maximal at the frequency shell $p=k_e$ of the transverse
waves, and small when $p\gg k$. From Eq.~(\ref{SC2}) we obtain a closed equation for $\sigma$,

\begin{equation}\label{SC4}
  \sigma(k) = \frac{1}{3} \sum_\mathbf{p} \frac{p^2 J_0^D(k,p)}{1 + (\sigma_c/\sigma) A(k,p) }
\end{equation}

The Kubo formula attributes a large weight to large $p$, nevertheless the integral converges for all  $\sigma > 0$. If $\sigma > \sigma_c$ large wave vectors
$p$ are not relevant in the denominator since $J_0^D(k,p)= 4 (\mathrm{Im}\, G_T(k,p))^2$
decays rapidly with $p$. The integral is dominated by $\mathbf{p}$ near the frequency shell $p\approx k_e$ so that
\begin{eqnarray}\label{SC5}
     \sigma(k) &\approx & \frac{\sigma^D (k)}{1 + (\sigma_c/\sigma)}
    \Rightarrow \sigma(k) = \sigma^D(k) \left( 1 - \frac{\sigma_c}{\sigma_D} \right) \nonumber \\
    &=& \sigma^D(k)  \left( 1 - \frac{3}{\pi} \frac{1}{(k_e\ell)^2}\right)
\end{eqnarray}
where $q_m= 1/\ell$ has been chosen.  This result, when extrapolated,
locates the mobility edge at $k_e\ell = 0.977$.

For $\sigma < \sigma_c$ however, the $p$-dependence of the denominator
shifts the integral over $p$ to larger values for  $p$.
At the
mobility edge $\sigma=0$  and
\begin{equation}\label{MEdiv}
    \sigma_c = \frac{4}{3}\sum_\mathbf{p} p^2 |G_T(k,p)|^2
\end{equation}
 {This involves an integral that diverges as } $\sum_\mathbf{p} 1/p^2$,  {or equivalently as} $1/r$ as $r\rightarrow 0$,  {a singularity that is
not to be confused with the diverging integral over diffuse modes } $\mathbf{Q} $ in
Eq.~(\ref{SC1})  {that is a clear artifact of the diffusion approximation at small length scales } {and}  { repaired by the cut-off} $q_m$.
 {This present divergency at  large} $p$ is likely to be artificial and caused
by one of the above approximations inherent of the self-consistent theory to go from} Eq.~(\ref{BSJ2}) to Eq.~(\ref{SC1}).  {In standard approaches
of the self-consistent theory} \cite{vw,sheng}  {this problem is avoided by assuming} $J(\mathbf{p},\mathbf{q})$ to be ``strongly peaked near the frequency shell'' $p\approx k_e$.
 {The neglect of the third term on the righthand side of} Eq.~(\ref{BSJ2})  {is no longer justified for large $p$ since $Q$ then also becomes large and one would need to generalize }
Eq.~(\ref{MC}) { beyond the diffusion approximation.}

In this work we will ignore this complication.
When we subtract the  singularity $\sum_\mathbf{p} 1/p^2$ by hand, assuming it cancels against other terms that have been ignored, we get
\begin{eqnarray}\label{SC6}
  \sigma_c &= &\frac{4}{3} \sum_\mathbf{p} \left( p^2|G_T(k, p)|^2 -\frac{1}{p^2} \right) \nonumber \\
  &=&
  \frac{k^2_e\ell}{3 \pi} \left(1 - \frac{3}{4(k_e\ell)^2}\right)
\end{eqnarray}
This locates the mobility edge at $k_e\ell = 0.866$ with the choice $q_m = 1/\ell$. This is close to the extrapolated value above, and
we could argue that the extrapolation in Eq.~(\ref{SC5}) is satisfactory up to the mobility edge and  consistent with both previous theory
\cite{zhang,sheng}  and numerical simulations of scalar dipoles \cite{sergeyscalar}. It is nevertheless tempting to speculate  that this divergency highlights
 a true breakdown of the self-consistent theory
and that a more rigorous regular solution  may actually exhibit a critical
exponent different from one, the value predicted by the extrapolation~(\ref{SC5}).

\subsection{Inclusion of longitudinal modes}

In this section we give a simplified description of how the self-consistent theory is extended when the other 3 diffusion modes are included.
Let us start with Eq.~(\ref{Jgeneral}) and write  the diffusion current tensor as
\begin{eqnarray}\label{Jgeneral2}
 {J}_{ij}(\mathbf{p},\mathbf{q}) &=& \sum_{n=0}^3 J_n (p)  {\chi}^{n}_{ij}(\mathbf{p},q)
\end{eqnarray}
Let us set $ U^{MC}_{ij;kl} =  (U/\sigma)  \delta_{kj} \delta_{il}$ with
$U= (\mathrm{Im}\, \Sigma)^2 \sigma_c$ (with dimension $1/m^5$) and
$\sigma = \pi N D$ the conductivity (with dimension $1/m$).
We can check that,
\begin{eqnarray*}
 && G_{ni}  G^*_{jm}   \delta_{kj} \delta_{il}  {\chi}^{0}_{kl}  = |G_T|^2{\chi}^{0}_{nm}  \\
 &&   G_{ni}  G^*_{jm}  \delta_{kj} \delta_{il}  {\chi}^{1}_{kl}=    |G_L|^2 {\chi}^{1}_{nm}\\
 && G_{ni}  G^*_{jm}  \delta_{kj} \delta_{il} {\chi}^{2}_{kl}= {R} ({\chi}^{2}_{nm}- 2{\chi}^{1}_{nm} ) + I   {\chi}^{3}_{nm} \\
   && G_{ni} G^*_{jm}  \delta_{kj} \delta_{il}  {\chi}^{3}_{kl}   =  -R {\chi}^{3}_{nm} + I  ({\chi}^{2}_{nm}- 2
   {\chi}^{1}_{nm}) \\
\end{eqnarray*}
where we abbreviated $R(p)= \mathrm{Re} \, G_L\bar{G}_T$ and $I(p) =  \mathrm{Im} \, G_L\bar{G}_T$. This gives the following self-consistent set of equations
\begin{eqnarray}\label{scvector}
 && J_0(p) = J_0^D(p) -   \left( \frac{U}{\sigma}  |G_T(p)|^2  +  \frac{{0.774}}{k_e\ell } \right)   J_0(p) \nonumber \\
 && \left( 1 + |G_L|^2\frac{U}{\sigma} \right) J_1(p) = J_1^D(p) + 2 R(p)  \frac{U}{\sigma}  J_2(p)  \nonumber \\
   & & \ \  \ \ \ +  2  I(p) \frac{U}{\sigma} J_3(p) \nonumber \\
 && \left( 1+ R(p) \frac{U}{\sigma} \right)J_2(p) + I(p) \frac{U}{\sigma} J_3(p) = J_2^D(p)   \nonumber \\
  && \ \ \nonumber  \\
 &&  I(p) \frac{U}{\sigma} J_2(p) + \left(1 -  R(p)  \frac{U}{\sigma} \right)  J_3(p) =  J_3^D(p)
\end{eqnarray}
The  equation for the transverse mode $J_0$ discussed in the previous section is not altered and decouples from the others. We have added the weak-localization contribution
caused by 2 dipoles found in Eq.~(\ref{WL0}), since it is not covered by the diffusion approximation, and assumed it enters just as a number in the equation for
$J_0(p)$. This is a clear oversimplification but has no huge consequences for what follows.
The purely longitudinal diffusion current $J_1$ is known once the others are known, but is not relevant for Poynting vector and can likewise
be ignored.
The modes $J_2$ and $J_3$ however, couple and the solution for $J_2$ is
\begin{equation}\label{scJ2}
    J_2(p)= \frac{J_2^D (p) + (U/\sigma) C_2(p)}{1- U^2 |G_L(p)G_T(p)|^2/\sigma^2}
\end{equation}
We recall from Eq.~(\ref{LT}) that $J_2^D(p)= -2\mathrm{Im}\, G_L \mathrm{Im}\, G_T <0 $ and $J_3^D (p) = -I(p)$. Thus, with $K= k_e + i/2\ell$ the complex pole of $G_T(p)$,
the function $C_2(p)$ is given by
\begin{eqnarray}\label{Fp}
    C_2(p) &=& I(p)^2-J_2^D(p)R(p) \nonumber \\
    &= &  \left( \frac{k_e}{\ell}\right)^2 \frac{|G(p)|^2 + |K|^4|G(p)|^4 }{|K|^8}
\end{eqnarray}
which is strictly positive.

Before calculating diffusion constant  we first discuss these results.
Since the wave number  integral
of $J_2(p)$ contributes to the diffusion constant  via Eq.~(\ref{SC2}), its denominator cannot possess any non-integrable singularity.
This implies that
\begin{equation}\label{limit1}
    \sigma(k) > U |G_L(p) G_T(p)|
\end{equation}
to be valid for all $p$. This inequality excludes \emph{de facto} a mobility edge. It is most stringent near the transverse frequency shell
$p \approx  k_e$ (more precisely $p^2 = \mathrm{Re}\, K^2 = k_e^2 - 1/4\ell^2$, positive as long as $k_e\ell > 1/2$) where
$G_T = 1/ (-i \mathrm{Im}\, \Sigma) $. Furthermore, since we neglect $p$-dependence in self-energies we set $|G_L| =  1/|K|^2 $ and neglect the fact
that near the transverse shell the complex wave numbers of longitudinal and transverse modes are not necessarily equal.
Recalling that $U = (\mathrm{Im}\, \Sigma)^2 \sigma_c$ and setting $q_m = q/\ell$, with $q$ of order unity, the minimal possible electromagnetic conductivity is
given by
\begin{equation}\label{limit2}
    \sigma(k) > c_2(k_e\ell)  \sigma_D(k)
\end{equation}
with $c_2(x) =(3q/\pi x) (x^2 + \frac{1}{4})^{-1}$ for $k_e\ell > 1/2$.
Equivalently,  if the transport mean free path $\ell^*$ is defined as usual via $ \sigma = k_e^2\ell^*/6\pi$ \cite{LB}, then
\begin{equation}\label{limit3}
   k_e \ell^* > \frac{3q }{\pi}  \frac{1}{(k_e \ell)^2 + \frac{1}{4}}
\end{equation}
for $k_e\ell > 1/2$. For $k_e\ell < 1/2$ the maximum occurs at $p=0$ and we find
\begin{equation}\label{limit4}
   k_e \ell^* > \frac{3q }{\pi}  \frac{(k_e\ell)^2}{[(k_e \ell)^2 + \frac{1}{4}]^3}
\end{equation}

The very existence of this minimum conductivity for vector waves is determined by scattering properties of longitudinal and transverse waves near the frequency shell
and not by large wave numbers $p$ that are subject to  uncertain regularization procedures. It nevertheless relies on our choice for $q$, and the
approximation that $K_L (p) = K_T(p) = K$.
The above lower bound becomes stringent for $k_e\ell \approx 1$ where one would have expected a mobility edge. In this regime the maximum occurs at $p < k $, so that setting
$K_L (p) = K_T(p) = K$  may not be a bad approximation, knowing that for $p \ll k$ it is valid (see for instance the $p$-dependent self-energies in Fig.~\ref{boom}).
If $q=1$, we find for $k_e\ell = 1$, $  k_e \ell^*   > 0.76 $,
and upon entering the evanescent regime $k_e\ell = 1/2$, $  k_e \ell^*   > 2.19 $.
For $k_e\ell= 0.35$ the minimum value is $2.26$.

\begin{figure}[t]
 \includegraphics[width=0.95\columnwidth]{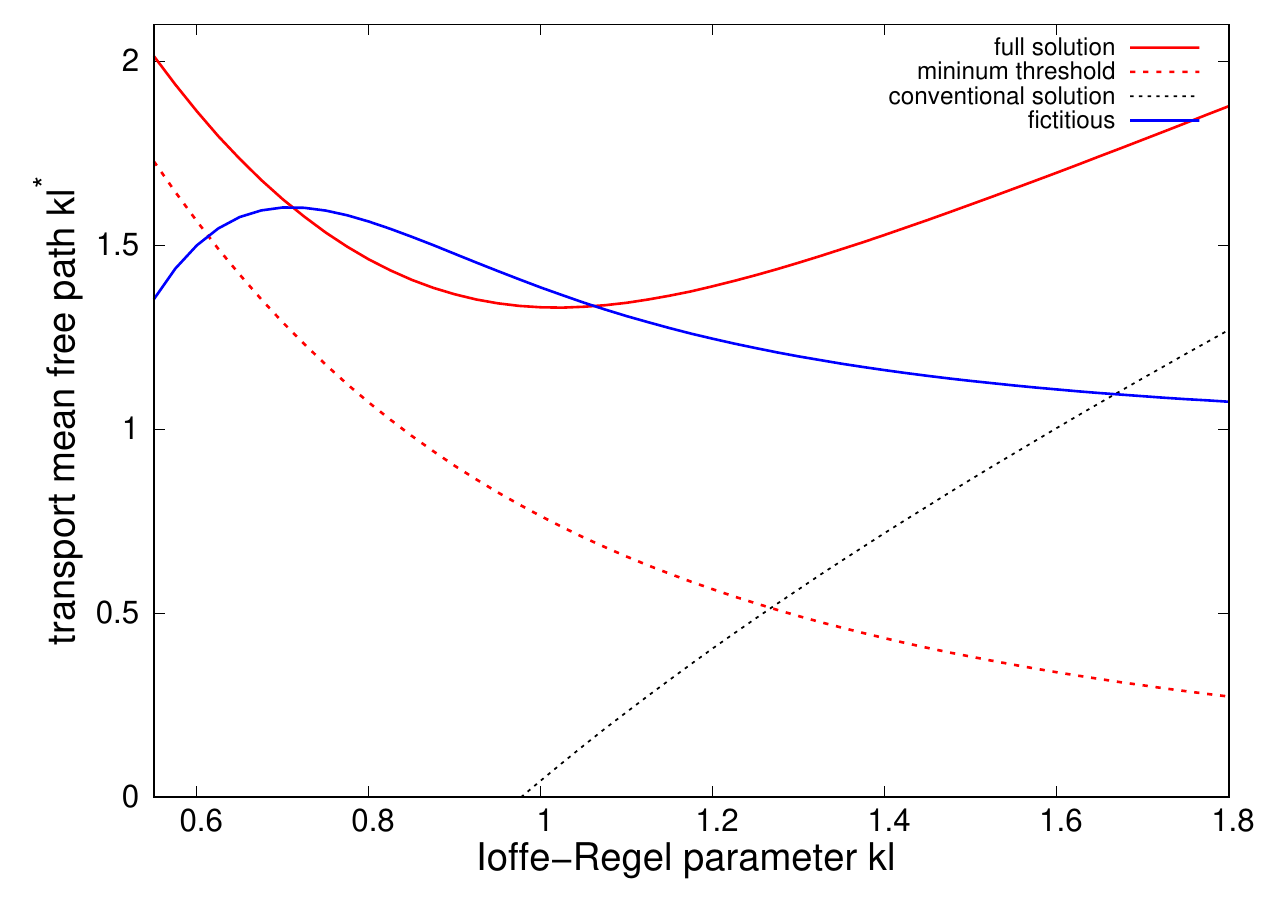}
  \caption{The self-consistent solution for the electromagnetic transport mean free path $\ell^*$ defined by $\sigma = k_e^2 \ell^*/6\pi$.
  Shown are the values for $k_e\ell^*$ for the full solution in this section, the conventional
  picture described by Eq.~(\ref{SC2}) with only the transverse mode $J_0$ considered, with a mobility edge predicted around $k_e\ell \approx 1$,
  the lower threshold imposed  by the existence of the diffusion modes $J_2$ and $J_3$, as well as $k_e\ell^*$
  associated with the fictitious conductivity and $J_3$. We used a cut-off $q_m = 1/\ell$.}\label{MEfig}
\end{figure}


We next calculate the electromagnetic conductivity, which is the sum of the conductivities of the  two channels,
 $\hat{\sigma} \equiv \sigma / \sigma_D= \hat{\sigma}_0 + \hat{\sigma}_2 $.
Since the mobility no longer vanishes, the transverse diffusion mode $J_0$,
which decouples from the others, can be given the same treatment as done in Eq.~(\ref{SC5}), with the denominator removed and taken outside at its maximum value.
This gives the first equation  for the conductivity of the transverse channel,
\begin{eqnarray}\label{scvector1}
  \hat{\sigma}_0 =   \frac{1}{1 + c_1(k_e\ell)  /\hat{\sigma} + {0.774}/k_e\ell}
\end{eqnarray}
with $c_1(x) = 3q/\pi x^2$. We can apply the same procedure for the diffusion current $J_2$. However, as
was seen in Eq.~(\ref{Dlong}) to be the case for the Drude component, the remaining integral suffers from a divergence at large $p$,
again of the kind~(\ref{MEdiv}). The regularization
proposed in Eq.~(\ref{SC6}) is not satisfactory here since it changes sign at $k_e\ell = 0.86$ and would produce a negative Drude conductivity in the $TL$-channel, arguably not physical.
In Sec.~\ref{drudesection} we found that for $p > k_e^2\ell$ the diffusion theory in the $J_2$ channel breaks down so that the present
theory is not valid for too large $p$.
We therefore propose a regularization
\begin{equation*}
     \sum_\mathbf{p}  p^2|G_T(p)|^2 \rightarrow |K|^2 \sum_\mathbf{p}  |G_T(p)|^2 = |K|^2 \frac{\ell}{4\pi}
\end{equation*}
with $K= k_e + i/2\ell$ the transverse complex wave number. In real space is $G_T({r}) = -\exp(iKr)/4\pi r$ and this regularization comes down to
 \begin{equation*}
     \int d^3\mathbf{r}\,    \left| \bm{\nabla} \left(\frac{\exp(iKr) }{-4\pi r} \right)\right|^2 \rightarrow |K|^2 \int d^3\mathbf{r}\,
     \left|\frac{\exp(iKr) }{-4\pi r}\right|^2
\end{equation*}
meaning that the regularization only considers the far field when taking the spatial derivative.
In particular this leads to the  Drude diffusion constant of channel $J_2$,
 \begin{eqnarray*}
   D_2^D &=&   -\frac{1}{3\pi N(k)}\sum_\mathbf{p}  p^2J_2^D(p) = \frac{4\pi }{3} \frac{v_E}{|K|^4 \ell^2 }\sum_\mathbf{p}  p^2 |G_T(p)|^2
   \\   &\rightarrow & \frac{1}{3} v_E \frac{1}{|K|^2 \ell}
\end{eqnarray*}
This is a satisfactory, positive extrapolation of the result $\frac{1}{3}v_E/k^2\ell$  obtained in Eq.~(\ref{SC2}) for low density, and where the divergence was seen to cancel explicitly.
If we adopt this regularization, we find in the $J_2$-channel,
\begin{eqnarray}\label{scvector2}
   \hat{\sigma}_2 &=& \frac{{F(\delta)} c_3(k_e\ell) }{1-(c_2(k_e\ell) /\hat{\sigma})^2} \left( 1 - \frac{c_4(k_e\ell)}{\hat{\sigma}}\right)
\end{eqnarray}
with $c_2(x)$ defined earlier, $c_3(x)= ( x^2 + 1/4 )^{-1}$
and $c_4(x) =  (3q/2\pi)(9/8 + x^2/2)( x^2 + 1/4 )^{-2} $. We recall that $F(\delta)$ is the function that describes the \emph{explicit} dependence on detuning of the diffusion constant
in the channel $J_2$,
shown in Fig.~\ref{DTLfig}.

Equations (\ref{scvector1}) and (\ref{scvector2}) lead to a cubic equation for $\hat{\sigma}$ that can be solved analytically. The resulting formula is quite lengthy and we do not present it here. The solution for $k\ell^* = \hat{\sigma} \times k\ell$
is shown in Fig.~\ref{MEfig}. We have put  {$F(\delta) = 1$}, its role will be discussed  later, in which case the self-consistent theory has only one parameter, the product $k_e\ell$, as in the scalar case. According to Eqs.~(\ref{scvector1}) and~(\ref{scvector2}) the traditional weak localization correction $\delta \sigma_0= - c_1$
in the transverse channel is partially compensated by the positive conductivity $\delta\sigma_2 = c_3$ of the $J_2$ channel, and even exactly when $q \approx 1$.
This explains why $k_e\ell^*$ is well in excess of the traditional prediction~(\ref{SC5}), for values as small as  $k_e\ell = 1.8$, and close to the Drude value $k_e\ell$ of the transverse channel.
The term containing $c_4 >0$ tends to suppress diffusion in the $J_2$ mode as $1/(k_e\ell)^4$ but the   coupling to $J_3$ described by $c_2$ reverses this trend.
Around the region $k_e\ell \approx 1$ where the conventional picture would locate the mobility edge,
the minimum conductivity starts to impose itself, and the total conductivity rises.

We recall that the  fictitious diffusion is determined by $J_3$, as described by Eq.~(\ref{DIM}). The self-consistent solution is given by
\begin{eqnarray}\label{J3SC}
    J_3(p) &=& \frac{ -\mathrm{Im}\, G_L(p) G_T(p) }{1 - U^2|G_L(p)G_T(p|^2/\sigma^2} \nonumber \\
    &\times& \left[ 1 + \frac{U}{\sigma} \mathrm{Re}\, G_L(p)G_T(p) \right]
\end{eqnarray}
and upon inserting this into  Eq.~(\ref{DIM}), the same procedure as above provides an expression for the ``fictitious'' conductivity
\begin{equation}\label{DIMSC}
    \hat{\sigma}_I = \frac{1}{k_e\ell}  + \frac{1}{1-(c_2(k_e\ell)/\hat{\sigma})^2}\left( \frac{d_3(k_e\ell) }{\hat{\sigma}}  -\frac{ d_2(k_e\ell)}{\hat{\sigma}^2} \right)
\end{equation}
with the functions $d_2(x) = \frac{1}{2}(3q/\pi)^2 x^{-1} ( x^2 + 1/4)^{-3} $ and $d_3(x) = (3q/4\pi)x^{-1}(x^2 + 1/4)^{-2}$.
The transport mean free path associated with the fictitious diffusion  is also shown in Fig.~\ref{MEfig}. For $k_e\ell \sim 1$ fictitious diffusion
 is of same order as the real conductivity and has the same sign.

\begin{figure}
\includegraphics[width=0.7\columnwidth, angle=-90]{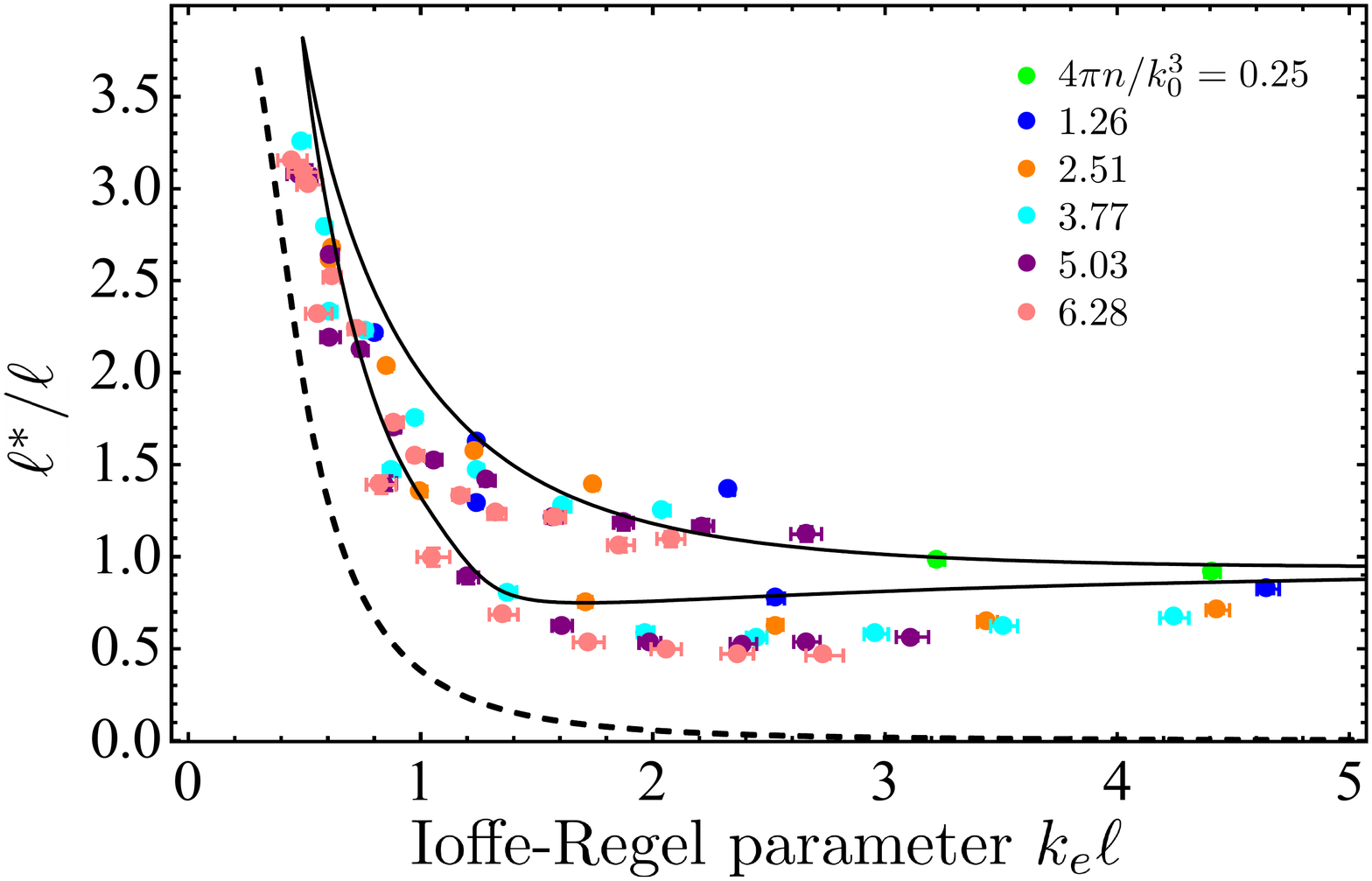}
\vspace*{-7mm}
\caption{
The ratio of transport and scattering mean free paths $\ell^*/\ell$ as a function of $k_e \ell$ compared to the self-consistent theory for $4\pi n/k_0^3 = 3.77$ and $q = 0.5$ (black solid line, for the two branches, see text for explanation). Points of different colors correspond to different scatterer number densities $n$ for detunings $\delta = (\omega-\omega_0)/\gamma \in [-3, 6.5]$ from the resonance. The dashed line is the lower bound for $\ell^*/\ell$ described by Eqs.~(\ref{limit3}) and (\ref{limit4}), again with $q=0.5$.
}\label{fig_comparison}
\end{figure}


For $q = 0$, Eqs.\ (\ref{scvector1}) and (\ref{scvector2}) simplify to the sum of the diffusion constants associated with one or two dipoles in the channels $J_0$ and $J_2$, without any cross-talk,
\begin{eqnarray}
\label{q0}
\hat{\sigma} = \frac{\ell^*}{\ell} = \frac{1}{1 + 0.774/k_e \ell} + \frac{{F(\delta)}}{(k_e \ell)^2 + \frac14}
\end{eqnarray}
For $k_e\ell < 1$ the second term from the $J_2$ channel starts dominating. If we ignore the explicit dependence on $\delta$ by putting $F(\delta) = 1$,
 this equation yields $\hat{\sigma} < 1$ for $k_e \ell > 1.73$  and below this value starts increasing monotonically. In the same limit of $q = 0$, we have $\hat{\sigma}_I = 1/k_e\ell$.

\subsection{Comparison with numerical simulations}

In Fig.\ \ref{fig_comparison} we compare the predictions of the self-consistent theory for electromagnetic waves developed above to numerical simulations in which we simulate the multiple scattering of light by an ensemble of dipolar resonant point scatterers. The results of the simulations allow us to estimate $k_e$, $\ell$ and $\ell^*$. Both the details of the simulations and the way in which we interpret their results are detailed in Appendix \ref{appB}. We repeat calculations for several atomic number densities $n$ and detunings $\delta = (\omega - \omega_0)/\gamma$; the resulting ratios $\ell^*/\ell$ are presented in Fig.\ \ref{fig_comparison} by circles of different colors as functions of the Ioffe-Regel parameter $k_e \ell$. The numerical results are bounded from below by Eqs.\ (\ref{limit3}) and (\ref{limit4}) for the minimum conductivity (dashed line). Equations  (\ref{limit3}) and (\ref{limit4}) impose a sharp rise of the ratio $\ell^*/\ell$ at small values of $k_e\ell$ where one would normally have expected a mobility edge. This rise is well reproduced by the numerical results.

A striking feature of the numerical results is the clear tendency of data to group together along two different ``branches''.
A careful inspection of Fig.\ \ref{fig_comparison} shows that the lower branch is composed of data corresponding to $\delta < 0$ whereas the upper branch corresponds to simulations with positive detunings $\delta > 0$. This means that -
apart from the absence of a localization transition - there is no one-parameter dependence with $k_e\ell$ either.
The double-branch structure actually follows from the explicit dependence of the $J_2$ channel on detuning $\delta$,
described by the factor $F(\delta)$, which is larger for positive detunings (see Fig.\ \ref{DTLfig}).
Figure \ref{fig_comparison} shows the prediction of the self-consistent theory for the ratio $\ell^*/\ell$ with the inclusion of the function $F(\delta)$
and with the dimensionless parameter $k_e\ell$ calculated from the averaged incident field  (see Appendix B)
for one fixed dipole density $4\pi n/k_0^3 = 3.77$ and for various detunings. Predictions for $k_e\ell$ corresponding to other densities are not shown since they all
exhibit the same overall appearance. Despite the fact that, strictly speaking,  $\ell^*/\ell$ is a function of two independent parameters ($k_e \ell$ and $\delta$ or
equivalently $k_e \ell$ and $4\pi n/k_0^3$), we see that all results for the quite wide explored density range  $4\pi n/k_0^3 = 0.25$--6.28 roughly
follow the same double-branch master curve that is close to the analytical result for the intermediate density $4\pi n/k_0^3 = 3.77$.
The agreement between numerical and analytical results is not perfect but we believe that it can be further improved by distinguishing explicitly between transverse and longitudinal complex wave numbers (see Sec.\ \ref{sectionDOS}), which are known to be different (see Appendix \ref{appA}, and Figs. \ref{fig_keff} and \ref{fig_tr}). This  can be done in future work.

\section{Conclusions and Outlook}

In this work we have included longitudinal excitations into a transport theory for electromagnetic waves propagating in a medium with randomly distributed dipolar electric
scatterers (dipoles).
We identify four diffuse modes, triggered by the gradient in electromagnetic energy, among which two carry a Poynting vector and contribute to the diffusion constant.
We have developed this theory by extending the independent scattering approximation (the elementary scattering unit is a single dipole) to include rigourously
recurrent scattering from two dipoles. This has led to the following results. 1) Longitudinal and transverse waves of the effective medium are characterized by different
complex wave numbers $K_{L}$ and $K_T$, respectively, and dominate near and far field in scattering. 2) The interference between longitudinal and transverse waves
creates a new diffuse
transport channel with a diffusion constant proportional to the number density of dipoles, to be compared to the usual diffusion constant that is
inversely proportional to this density. 3) Divergent terms appear at large wave numbers in the diffusion constant, in the longitudinal density of states and in the
 collision operator.
Many of them cancel, in particular for the electromagnetic Kubo diffusion constant all divergent terms cancel. We postulate that this cancelation holds
in all orders of perturbation theory.  4) When extending the self-consistent theory of localization, with all its usual assumptions, to include the four diffuse modes,
we find a minimum conductivity that prevents the onset of Anderson
localization of light, as also observed in numerical simulations \cite{sergey0}.
5) The predictions of the developed self-consistent theory are surprisingly close to the results of independent numerical simulations, including the explicit dependence of the new transport channel on frequency detuning from the dipolar resonance.
These findings demonstrate that, due to the presence of longitudinal, non-propagating  waves,  (weak) localization of light is fundamentally different from what was believed so far.

Early stages of this  work were supported by collaborations with Yvan Castin, Ad Lagendijk, Nicolas Cherroret and Dominique Delande. We thank Denis Basko
for useful discussions.


\appendix

\section{Longitudinal and transverse DOS of M electric dipoles }\label{appA}

In this Appendix we pose the problem of light scattering from $M$ point electric dipolar scatterers (``dipoles'' for short) in a volume $V$ and derive the DOS for both longitudinal and transverse excitations in the thermodynamic limit
$M,V\rightarrow \infty$ at constant density $M/V = n$.

The real-space Green's function $\mathbf{G}(\mathbf{r},\mathbf{r}')$ of $M$ point-like dipoles at positions $\{ \mathbf{r}_m \}$, $m = 1, \ldots, M$, is defined in terms of their collective $T$-matrix as
\begin{equation}\label{TmatrixM}
    \mathbf{G}(\mathbf{r},\mathbf{r}') = \mathbf{G}_0(\mathbf{r}-\mathbf{r}') + \sum_{mm'} \mathbf{G}_0(\mathbf{r}-\mathbf{r}_m) \cdot \mathbf{T}_{mm'}\cdot \mathbf{G}_0(\mathbf{r}_{m'}-\mathbf{r}')
\end{equation}
If we impose that all dipoles be impenetrable, we must have $ \mathbf{G}(\mathbf{r}_n,\mathbf{r}') = 0$ for all $\mathbf{r}'$  outside the dipoles. Thus,
 \begin{equation*}
    0 = \sum_{m'} \left[ \delta_{nm'} + \sum_{m} \mathbf{G}_0(\mathbf{r}_n -\mathbf{r}_m) \cdot \mathbf{T}_{mm'}\right] \cdot \mathbf{G}_0(\mathbf{r}_{m'}- \mathbf{r}')
 \end{equation*}
For this to be true for all $\mathbf{r}'$, the $3M \times 3M$ matrix between square brackets must vanish,
 \begin{equation}\label{TGmin1}
\left\{\mathbf{ T}_{mm'} \right\} =  -  \left( \left\{ \mathbf{G}_{0}(\mathbf{r}_m-\mathbf{r}_{m'}) \right\}\right)^{-1}
 \end{equation}
We can split off the singular diagonal elements and use the fact that the $t$-matrix of one single dipole is $\mathbf{t} = -\mathbf{G}_0^{-1}(0)$ and here proportional to the $3 \times 3$ identity matrix,
 \begin{equation}\label{TGmin2}
\left\{\mathbf{ T}_{mm'} \right\} =   t\left( \left\{ \mathbf{1}\delta_{mm'} - t  \mathbf{G}_{0}(\mathbf{r}_m-\mathbf{r}_{m'} \neq 0) \right\}\right)^{-1}
 \end{equation}
This matrix is regular as long as the  dipoles do not overlap. For any point source $\mathbf{s}(\mathbf{r}')$ located at $\mathbf{r}'$ the electric field anywhere in the medium is given by $\mathbf{E}(\mathbf{r}) = \mathbf{G}(\mathbf{r},\mathbf{r}')\cdot
\mathbf{s}(\mathbf{r}')$, and the incident field is $\mathbf{E}_0(\mathbf{r}) = \mathbf{G}(\mathbf{r},\mathbf{r}')\cdot
\mathbf{s}(\mathbf{r}')$.  If the source  is  located in the far field of the $M$ dipoles, and the origin $\mathbf{r }= 0$ is chosen inside the scattering medium, we have $|\mathbf{r}_m| \ll |\mathbf{r}'|$, and we can approximate the incident field inside the medium as
$\mathbf{E}_0(\mathbf{r}) = [-\exp(ikr)/4\pi r] \mathbf{ \Delta}_{\mathbf{r}'} \cdot \mathbf{s}(\mathbf{r}') \exp(-ik\hat{\mathbf{r}'}\cdot \mathbf{r} ) $ and  equal to a transverse plane wave with wave vector $\mathbf{k}=-k\hat{\mathbf{r}'}$.
It follows that
\begin{equation}\label{Efield1}
    \mathbf{E}(\mathbf{r})=  \mathbf{E}_0(\mathbf{k}, \mathbf{r}) +  \sum_{mm'} \mathbf{G}_0(\mathbf{r}-\mathbf{r}_m) \cdot \mathbf{T}_{mm'}\cdot  \mathbf{E}_0(\mathbf{k}, \mathbf{r}_{m'})
\end{equation}
The fields $\{\mathbf{E}(\mathbf{r}_n)\}$ vanish because the $T$-matrix has earlier been designed to do so. For $\mathbf{r}=\mathbf{r}_n$ we can extract the singular term  $\mathbf{G}_0(\mathbf{r}_n,\mathbf{r}_n) = -1/t$ and
define the ``macroscopic field'' $\mathbf{\tilde{E}}(\mathbf{r}_n)$ in the vicinity of the dipole $n$ as the one scattered from all others,
\begin{eqnarray}\label{Efield2}
    \mathbf{\tilde{E}}(\mathbf{r}_n)&=&  \mathbf{E}_0(\mathbf{k}, \mathbf{r}_n)
     \nonumber \\
    &+&  \sum_{m\neq n;  m'} \mathbf{G}_0(\mathbf{r}_n,\mathbf{r}_m) \cdot \mathbf{T}_{mm'}\cdot  \mathbf{E}_0(\mathbf{k}, \mathbf{r}_{m'})
    \nonumber \\
    &=&  \mathbf{E}_0(\mathbf{k}, \mathbf{r}_n) +  \sum_{m\neq n} t\mathbf{G}_0(\mathbf{r}_n,\mathbf{r}_m) \cdot  \mathbf{\tilde{E}}(\mathbf{r}_m) \nonumber \\
    &=& \frac{1}{t}\sum_{m} \mathbf{ T}_{nm}\cdot \mathbf{E}_0(\mathbf{r}_{m})
\end{eqnarray}
The omission of the diagonal term $m=n$ gives a good impression of the electric field inside the medium and the solution of Eq.~(\ref{Efield2})
is equivalent to the calculation of the matrix
$\mathbf{T}_{mm'}$ as is apparent from the last identity. However, it  misses completely the singular field scattered by the dipole at $\mathbf{r}_n$.
At a small distance  $\mathbf{x}$ from dipole $n$  the  relation between the fields $\mathbf{E}(\mathbf{r}_n)$ and  $\mathbf{\tilde{E}}(\mathbf{r}_n)$ is
\begin{eqnarray}\label{Efield3}
    \mathbf{{E}}(\mathbf{r}_n+ \mathbf{x})  = \mathbf{\tilde{E}}(\mathbf{r}_n +\mathbf{x}) + t \mathbf{G}_0(\mathbf{x})\cdot \mathbf{\tilde{E}}(\mathbf{r}_n)
\end{eqnarray}
Especially the longitudinal part is strongly singular as $\mathbf{x} \rightarrow 0$ but carries no energy flux. The transverse part also diverges as $1/x$ but
carries a finite energy flux and poses less a problem.

To illustrate this consider first one electric dipole located at $\mathbf{r}=0$, and for which only one diagonal term exists. Equation~(\ref{TmatrixM}) reduces to
$\mathbf{G}(k, \mathbf{r},\mathbf{r}) = \mathbf{G}_0(k, 0) + t \mathbf{G}_0(k,\mathbf{r})^2$.
According to the analysis that has led to Eq.~(\ref{DOS3}), the local density of states (per unit volume, here per interval $dk= d\omega/c_0$) at position $\mathbf{r}$ is given by
 $N(k, \mathbf{r})= -(k/\pi) \mathrm{Im \, Tr} \, \mathbf{G}(k,\mathbf{r},\mathbf{r})$.  Formally, $N(k,0) =0$, but for $\mathbf{r}\neq 0$ we can identify longitudinal states close to the dipole, and transverse states far away.
The total extra number of states due to the presence of the dipole is
\begin{eqnarray*}
    d\Pi (k) &=& dk \int d^3\mathbf{r}\, \left[ N(k,\mathbf{r}) - N_0(k, \mathbf{r}) \right] \\
&=&   -\frac{k dk}{\pi} \mathrm{Im } \, t(k) \, \mathrm{Tr} \sum_{\mathbf{p}} \mathbf{G}^2_0(k,\mathbf{p}) \\
&=&  -\frac{k dk }{\pi } \mathrm{Im } \, t(k) \sum_{\mathbf{p}} \left[ \frac{1}{k^4} +  \frac{2}{ [(k+i0^+)^2 -p^2]^2} \right]
\end{eqnarray*}
This clearly separates into  a strongly diverging longitudinal and a regular transverse component.
Regularizing the first to $Q_0k_0^3/2\pi$, with $Q_0 =
k_0c_0/\gamma $ the quality factor,  as
proposed in Sec.~\ref{section1D}, we find that for $k$ within a line width of $k_0$
\begin{equation}\label{DOS1D}
        d\Pi (k)  =  \frac{dk }{2\pi^2 } \left[ - Q_0\mathrm{Im }\, t(k) + \frac{1}{2} \mathrm{Re}\, t(k) \right]
\end{equation}
The first term is missed by ignoring divergencies, and thus difficult to
capture by a numerical simulation. It largely dominates near resonance and is Lorentzian as is the cross-section. The second term describes the modification of transverse energy density and can be interpreted
as the change in local refractive index due to the presence of the dipole.
Upon integrating over the entire resonance, using that
\begin{equation*}
    \int_{-\infty}^{\infty} dk \, t(k) = -\frac{6\pi^2 i }{Q_0}
\end{equation*}
we find that only the first term survives and giving a total number of extra states per dipole is $\int d\Pi(k) = 3$, equal to the number of degrees of freedom associated with the optical
polarization.

The analysis above can be straightforwardly generalized to $M$ dipoles. This yields expressions such as Eq.~(\ref{allloop}) for the
longitudinal complex wave number $K_L(\infty)$ and associated with the DOS of longitudinal states, and a similar one for transverse waves.
From  the ensemble-averaged Dyson Green's function $\mathbf{G}(\mathbf{r}, \mathbf{r})$ in the unbounded medium, given in Eq.~(\ref{GL}),
upon splitting off terms, singular as $\mathbf{x} \rightarrow 0$, we obtain
\begin{eqnarray}\label{dysonApp}
    \mathrm{Tr}\, \langle \mathbf{G}(0) \rangle &=& \frac{\delta(0)}{K_L^2(\infty)}
     + \mathrm{Tr}\, \mathbf{D}(0)
    \nonumber \\ &+& \sum_\mathbf{p} \frac{ 2 }{p^2} \frac{K^2_T(p)}{K_T^2(p) -p^2} -  \sum_\mathbf{p} \frac{2}{p^2}
\end{eqnarray}
The Lorentz contact term of the effective medium emerges as $\mathrm{Tr} \, \mathbf{G}_L(0) = \delta(0)/K_L^2(\infty)$ that can be regularized as before.
Since $\mathbf{D}(0)$ is a finite longitudinal contribution, it will be neglected.
The transverse divergence described by the last term is real-valued and plays no role for DOS and is also independent of dipole density.
The integral over transverse wave numbers
can be defined as $-iK_T/2\pi$, which would be the value if $K_T(p)$ were independent on $p$.
Before ensemble averaging, the Green's function satisfies Eq.~(\ref{TmatrixM}).
Let us first focus on the trace of the diagonal terms that average to
\begin{eqnarray*}
&&\left\langle \mathrm{Tr}\,  \sum_{m=1}^M \mathbf{G}_0(\mathbf{r}-\mathbf{r}_m)\cdot \mathbf{T}_{mm}\cdot \mathbf{G}_0(\mathbf{r}_{m}-\mathbf{r} )
\right \rangle = \\
&&\mathrm{Tr}\, \sum_{m=1}^M \frac{1}{V} \int d^3\mathbf{r} \mathbf{G}_0(\mathbf{r}-\mathbf{r}_m)\cdot \left\langle \mathbf{T}_{mm}\right \rangle
\cdot \mathbf{G}_0(\mathbf{r}_{m}-\mathbf{r} )
\end{eqnarray*}
In the thermodynamic limit, the average $ \left\langle \mathbf{T}_{mm}\right \rangle $ should be  independent of the dipole $m$ and its position
 $\mathbf{r}_m$. Hence the integral over $\mathbf{r}$ can be converted to Fourier space to become

\begin{eqnarray*}
n \mathrm{Tr}\sum_\mathbf{p} \mathbf{G}_0(\mathbf{p})^2  \cdot \left\langle \mathbf{T}_{mm}\right \rangle = n
\left[ \frac{\delta(0)}{3k^4} + \frac{i}{12\pi k } \right] \mathrm{Tr}\, \left\langle \mathbf{T}_{mm}\right \rangle
\end{eqnarray*}
The contact term that occurs in this expression must be identified
with the one in Eq.~(\ref{dysonApp}), so that
\begin{equation}\label{bewijsL}
    \frac{1}{K_L^2(\infty)} = \frac{1}{k^2} +\frac{n }{3k^4} \mathrm{Tr}\, \left\langle \mathbf{T}_{mm}(k)\right \rangle
\end{equation}
valid as $M,V \rightarrow \infty$, at constant density $n= M/V$.
Since we expect  $\left\langle \mathbf{T}_{mm}(k)\right \rangle \propto \mathbf{1}$ the trace compensates the factor $3$ in the denominator.

The off-diagonal elements
$m\neq m'$ in Eq.~(\ref{TmatrixM}) are negligible for the longitudinal states [and identified as $\mathbf{D}(0)$], but not for the transverse waves,
\begin{eqnarray*}
    \int d^3\mathbf{r} \, \mathbf{G}_{0,T}(\mathbf{r}_{m'}-\mathbf{r })&\cdot& \mathbf{G}_{0,T}(\mathbf{r}-\mathbf{r}_m) \\
    = \sum_\mathbf{p } \mathbf{G}_{0,T}(\mathbf{p})^2  && \exp[i\mathbf{p}\cdot (\mathbf{r}_m - \mathbf{r}_{m'})] \equiv \frac{1}{4\pi k } \mathbf{H}_T(k\mathbf{r})
\end{eqnarray*}
with
\begin{eqnarray*}
     \mathbf{H}_T(\mathbf{y}) = \frac{1}{2}
     \left\{ i{e^{iy}}\mathbf{\Delta}_\mathbf{y} - \frac{d}{dy}\left(\frac{e^{iy}}{iy} + \frac{e^{iy}-1}{y^2} \right) (1-3\mathbf{\hat{y}}\mathbf{\hat{y}})\right\}
     \end{eqnarray*}
which is  regular ($\mathbf{H}_T(0) = i/3$). The off-diagonal elements become
\begin{equation*}
    \frac{n}{4\pi k} \mathrm{Tr } \sum_{m'\neq m} \left\langle \mathbf{T}_{mm'} \cdot \mathbf{H}_T(k\mathbf{r}_{mm'}) \right\rangle
\end{equation*}
Again, we  suppose that the sum over $m'$ not to depend on $m$ in the thermodynamic limit. Comparing to Eq.~(\ref{dysonApp})
this gives the following expression for the complex transverse wave number
\begin{eqnarray}\label{bewijsT}
    K_T = k &-& \frac{n}{6k}   \mathrm{Tr}\, \left\langle \mathbf{T}_{mm}\right \rangle \nonumber \\ &+&
    i  \frac{n}{2 k} \mathrm{Tr }\, \sum_{m'\neq m} \left\langle \mathbf{T}_{mm'} \cdot \mathbf{H}_T(k\mathbf{r}_{mm'}) \right \rangle
\end{eqnarray}
If we neglect any recurrent scattering (r.s.) from two or more dipoles, we have $\mathbf{T}_{mm} = t \mathbf{1}$ and $\mathbf{T}_{mm'} = t^2 \mathbf{G}_0(\mathbf{r}_{mm'})$.
Converting the sum over $m'$ into the  integral
$n\int d^3\mathbf{r }\, \mathrm{ Tr}\, \mathbf{G}_0(\mathbf{r})\cdot \mathbf{H}(\mathbf{r}) = in/4k^2$ gives
\begin{eqnarray}\label{Klown}
    &&\frac{1}{K_L^2(\infty)} = \frac{1}{ k^2} + \frac{nt}{k^4} + \mathrm{r.s.} \nonumber \\
     && K_T^2 =\left( k- \frac{nt}{2k} - \frac{n^2t^2}{8k^3 }+ \mathrm{r.s.}\right)^2 = k^2 - nt +\mathrm{r.s.}
     \nonumber
\end{eqnarray}
and we recover the ISA approximation. In particular, the off-diagonal elements in Eq.~(\ref{bewijsT}) are not negligible.

We emphasize that the complex wave numbers $K_T$ and $K_L$ relate to transverse and longitudinal DOS and should not be interpreted as effective medium parameters of
electromagnetic excitations.

\section{Numerical simulation of scattering and transport mean free paths}
\label{appB}

\begin{figure}[t]
\includegraphics[width=0.7\columnwidth, angle=-90]{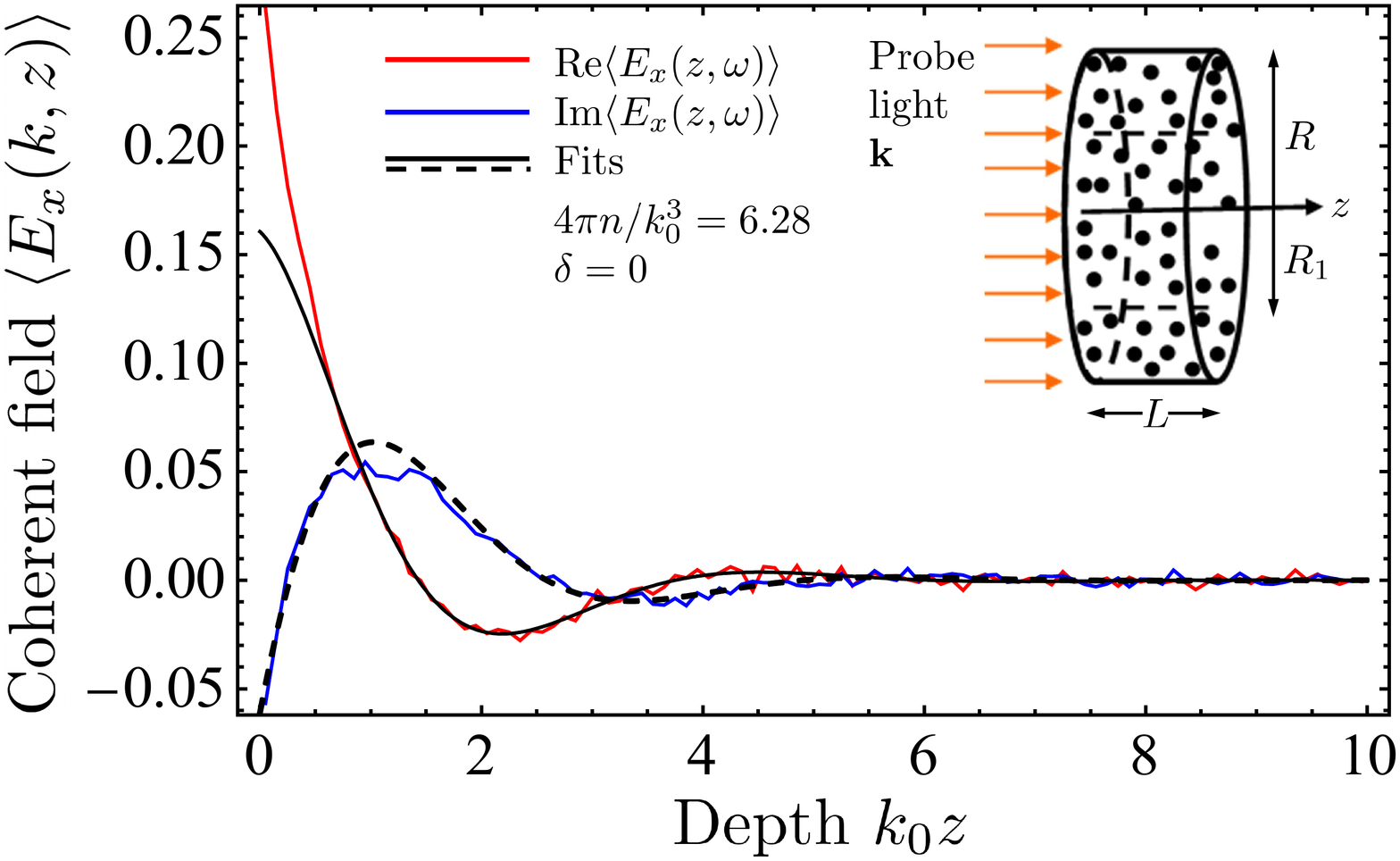}
\vspace*{-10mm}
\caption{\label{fig_field}
Average, on-resonance, electric field in the most dense sample. Red and blue lines show the real and imaginary parts of the field, respectively. The black solid line is the fit of Eq.\ (\ref{eqnmfp}) to the numerical data; the dashed line is the result for the imaginary part obtained from the fit to the real part.
}
\end{figure}

We consider a sample having the shape of a cylinder of radius $R$ and thickness $L$ parallel to the $z$ axis of the reference frame and confined between the planes
$z = 0$ and $z = L$ (see the inset of Fig.\ \ref{fig_field}). The sample is made of $M$ point-like resonant scatterers described by Eq.~(\ref{tEDreso}) with a resonant frequency
$\omega_0= k_0c_0$ and a decay rate $\gamma \ll \omega_0$ of the excited state. The scatterers are located at random positions $\mathbf{r}_j$, $j = 1, \ldots, M$, inside the sample.
The scatterer number density is $n = M/V$ with $V = \pi R^2 L$ being the volume of the sample.
In the following, we set $k_0 L = 10$ and $k_0 R = 30$, which implies $M = 2827$--14137 for $n/k_0^3 = 0.1$--0.5.
We have also performed calculations for a relatively low density $n/k_0^3 = 0.02$ at which we set $k_0 L = 30$, $k_0 R = 60$ and $M = 6786$.

It is convenient to introduce dimensionless quantities and to neglect the frequency dependence of the Green's tensor over the bandwidth of interest that is assumed to be much less than $\omega_0$ though can exceed $\gamma$ considerably. Thus we put $\mathbf{G}_0(k) = \mathbf{G}_0(k_0)$.
Given an incident wave $\mathbf{E}_0(k, \mathbf{r})$ and using Eqs.~(\ref{Efield1}) and (\ref{Efield2}), we obtain equations for the  electric field ${\mathbf{E}}(k, {\mathbf{r}})$ at any point in space:
\begin{eqnarray}
\mathbf{E} ({k}, \mathbf{r}) = \mathbf{E}_0(k, \mathbf{r}) + {t}(k) \sum\limits_{j=1}^M \mathbf{G}_0(k_0, \mathbf{r}-\mathbf{r}_j)
\cdot  \mathbf{\tilde{E}}(k, \mathbf{r}_j) \nonumber \\
\label{eqne}
\end{eqnarray}
with
\begin{eqnarray}
\mathbf{G}_{0}(k_0, \mathbf{r}) = -
\frac{e^{i k_0 r}}{4 \pi r}
\left[ P(k_0 r) \mathbf{\Delta_r} + \ Q(k_0r) \mathbf{\hat{r}}\mathbf{\hat{r}}  \right]
\label{greene}
\end{eqnarray}
the free-space Green's tensor for $\mathbf{r} \ne 0$. Here $P(x) = 1 - 1/i x - 1/x^2$ and $Q(x) = 2/i x + 2/x^2$.

The magnetic field $\mathbf{B}(k, \mathbf{r})$ can be found by applying $\mathbf{B}(k, \mathbf{r}) = \bm{\nabla} \times \mathbf{E}(k, \mathbf{r})/(i k_0)$ to Eq.\ (\ref{eqne}):
\begin{eqnarray}
\mathbf{B}({k},  \mathbf{r} ) = \mathbf{B}_0({k}, \mathbf{r}) - i{t}(k) \sum\limits_{j=1}^M \mathbf{G}_0^{(B)}(k_0, \mathbf{r} -\mathbf{r}_j) \cdot \mathbf{\tilde{E}}
(k, \mathbf{r}_j) \nonumber \\
\label{eqnb}
\end{eqnarray}
where
\begin{eqnarray}\label{eqgb}
\mathbf{G}_0^{(B)}(k_0, \mathbf{r}) = - \frac{k_0}{4\pi }(\epsilon\cdot \mathbf{\hat{r}})
{e^{i k_0 r}}   \left[ 1 - P(k_0 r) \right]
\end{eqnarray}
with the transverse, antisymmetric matrix
\begin{eqnarray}
(\epsilon\cdot \mathbf{\hat{r}}) =
\begin{bmatrix}
0 & z/r & -y/r\\
-z/r & 0 & x/r\\
y/r  & -x/r & 0
\end{bmatrix} \nonumber \\
\end{eqnarray}
and $\mathbf{r} = x  \mathbf{\hat{x}} + y \hat{\mathbf{y}} + z \hat{\mathbf{z}}$. Note that $\mathbf{G}_0^{(B)}(k_0, \mathbf{r})$ only diverges as $1/r^2$ for small $r$, whereas
$\mathbf{G}_0(k_0, \mathbf{r})$ diverges as $\delta(\mathbf{r}) + 1/r^3$. {In addition, the angular integral of $\mathbf{G}_0^{(B)}(k_0, \mathbf{r})$ vanishes.} For these reasons, $\mathbf{B}(k, \mathbf{r})$ suffers from less fluctuations and is better suitable for numerical studies.

Equations (\ref{eqne}) and (\ref{eqnb}) can be written for the fields $\mathbf{\tilde{E}}(k, \mathbf{r}_m)$ and $\mathbf{\tilde{B}}(k, \mathbf{r}_m)$ at each scatterer, excluding the fields scattered by themselves:
\begin{eqnarray}
\mathbf{\tilde{E}}(k, \mathbf{r}_m) &=& \mathbf{E}_0(k, \mathbf{r}_m)
\nonumber \\
&+& {t}(k) \sum\limits_{j \ne m}^M \mathbf{G}_0(k_0, \mathbf{r}_m-\mathbf{r}_j)
\cdot  \mathbf{\tilde{E}}(k, \mathbf{r}_j)
\label{eqne2}
\\
\mathbf{\tilde{B}}(k, \mathbf{r}_m) &=& \mathbf{B}_0(k, \mathbf{r}_m)
\nonumber \\
&-& i t(k) \sum\limits_{j \ne m}^M \mathbf{G}_0^{(B)}(k_0, \mathbf{r}_m -\mathbf{r}_j) \cdot \mathbf{\tilde{E}}
(k, \mathbf{r}_j)\;\;\;\;\;\;
\label{eqnb2}
\end{eqnarray}

We solve the system of equations (\ref{eqne2}) for $\mathbf{\tilde{E}}(k, \mathbf{r}_m)$ ($m = 1, \ldots M$) assuming that the sample is
illuminated by an incident linearly polarized plane wave: $\mathbf{E}_0(k, \mathbf{r}) = \hat{\mathbf{x }}\exp(i k_0 z)$.
Magnetic fields on the scatterers and electric and magnetic fields everywhere in space can be then found from Eq.\ (\ref{eqnb2}) and Eqs.\ (\ref{eqne}), (\ref{eqnb}), respectively.
 {Despite the singular behavior of the Green's function at small} $\mathbf{r}$  {our simulation did not suffer from inaccuracies when two dipoles happened to
come close. Such inaccuracy is probably reduced by recurrent
scattering between two dipoles.}

\subsection{Scattering mean free path and effective wave number of coherent wave}
\label{secmfp}

To determine the scattering mean free path $\ell$, the solution of Eqs.\ (\ref{eqne2}) is averaged over many (up to $10^3$) independent configurations of scatterers
inside the sample, over slices of width $\Delta z = L/100$ along the $z$ axis, and over the central part of the cylinder with radius $R_1 = R - L$ (see the inset of Fig.\ \ref{fig_field}). The average field $\langle \mathbf{E}(k, z) \rangle$ obtained in this way should mimic the average field in a slab of infinite lateral extent ($R \to \infty$).
A typical result obtained from these calculations is illustrated in Fig.\ \ref{fig_field}.

\begin{figure}[t]
\includegraphics[width=0.7\columnwidth, angle=-90]{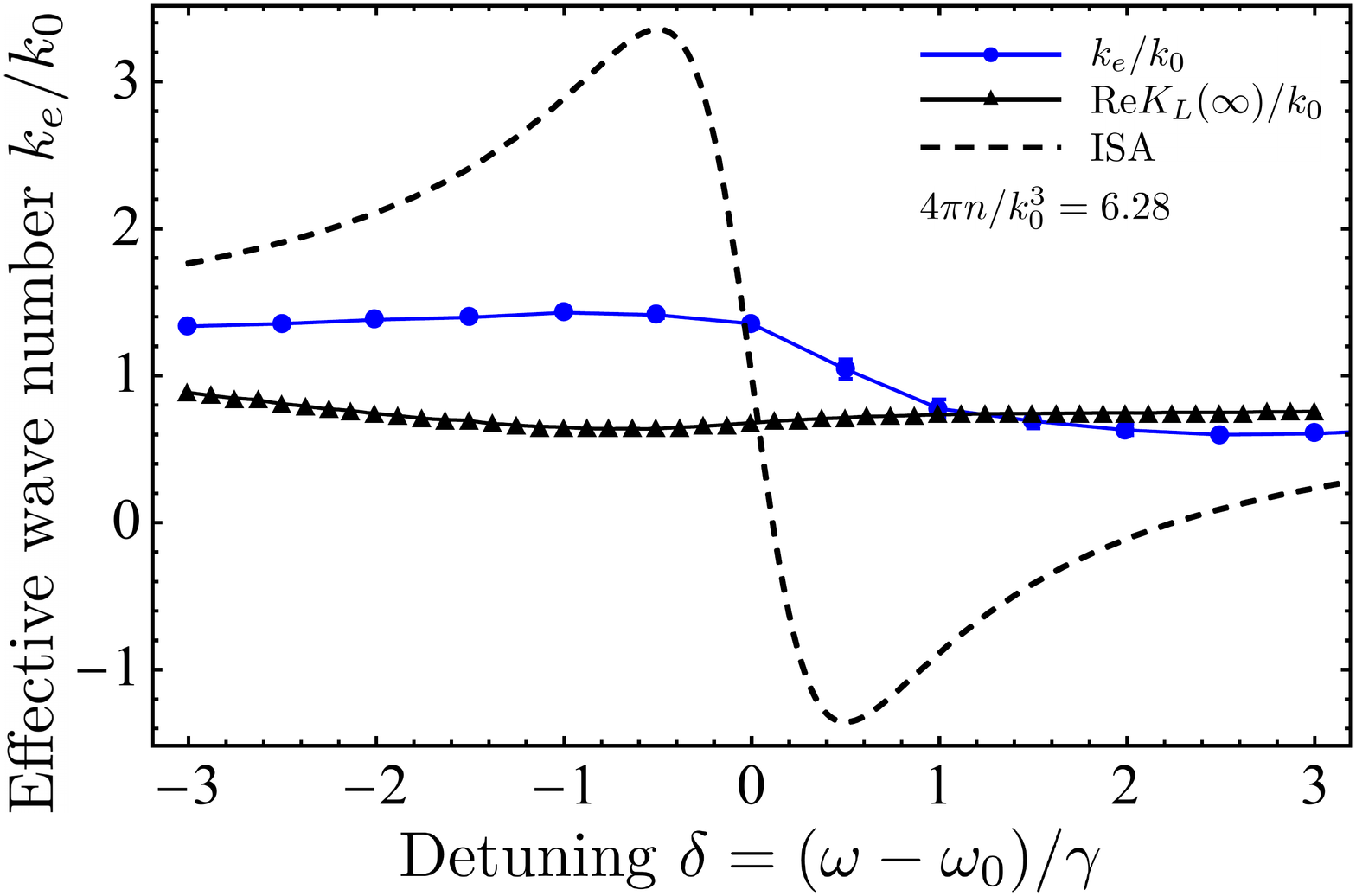}
\vspace*{-5mm}
\caption{\label{fig_keff}
The effective wave number obtained from the fit to the coherent field (see Fig.\ \ref{fig_field}), the real part of $K_L(\infty)$ calculated using Eq.\ (\ref{bewijsL}), and the ISA result.
}
\end{figure}

\begin{figure}
\includegraphics[width=0.7\columnwidth, angle=-90]{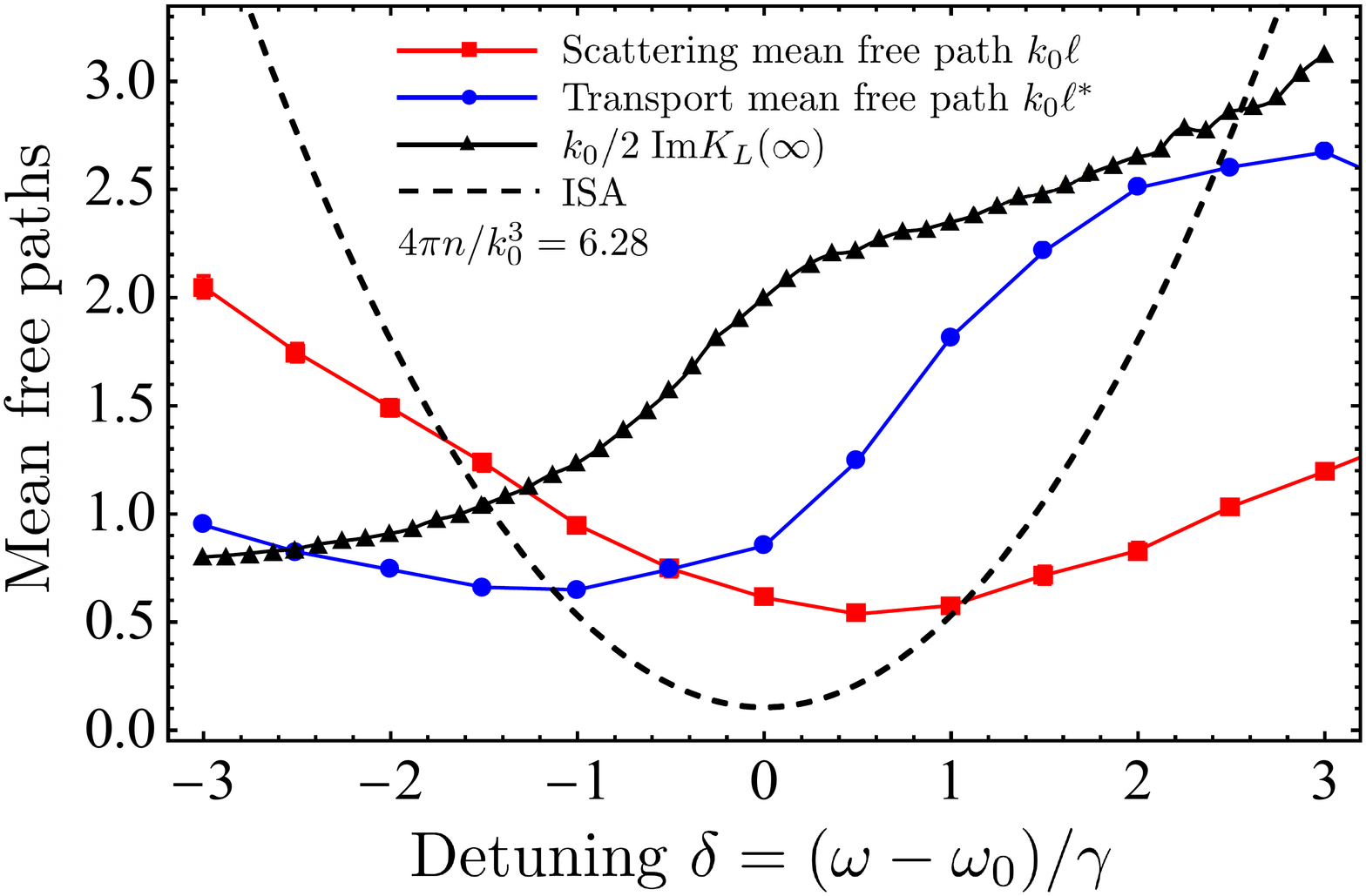}
\vspace*{-5mm}
\caption{\label{fig_tr}
Scattering and transport mean free paths for the most dense sample. We also show the scattering length $1/2\; \mathrm{Im} K_L(\infty)$ associated with longitudinal waves.
}
\end{figure}

To determine the scattering mean free path $\ell$ and the effective wave number $k_e$ of the transverse waves, we fit the results for the real part of $\langle E_x(k, z) \rangle$ to the expression
\begin{eqnarray}
\mathrm{Re} \langle E_x(k, z) \rangle &=& A \cos(k_e z + \phi ) \exp\left(-\frac{z}{2 \ell} \right)
\label{eqnmfp}
\end{eqnarray}
where $\ell$, $k_e$, $\phi$ and $A$ are free fit parameters (see Fig.\ \ref{fig_field}). In order to reduce the influence of boundary effects, we ignore the data corresponding to $z < L/10$ and $z > (9/10)L$ in the fits. The resulting effective wave number and scattering mean free path are shown in Figs.\ \ref{fig_keff} and \ref{fig_tr} (red line) as functions of detuning $\delta = (\omega-\omega_0)/\gamma$ for $n/k_0^3 = 0.5$.

\subsection{Diffuse field}

We compute the average energy density of light inside the sample $\tilde{\rho}(k, z)$ by averaging the square of the macroscopic magnetic field $\mathbf{\tilde{B}}(k, \mathbf{r}_m)$ on the scatterers. The diffuse energy density is obtained by subtracting  the coherent intensity:
\begin{eqnarray}
\tilde{\rho}(k, z) &=& \frac{c_0}{8\pi} \langle | \mathbf{\tilde{B}}(k, \mathbf{r}_m) |^2 \rangle
\label{eqnrho}
\\
\tilde{\rho}_{\mathrm{dif}} (k, z)  &=& \tilde{\rho}(k, z) - \frac{c_0}{8\pi} | \langle \mathbf{\tilde{B}}(k, \mathbf{r}_m) \rangle |^2
\label{eqnrhodif}
\end{eqnarray}
where, as previously, the averaging $\langle \ldots \rangle$ is done over scatterer configurations as well as over the central part of the cylindrical sample.  Because of equipartition, electric and magnetic energies should be equal on average.  Nevertheless, because we calculate $\tilde{\mathbf{B}}$, and not $\mathbf{B}$, the magnetic energy density still misses the singular stored energy inside the dipole.

\begin{figure}[t]
\includegraphics[width=0.7\columnwidth, angle=-90]{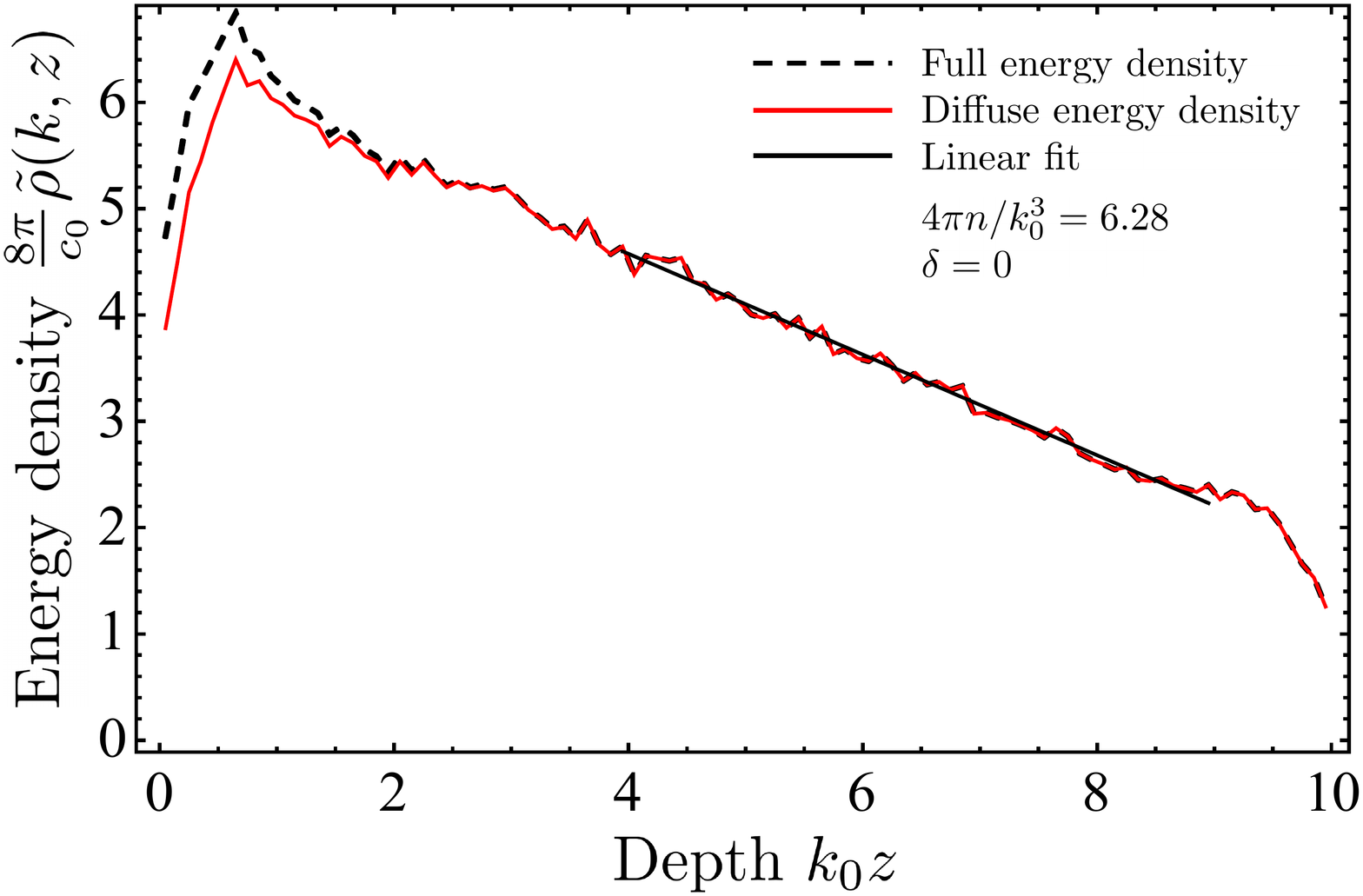}
\vspace*{-5mm}
\caption{\label{fig_energy}
The average energy density, consisting of a coherent and a diffuse part. The solid straight line is a linear fit to the diffuse
energy density for $k_0 z = 4$--9.
}
\end{figure}

\begin{figure}[t]
\includegraphics[width=0.7\columnwidth, angle=-90]{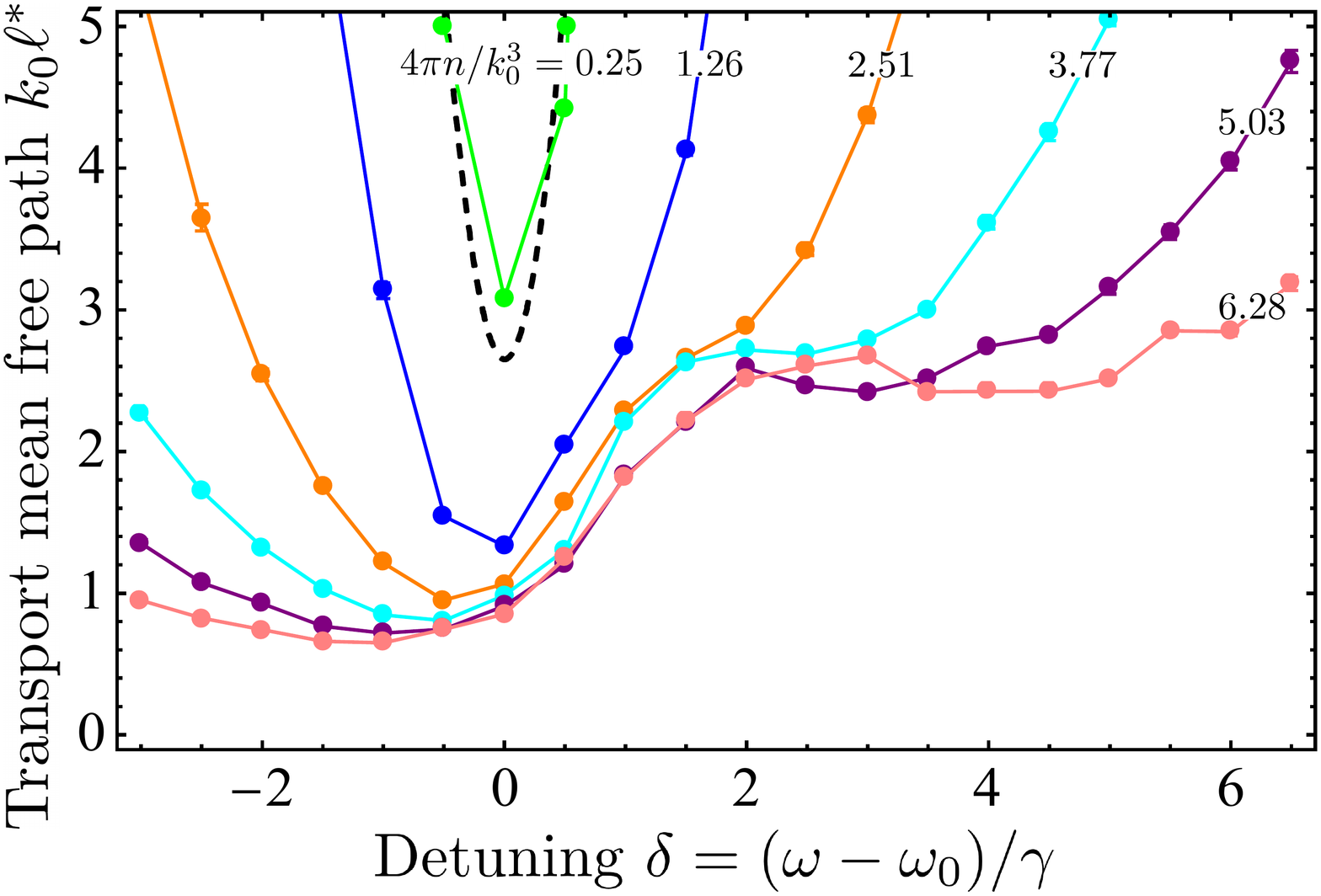}
\vspace*{-5mm}
\caption{\label{fig_tr_all}
Transport mean free paths for different densities $n$. The dashed line shows the ISA results for the lowest density $4 \pi n/k_0^3 = 0.25$.
}
\end{figure}

A typical profile of energy density inside the sample is shown in Fig.\ \ref{fig_energy}. We fit $\tilde{\rho}_{\mathrm{dif}}(k, z)$ by a linear function for $k_0 z$ between 4 and 9 to determine its gradient $d\tilde{\rho}_{\mathrm{dif}}(k, z)/dz$.

The average energy flux is given by the average Poynting vector
\begin{eqnarray}
\langle \mathbf{K}(k, z) \rangle &=& \frac{c_0}{8\pi } \mathrm{Re} \langle \mathbf{E}(k, \mathbf{r}) \times \mathbf{\bar{B}}(k, \mathbf{r}) \rangle
\label{eqnpoy}
\end{eqnarray}
We compute the $z$ component of $\langle \mathbf{K} \rangle$ outside scatterers using Eqs.\ (\ref{eqne})--(\ref{eqgb}) and extend the calculation to the space in front and behind
the sample. Inside the sample, the calculation fails to average because of the large fluctuations stemming from the near fields of the $M$ scatterers. However, the calculation converges  very well when averaging the Poynting vector calculated outside the sample. However, only in a slab of infinite lateral extent ($R \to \infty$) without lateral leakage, the energy flux $\langle K_z \rangle$ would be independent of $z$ and equal in- and outside the sample. For finite $R$, this equality  is only valid approximately.
To correct for this, we perform a linear fit of $\langle K_z(k, z) \rangle$ calculated at $z/L \in [-0.3, -0.1]$ in front of  the sample, and $z/L \in [1.1,1.3] $ just behind the sample, and use the fit to find the value of $\langle K_z(k, z) \rangle$ at $z/L = 0.65$ as the best estimate of $\langle K_z \rangle$ for the slab of infinite lateral extent. The point $z/L = 0.65 $ is chosen in the middle of the depth range where $\tilde{\rho}_{\mathrm{dif}}(k, z)$ is seen to exhibit a clear linear decay (see Fig.\ \ref{fig_energy}).

The transport mean free path $\ell^*$ is obtained by using the Fick's law
\begin{eqnarray}
\langle K_z \rangle = -D \frac{d}{dz} \tilde{\rho}_{\mathrm{dif}}(k, z)
\label{fick}
\end{eqnarray}
where $D = (c_0^2/v_p) \ell^*/3$ is the diffusion coefficient and $v_p = c_0 k_0/k_e$ is the phase velocity. No energy velocity appears here since $\tilde{ \rho}_{\mathrm{dif}}(k, z)$ does not count the stored energy. Expressing $\ell^*$ from this equation yields
\begin{eqnarray}
\ell^* &=& -3\frac{k_0}{k_e} \times \frac{\langle K_z \rangle}{d\tilde{\rho}_{\mathrm{dif}}(k, z)/dz}
\label{eqntr}
\end{eqnarray}
The results following from this equation are shown in Figs.\ \ref{fig_tr} (blue line) and \ref{fig_tr_all}. The comparison of these results with the analytic theory is presented in Fig.\ \ref{fig_comparison} and is discussed in the main text.

Figure\ \ref{fig_tr} shows that the transport mean free path differs significantly from any of the scattering mean free paths,  including the scattering length $1/2\; \mathrm{Im} K_L(\infty)$ associated with longitudinal waves. The transport mean free path is an asymmetric function of the detuning from the resonance, and it is larger for positive than for negative detunings.


\begin{thebibliography}{99}
\bibitem{LB} The electromagnetic conductivity is defined by the left hand side of Eq.~(\ref{Kubo}), in analogy with electric conductivity.
For a slab with depth $L$ the total average transmission of one channel $a$ is $\langle \sum_b T_{ab}\rangle = 4\ell^*/3L$.
The number of channels is equal to $N = 2 \times k_e^2A/4\pi$, with $A$ the lateral surface, $k_e$ the effective medium wave number,
 and the factor two stemming from the two polarizations. This gives
$\langle \sum_b T_{ab}\rangle = 4\pi N(k) D(k) A/L= 4 \sigma A/L$, with density of states $N(k) = k_e^2/2\pi^2 v_E$ and
diffusion constant $D(k) = \frac{1}{3} v_E \ell^*$ as discussed in the text.
\end{thebibliography}
\end{document}